\pdfoutput=1

\PassOptionsToPackage{final}{changes}

\documentclass[a4paper, 10pt]{article}

\newcommand{\papertitle}{
The Effectiveness of Strategies to Contain SARS-CoV-2: Testing, Vaccinations, and NPIs
}

\usepackage{ifxetex}  

\usepackage{ifthen}

\usepackage[utf8]{inputenc}  

\ifxetex
	\usepackage[euenc]{fontspec}
\else
	\usepackage[LGR, T1]{fontenc}  
\fi

\usepackage[ngerman, american, USenglish]{babel}  
\usepackage[ngerman, USenglish]{isodate}

\usepackage[babel, german=quotes]{csquotes}  

\usepackage{calc}  
\usepackage{fp}  

\usepackage{etoolbox}  
\usepackage{xpatch}  
\usepackage{letltxmacro}
\usepackage{xparse}

\usepackage{ragged2e}

\usepackage{import}	 

\usepackage{placeins}  

\usepackage{amsmath}
\MakeRobust{\eqref}
\renewcommand{\eqref}[1]{(\ref{#1})}  
\usepackage{amsthm}	
\usepackage{mathtools}
\newcommand{\succeqq}{%
  \mathrel{%
    \vcenter{\offinterlineskip
      \ialign{##\cr$\succ$\cr\noalign{\kern 1pt}$=$\cr}%
    }%
  }%
}

\usepackage[
	colorlinks=true,
	linkcolor=UBonnBlue,
	citecolor=UBonnBlue,
	filecolor=black,
	urlcolor=UBonnBlue,
	bookmarks=true,
	bookmarksnumbered=true,
	bookmarksopenlevel=2,
	pdfstartview=Fit,
	pdfpagelayout=SinglePage,
	plainpages=false,
	pdfpagelabels=true
]{hyperref}
\newcommand{\email}[1]{\href{mailto:#1}{\nolinkurl{#1}}}
\addto\extrasUSenglish{%
}
\addto\extrasamerican{%
}
\addto\extrasngerman{%
}

\usepackage{verbatim}

\usepackage{graphicx}
\usepackage{epstopdf}  

\usepackage[table]{xcolor}
\definecolor{UBonnBlue}   {RGB}{0007,0082,0154}
\definecolor{darkblue}    {rgb}{0.00,0.20,0.40}
\definecolor{darkred}     {rgb}{0.80,0.00,0.00}
\colorlet   {darkred25}   {darkred!25!white}
\definecolor{customgreen} {rgb}{0.15,0.55,0.00}
\definecolor{custompurple}{rgb}{0.15,0.00,0.75}

\usepackage{pgf, pgfarrows, pgfnodes, pgfshade}
\usepackage{pgfplots}

\usepackage{tikz}
\usetikzlibrary{mindmap, trees, patterns}

\usepackage{pdflscape}

\usepackage{afterpage}

\usepackage[full]{textcomp}  
\usepackage{xfrac}	

\usepackage{soul}  

\usepackage{relsize}

\ifxetex
	\renewcommand{\caps}[1]{\textscale{0.96}{\addfontfeature{LetterSpace=5}\MakeUppercase{#1}}}
\else
	\renewcommand{\caps}[1]{\textscale{0.96}{\textls[35]{\MakeUppercase{#1}}}}
\fi

\usepackage{printlen}  

\usepackage[math]{blindtext}
\makeatletter
\def\blindtext@american{}
\renewcommand{\blindmathpaper}{%
	\blindtext
	\blindtext@formula\par
	\blindtext
	\blindtext@formula
	\blindtext
	\blindtext@formula\par
	\blindtext
	\blindtext@formula
	\blindtext
	\blindtext@formula\par
	\blindtext\relax%
}
\makeatother
\LetLtxMacro{\blindtextblindtext}{\blindtext}
\LetLtxMacro{\blindtextBlindlist}{\Blindlist}
\LetLtxMacro{\blindtextBlindtext}{\Blindtext}
\RenewDocumentCommand{\blindtext}{O{\value{blindtext}}}{%
	\begingroup%
	\iflanguage{USenglish}{\selectlanguage{american}}{}%
	\blindtextblindtext[#1]%
	\endgroup%
}
\RenewDocumentCommand{\Blindtext}{O{\value{blindtext}} O{\value{Blindtext}}}{%
	\begingroup%
	\iflanguage{USenglish}{\selectlanguage{american}}{}%
	\blindtextBlindtext[#1][#2]%
	\endgroup%
}
\RenewDocumentCommand{\Blindlist}{m O{\value{blindlist}}}{%
	\begingroup%
	\iflanguage{USenglish}{\selectlanguage{american}}{}%
	\blindtextBlindlist{#1}[#2]%
	\endgroup%
}

\makeatletter
\newcommand{\showfont}{{%
	\color{magenta}
	\textit{Encoding:} \f@encoding{},
	\textit{family:}   \f@family{},
	\textit{series:}   \f@series{},
	\textit{shape:}    \f@shape{},
	\textit{size:}     \f@size{}.
}}
\newcommand{\showfamily}{\f@family{}}
\makeatother

\makeatletter
\newcommand*{\checkgreekletters}{%
	\@for\@tempa:=%
	alpha,beta,gamma,delta,epsilon,varepsilon,zeta,eta,theta,vartheta,iota,kappa,lambda,mu,nu,xi,%
	omicron,pi,varpi,rho,varrho,sigma,varsigma,tau,upsilon,phi,varphi,chi,psi,omega,digamma,%
	Alpha,Beta,Gamma,Delta,Epsilon,Zeta,Eta,Theta,Iota,Kappa,Lambda,Mu,Nu,Xi,%
	Omicron,Pi,Rho,Sigma,Tau,Upsilon,Phi,Chi,Psi,Omega,Digamma%
	\do{$\csname\@tempa\endcsname,$ }%
}
\makeatother

\usepackage{fonttable}
\makeatletter
\renewcommand*\f@placedecimal[2]{#1\ {\color{gray}\tiny #2}}
\renewcommand*{\f@toct}[1]{\hbox{\color{gray}\rmfamily\'{}\kern-.2em\itshape#1\/\kern.05em}} 
\renewcommand*{\f@thex}[1]{\hbox{\color{gray}\rmfamily\H{}\ttfamily#1}} 
\makeatother

\hypersetup{
    pdftitle={\papertitle},
    pdfauthor={Janoś Gabler, Tobias Raabe, Klara Röhrl, Hans-Martin von Gaudecker}
}

\newlength{\origbaselineskip}
\ifdef{\linesperpagedesired}
	{}
	{\newcommand{\linesperpagedesired}{42}}

\makeatletter
\newcommand{\newbaselinestretch}{1.1}
\makeatother
\newlength{\newbaselineskip}
\newlength{\newparindent}
\AtBeginDocument{%
	\setlength{\baselineskip}{\newbaselineskip}%
	\setlength{\parindent}{\newparindent}
	\setlength{\lineskiplimit}{0pt}
}
\newlength{\abovedisplayauxskip}
\newlength{\belowdisplayauxskip}
\newcommand{\predisplaycmd}{%
	\ifvmode\else\unskip\fi%
	\nopagebreak[2]%
	\vspace{\abovedisplayauxskip}%
}
\makeatletter
\def\@itemize@name{itemize}
\def\@enumerate@name{enumerate}
\def\@description@name{description}
\newcommand{\postdisplaycmd}{%
	\ifx\@currenvir\@itemize@name%
	\else%
		\ifx\@currenvir\@enumerate@name%
		\else%
			\ifx\@currenvir\@description@name%
			\else%
				\vskip\belowdisplayauxskip%
			\fi%
		\fi%
	\fi%
	\noindent%
}
\makeatother
\AtBeginEnvironment {align}      {\predisplaycmd}
\AfterEndEnvironment{align}      {\postdisplaycmd}
\AtBeginEnvironment {align*}     {\predisplaycmd}
\AfterEndEnvironment{align*}     {\postdisplaycmd}
\AtBeginEnvironment {alignat}    {\predisplaycmd}
\AfterEndEnvironment{alignat}    {\postdisplaycmd}
\AtBeginEnvironment {alignat*}   {\predisplaycmd}
\AfterEndEnvironment{alignat*}   {\postdisplaycmd}
\AtBeginEnvironment {displaymath}{\predisplaycmd}
\AfterEndEnvironment{displaymath}{\postdisplaycmd}
\AtBeginEnvironment {eqnarray}   {\predisplaycmd}
\AfterEndEnvironment{eqnarray}   {\postdisplaycmd}
\AtBeginEnvironment {eqnarray*}  {\predisplaycmd}
\AfterEndEnvironment{eqnarray*}  {\postdisplaycmd}
\AtBeginEnvironment {equation}   {\predisplaycmd}
\AfterEndEnvironment{equation}   {\postdisplaycmd}
\AtBeginEnvironment {equation*}  {\predisplaycmd}
\AfterEndEnvironment{equation*}  {\postdisplaycmd}
\AtBeginEnvironment {flalign}    {\predisplaycmd}
\AfterEndEnvironment{flalign}    {\postdisplaycmd}
\AtBeginEnvironment {flalign*}   {\predisplaycmd}
\AfterEndEnvironment{flalign*}   {\postdisplaycmd}
\AtBeginEnvironment {gather}     {\predisplaycmd}
\AfterEndEnvironment{gather}     {\postdisplaycmd}
\AtBeginEnvironment {gather*}    {\predisplaycmd}
\AfterEndEnvironment{gather*}    {\postdisplaycmd}
\AtBeginEnvironment {multiline}  {\predisplaycmd}
\AfterEndEnvironment{multiline}  {\postdisplaycmd}
\AtBeginEnvironment {multiline*} {\predisplaycmd}
\AfterEndEnvironment{multiline*} {\postdisplaycmd}

\newcommand{\thickmu}  {\thickmuskip=10mu \medmuskip=5mu}
\AtBeginEnvironment{align}      {\thickmu}
\AtBeginEnvironment{align*}     {\thickmu}
\AtBeginEnvironment{alignat}    {\thickmu}
\AtBeginEnvironment{alignat*}   {\thickmu}
\AtBeginEnvironment{displaymath}{\thickmu}
\AtBeginEnvironment{eqnarray}   {\thickmu}
\AtBeginEnvironment{eqnarray*}  {\thickmu}
\AtBeginEnvironment{equation}   {\thickmu}
\AtBeginEnvironment{equation*}  {\thickmu}
\AtBeginEnvironment{flalign}    {\thickmu}
\AtBeginEnvironment{flalign*}   {\thickmu}
\AtBeginEnvironment{gather}     {\thickmu}
\AtBeginEnvironment{gather*}    {\thickmu}
\AtBeginEnvironment{multiline}  {\thickmu}
\AtBeginEnvironment{multiline*} {\thickmu}

\usepackage{bm}

\makeatletter
\predisplaypenalty=\@medpenalty
\makeatother

\usepackage[all]{nowidow}

\AtBeginDocument{%
	
	\frenchspacing
	\sloppy
}

\ifxetex
	\usepackage[protrusion=true, expansion=false]{microtype}
\else
	\usepackage[protrusion=true, expansion=false, kerning=true]{microtype}
\fi

\spaceskip=\fontdimen2\font plus \fontdimen3\font minus 0.75\fontdimen4\font

  {\list{}{\leftmargin=\parindent \rightmargin=\parindent}%
   \linespread{\newbaselinestretch}\item[]\itshape\small}%
  {\endlist}
  {\list{}{\leftmargin=\parindent \rightmargin=\parindent
           \listparindent=\parindent \parsep=0pt}%
   \item[]}%
  {\endlist}

\usepackage{csquotes}

\usepackage{nameref}
\makeatletter
\newcommand*{\currentname}{\@currentlabelname}
\makeatother

\usepackage[multiple, bottom, norule, splitrule, marginal]{footmisc}
\preto{\footnote}{\setlength{\parindent}{\newparindent}}
\let\oldfootnoterule\footnoterule
\AtBeginDocument{%
	\renewcommand{\footnoterule}{\oldfootnoterule\medskip}%
}

\usepackage{xstring}
\newlength{\textparindent}
\newlength{\templength}
\makeatletter
\let \@makefntextorig \@makefntext
\newcommand{\@makefntextcustom}[1]{%
	\parindent 2\textparindent%
	\hspace{-\textparindent}%
	\settowidth{\templength}{0}%
	\ifnum\value{footnote}<10 \hspace{\templength}\else\fi%
	\thefootnote.\enskip #1%
}
\renewcommand{\@makefntext}[1]{\@makefntextcustom{#1}}
\makeatother

\DefineFNsymbols*{star}{%
	{$\mathrm{\star}$}{$\mathrm{\star\star}$}{$\mathrm{\ddagger}$}%
	{$\mathrm{\ddagger\ddagger}$}{\S}{\S\S}{\P}{\P\P}{$\mathrm{\|}$}{$\mathrm{\|\|}$}%
}
\setfnsymbol{star}

\usepackage[singlelinecheck=on]{caption}
\DeclareCaptionLabelSeparator{periodlargespace}{.\:\:}
\usepackage{subcaption}

\usepackage{booktabs}
\usepackage{tabularx}	
\newcolumntype{C}{>{\centering\arraybackslash}X}
\newcolumntype{J}{>{\arraybackslash}X}
\newcolumntype{L}{>{\RaggedRight\arraybackslash}X}
\newcolumntype{R}{>{\RaggedLeft\arraybackslash}X}

\LetLtxMacro{\oldtabular}{\tabular}
\LetLtxMacro{\endoldtabular}{\endtabular}
\RenewDocumentEnvironment{tabular}{O{c} m}{%
	\oldtabular[#1]{@{}#2@{}}%
}{%
	\endoldtabular%
}
\LetLtxMacro{\oldtabularx}{\tabularx}
\LetLtxMacro{\endoldtabularx}{\endtabularx}
\RenewDocumentEnvironment{tabularx}{m O{c} m}{%
	\oldtabularx{#1}[#2]{@{}#3@{}}%
}{%
	\endoldtabularx%
}

\usepackage{siunitx}
\newcolumntype{T}[1]{@{}S[table-format = #1, table-space-text-pre = {***}, table-space-text-post = {***}]}
\ExplSyntaxOn\makeatletter
\newcommand{\thisseries}{\f@series}
\cs_set_protected:Npn \__siunitx_detect_font_weight_text: {%
	\let\origmdseries\mdseries@sf%
	\let\origbfseries\bfseries@sf%
	\let\currentseries\f@series%
	\edef\XcurrentseriesX{/\f@series/}%
	\tl_if_in:noTF
		{ /sb/ /b/ /bx/ /eb/ /ub/ /bold/ /extrabold/ /ultrabold/ /heavy/ /black/ /demibold/ /semibold/ }
		{ \XcurrentseriesX }
		{
			\cs_set:Nn \__siunitx_font_weight: {%
				\boldmath%
				\let\bfseries@sf\currentseries%
				\fontseries{\bfseries@sf}\selectfont%
			}%
			\let\bfseries@sf\origbfseries
		}
		{
			\cs_set:Nn \__siunitx_font_weight: {%
				\unboldmath
				\let\mdseries@sf\currentseries%
				\fontseries{\mdseries@sf}\selectfont%
			}%
			\let\mdseries@sf\origmdseries
		}%
}
\makeatother\ExplSyntaxOff

\usepackage[restart, breakwithin, indentafter]{parnotes}

\usepackage{makecell}

\usepackage{longtable}

\usepackage{multirow}

\makeatletter
\let\oldfigure\figure
\def\figure{\@ifnextchar[\figure@i \figure@ii}
\def\figure@i[#1]{\oldfigure[#1]\centering}
\def\figure@ii{\oldfigure\centering}
\makeatother

\usepackage[font={sf, small}]{floatrow}	
\usepackage{placeins} 

\usepackage{verbatim}

\usepackage{longtable}

\usepackage[inline]{enumitem}
\setlist{leftmargin=\parindent, listparindent=\parindent, itemsep=\smallskipamount, parsep=0pt}
\setlist[enumerate]{leftmargin=\parindent, labelsep=*}
\setlist[enumerate, 1]{label=(\arabic*), labelindent=-0.5pt}
\setlist[enumerate, 2]{label=\alph*., align=right}
\setlist[enumerate, 3]{label=\roman*., align=right, widest*=3, labelsep=0.3\parindent}
\setlist[itemize]{labelsep=0.435\parindent}
\setlist[description]{font=\rmfamily\normalsize}

\newtheoremstyle{Standard}
	{\topsep}    
	{\topsep}    
	{\itshape}   
	{}           
	{\bfseries}  
	{.}          
	{.5em}       
	{\thmname{#1}\thmnumber{\:#2}\thmnote{\bfseries\upshape\ (#3)}}

\theoremstyle{Standard}

\theoremstyle{definition}

\makeatletter
\@ifclassloaded{book}{%
	\numberwithin{theorem}{chapter}%
	\numberwithin{hypothesis}{chapter}%
	\numberwithin{result}{chapter}%
}
\makeatother

\usepackage{mfirstuc-english}
\gMFUnocap{about}
\gMFUnocap{at}
\gMFUnocap{against}
\gMFUnocap{around}
\gMFUnocap{between}
\gMFUnocap{by}
\gMFUnocap{from}
\gMFUnocap{on}
\gMFUnocap{over}
\gMFUnocap{per}
\gMFUnocap{to}
\gMFUnocap{versus}
\gMFUnocap{vs.}
\gMFUnocap{vis-\`a-vis}
\gMFUnocap{within}
\gMFUnocap{without}

\ifcase 0
	\renewcommand{\capitalisewords}[1]{#1}
	
	\renewcommand{\xcapitalisewords}[1]{#1}
\or
	\let\SavedContentsline\contentsline
	\renewcommand{\contentsline}[4]{%
		\SavedContentsline{#1}{\capitalisewords{#2}}{#3}{#4}%
	}
\else
	\let\SavedContentsline\contentsline
	\renewcommand{\contentsline}[4]{%
		\SavedContentsline{#1}{\capitalisewords{#2}}{#3}{#4}%
	}
	\LetLtxMacro{\SavedCaption}{\caption}
	\RenewDocumentCommand{\caption}{ O{\shortcaption} m }{%
		\def\shortcaption{%
			\xcapitalisewords{%
				#2%
			}%
		}%
		\SavedCaption[#1]{%
			\xcapitalisewords{%
				#2%
			}%
		}%
	}
\fi

\usepackage[title, titletoc]{appendix}  
\usepackage{chngcntr}  

\AtBeginEnvironment{appendices}{%
	\counterwithin{figure}{section}%
	\counterwithin{table}{section}%
	\counterwithin{equation}{section}%
}
\AtBeginEnvironment{subappendices}{%
	\counterwithin{figure}{section}%
	\counterwithin{table}{section}%
	\counterwithin{equation}{section}%
}

\makeatletter
\@ifclassloaded{book}{%
	\AfterEndEnvironment{appendices}{%
		\counterwithout{figure}{section}%
		\counterwithin{figure}{chapter}%
		\counterwithout{table}{section}%
		\counterwithin{table}{chapter}%
		\counterwithout{equation}{section}%
		\counterwithin{equation}{chapter}%
	}%
	\AfterEndEnvironment{subappendices}{%
		\counterwithout{figure}{section}%
		\counterwithin{figure}{chapter}%
		\counterwithout{table}{section}%
		\counterwithin{table}{chapter}%
		\counterwithout{equation}{section}%
		\counterwithin{equation}{chapter}%
	}%
}{}
\makeatother

\usepackage[sf, bf, raggedright, explicit]{titlesec}	

\let\dateorig\date
\renewcommand{\date}[1]{\dateorig{\small #1}}

\usepackage[style]{abstract}
\renewcommand{\abstitlestyle}[1]{\addcontentsline{toc}{section}{\abstractname}}

\usepackage[max6]{authblk}

\makeatletter
\renewcommand\AB@authnote[1]{\textsuperscript{{\kern.5pt}\textit{#1}}}
\renewcommand\AB@affilnote[1]{\textsuperscript{\textit{#1}}\,}
\makeatother

\newcommand{\appendixformat}[1][\currentname]{%
	\titleformat{\section}[hang]%
		{}{\textscale{1.3}{\textsf{\textbf{\appendixname~\thesection:\quad}}}}{0pt}{\textscale{1.3}{\textsf{\textbf{#1}}}}[]%
}
\AtBeginEnvironment{appendices}{\appendixformat}

\titleformat{name=\section}[hang]%
	{}{\textscale{1.3}{\textsf{\textbf{\thesection\quad}}}}{0pt}{\textscale{1.3}{\textsf{\textbf{#1}}}}[]
\titleformat{name=\section, numberless}[hang]%
	{}{}{0pt}{\textscale{1.3}{\textsf{\textbf{#1}}}}[]
\titleformat{name=\subsection}[hang]%
	{}{\textscale{1.15}{\textsf{\textbf{\thesubsection\quad}}}}{0pt}{\textscale{1.15}{\textsf{\textbf{#1}}}}[]
\titleformat{name=\subsection, numberless}[hang]%
	{}{}{0pt}{\textscale{1.15}{\textsf{\textbf{#1}}}}[]
\titleformat{name=\subsubsection}[runin]%
	{}{}{0pt}{\textsf{\textbf{\thesubsubsection\quad #1\iftextterm{#1}{}{\unskip.}}}}[]
\titleformat{name=\subsubsection, numberless}[runin]%
	{}{}{0pt}{\textsf{\textbf{#1\iftextterm{#1}{}{\unskip.}}}}[]
\titleformat{name=\paragraph, numberless}[runin]%
	{}{}{0pt}{\textbf{#1\iftextterm{#1}{}{\unskip.}}}[]
\titleformat{name=\subparagraph, numberless}[runin]%
	{}{}{0pt}{\textbf{#1\iftextterm{#1}{}{\unskip.}}}[]

\titlespacing*{\section}
	{0pt}{2\baselineskip plus 0.67\baselineskip}{1\baselineskip plus 0.5\baselineskip minus 0.33\baselineskip}
\titlespacing*{\subsection}%
	{0pt}{1.33\baselineskip plus 0.5\baselineskip minus 0.33\baselineskip}{0.67\baselineskip plus 0.33\baselineskip minus 0.17\baselineskip}
\titlespacing{\subsubsection}%
	{0pt}{1\baselineskip plus 0.67\baselineskip minus 0.33\baselineskip}{4\wordsep}[]
\titlespacing{\paragraph}
	{0pt}{0.5\baselineskip plus 0.5\baselineskip}{3\wordsep}[]
\titlespacing{\subparagraph}
	{\parindent}{0pt}{3\wordsep}[]

\usepackage[lining, scale=0.88]{FiraMono}  
\usepackage[book, semibold, lining, tabular, scale=0.88]{FiraSans}[2018/05/23]
\makeatletter
\@ifpackagelater{FiraSans}{2018/05/23}
	{}
	{\PackageError{FiraSans}
	     {Outdated 'FiraSans' package}
	     {Upgrade to FiraSans 2018/05/23 or newer!\MessageBreak
	      Otherwise the sans-serif font may not work properly, or the compilation might fail.\MessageBreak
	      This is a fatal error. I'm aborting now.}%
   }
\makeatother
\ifthenelse{\isundefined{\AfterPackage}}{\usepackage{afterpackage}}{}
\AfterPackage{siunitx}{%
  
}

\newcommand{\savesffamily}{\sfdefault}
\makeatletter
\newcommand{\savesfmdseries}{\mdseries@sf}
\newcommand{\savesfbfseries}{\bfseries@sf}
\makeatother

\usepackage[charter, greeklowercase=italicized, greekuppercase=italicized, sfscaled=false, ttscaled=false, euro=false]{mathdesign}
\let \savermfamily   \rmdefault
\let \savermmdseries \mddefault
\let \savermbfseries \bfdefault

\makeatletter
\@for\@tempa:=%
	alpha,beta,gamma,delta,epsilon,zeta,eta,theta,iota,kappa,lambda,mu,nu,xi,%
	omicron,pi,rho,sigma,varsigma,tau,upsilon,phi,chi,psi,omega,digamma,%
	Alpha,Beta,Gamma,Delta,Epsilon,Zeta,Eta,Theta,Iota,Kappa,Lambda,Mu,Nu,Xi,%
	Omicron,Pi,Rho,Sigma,Tau,Upsilon,Phi,Chi,Psi,Omega,Digamma%
	\do{%
		\expandafter\let\csname up\@tempa\expandafter\endcsname\csname\@tempa up\endcsname%
	}%
\makeatother

\ifx\nequiv\undefined
	\newcommand{\nequiv}{\not\equiv}
\fi

\usepackage[scaled=.96, lining, sups]{XCharter}  

\renewcommand{\textellipsis}{\mbox{.{\kern.09em}.{\kern.09em}.}}
\makeatletter
\newcommand{\@makefnmarkorig}{%
	\hbox{\sufigures\hspace*{.04em}\@thefnmark\hspace*{.04em}}%
}  
\makeatother

\ifxetex
\else
	\DisableLigatures[f]{family = {rm*, tt*}}
	\SetExtraKerning[unit=space]
		{encoding=*, family=*, series=*, size={*, normalsize, footnotesize}, font = */*/*/*/*}
		{\textemdash = {325, 325},
		           / = {100, 100},
		           : = { 50,   0},
		           ; = { 50,   0}}
\fi

\usepackage{xfrac}	

\AtBeginDocument{%
	\providecommand{\euro}[1]{\relax\ifmmode\text{\texteuro}#1\else\texteuro #1\fi}%
}

\makeatletter

\def\mweights@init{%
\ifdefined\mdseries@rm\else\edef\mdseries@rm{\mddefault}\fi
\ifdefined\bfseries@rm\else\edef\bfseries@rm{\bfdefault}\fi
\ifdefined\mdseries@sf\else\edef\mdseries@sf{\mddefault}\fi
\ifdefined\bfseries@sf\else\edef\bfseries@sf{\bfdefault}\fi
\ifdefined\mdseries@tt\else\edef\mdseries@tt{\mddefault}\fi
\ifdefined\bfseries@tt\else\edef\bfseries@tt{\bfdefault}\fi
\edef\rmdef@ult{\rmdefault}%
\edef\sfdef@ult{\sfdefault}%
\edef\ttdef@ult{\ttdefault}%
\edef\bfdef@ult{\bfdefault}%
\edef\mddef@ult{\mddefault}%
\edef\famdef@ult{\familydefault}%
}

\DeclareRobustCommand\bfseries{%
\mweights@init
\not@math@alphabet\bfseries\mathbf
\ifx\f@family\rmdef@ult\fontseries\bfseries@rm
\else\ifx\f@family\sfdef@ult\fontseries\bfseries@sf
\else\ifx\f@family\ttdef@ult\fontseries\bfseries@tt
\else\fontseries\bfdefault\fi\fi\fi\selectfont}%

\DeclareRobustCommand\mdseries{%
\mweights@init
\not@math@alphabet\mdseries\relax
\ifx\f@family\rmdef@ult\fontseries\mdseries@rm
\else\ifx\f@family\sfdef@ult\fontseries\mdseries@sf
\else\ifx\f@family\ttdef@ult\fontseries\mdseries@tt
\else\fontseries\mddefault\fi\fi\fi\selectfont}

\DeclareRobustCommand\rmfamily{%
\mweights@init
\not@math@alphabet\rmfamily\mathrm
\ifx\f@family\sfdef@ult
    \ifx\f@series\mdseries@sf\fontseries\mdseries@rm
    \else\ifx\f@series\bfseries@sf\fontseries\bfseries@rm
    \else\ifx\f@series\mddef@ult\fontseries\mdseries@rm
    \else\ifx\f@series\bfdef@ult\fontseries\bfseries@rm
    \else\fontseries\mdseries@rm
    \fi\fi\fi\fi
\else\ifx\f@family\ttdef@ult
    \ifx\f@series\mdseries@tt\fontseries\mdseries@rm
    \else\ifx\f@series\bfseries@tt\fontseries\bfseries@rm
    \else\ifx\f@series\mddef@ult\fontseries\mdseries@rm
    \else\ifx\f@series\bfdef@ult\fontseries\bfseries@rm
    \else\fontseries\mdseries@rm
    \fi\fi\fi\fi
\else\ifx\f@family\rmdef@ult
    \ifx\f@series\mdseries@rm\fontseries\mdseries@rm
    \else\ifx\f@series\bfseries@rm\fontseries\bfseries@rm
    \else\ifx\f@series\mddef@ult\fontseries\mdseries@rm
    \else\ifx\f@series\bfdef@ult\fontseries\bfseries@rm
    \else\fontseries\mdseries@rm
    \fi\fi\fi\fi
\else\fontseries\mdseries@rm
\fi\fi\fi\fontfamily\rmdefault\selectfont}

\DeclareRobustCommand\sffamily{%
\mweights@init
\not@math@alphabet\sffamily\mathsf
\ifx\f@family\rmdef@ult
    \ifx\f@series\mdseries@rm\fontseries\mdseries@sf
    \else\ifx\f@series\bfseries@rm\fontseries\bfseries@sf
    \else\ifx\f@series\mddef@ult\fontseries\mdseries@sf
    \else\ifx\f@series\bfdef@ult\fontseries\bfseries@sf
    \else\fontseries\mdseries@sf
    \fi\fi\fi\fi
\else\ifx\f@family\ttdef@ult
    \ifx\f@series\mdseries@tt\fontseries\mdseries@sf
    \else\ifx\f@series\bfseries@tt\fontseries\bfseries@sf
    \else\ifx\f@series\mddef@ult\fontseries\mdseries@sf
    \else\ifx\f@series\bfdef@ult\fontseries\bfseries@sf
    \else\fontseries\mdseries@sf
    \fi\fi\fi\fi
\else\ifx\f@family\sfdef@ult
    \ifx\f@series\mdseries@sf\fontseries\mdseries@sf
    \else\ifx\f@series\bfseries@sf\fontseries\bfseries@sf
    \else\ifx\f@series\mddef@ult\fontseries\mdseries@sf
    \else\ifx\f@series\bfdef@ult\fontseries\bfseries@sf
    \else\fontseries\mdseries@sf
    \fi\fi\fi\fi
\else\fontseries\mdseries@sf
\fi\fi\fi\fontfamily\sfdefault\selectfont}

\DeclareRobustCommand\ttfamily{%
\mweights@init
\not@math@alphabet\ttfamily\mathtt
\ifx\f@family\rmdef@ult
    \ifx\f@series\mdseries@rm\fontseries\mdseries@tt
    \else\ifx\f@series\bfseries@rm\fontseries\bfseries@tt
    \else\ifx\f@series\mddef@ult\fontseries\mdseries@tt
    \else\ifx\f@series\bfdef@ult\fontseries\bfseries@tt
    \else\fontseries\mdseries@tt
    \fi\fi\fi\fi
\else\ifx\f@family\sfdef@ult
    \ifx\f@series\mdseries@sf\fontseries\mdseries@tt
    \else\ifx\f@series\bfseries@sf\fontseries\bfseries@tt
    \else\ifx\f@series\mddef@ult\fontseries\mdseries@tt
    \else\ifx\f@series\bfdef@ult\fontseries\bfseries@tt
    \else\fontseries\mdseries@tt
    \fi\fi\fi\fi
\else\ifx\f@family\ttdef@ult
    \ifx\f@series\mdseries@tt\fontseries\mdseries@tt
    \else\ifx\f@series\bfseries@tt\fontseries\bfseries@tt
    \else\ifx\f@series\mddef@ult\fontseries\mdseries@tt
    \else\ifx\f@series\bfdef@ult\fontseries\bfseries@tt
    \else\fontseries\mdseries@tt
    \fi\fi\fi\fi
\else\fontseries\mdseries@tt
\fi\fi\fi\fontfamily\ttdefault\selectfont}

\AtBeginDocument{\mweights@init
\ifx\famdef@ult\rmdef@ult\rmfamily
\else\ifx\famdef@ult\sfdef@ult\sffamily
\else\ifx\famdef@ult\ttdef@ult\ttfamily
\fi\fi\fi}

\makeatother

\makeatletter
\newif\ifkp@upRm
\newif\ifkp@osm
\newif\ifkp@vosm
\makeatother

\DeclareMathVersion{normalup}
\DeclareMathVersion{boldup}
\DeclareMathVersion{sans}

\SetSymbolFont{operators}   {sans}{OT1}{mdbch}{m}{n}
\SetSymbolFont{letters}     {sans}{OML}{jkpss}{m}{it}
\SetSymbolFont{symbols}     {sans}{OMS}{jkp}  {m}{n}
\DeclareSymbolFont{extrasymbols}  {OMS}{cmbrs}{m}{n}
\SetSymbolFont{extrasymbols}{sans}{OMS}{cmbrs}{m}{n}

\SetMathAlphabet{\mathit} {sans}{T1}{\savesffamily}{\savesfmdseries}{it}
\SetMathAlphabet{\mathbf} {sans}{T1}{\savesffamily}{\savesfbfseries}{n}
\SetMathAlphabet{\mathtt} {sans}{OT1}{cmtl}{m}{n}
\SetMathAlphabet{\mathcal}{sans}{OMS}{ntxsy}{m}{n}
\SetSymbolFont{largesymbols} {sans}{OMX}{mdbch}{m}{n}

\DeclareMathVersion{sansup}
\SetSymbolFont{letters}  {sansup}{OML}{jkpss}{m}{it}
\SetSymbolFont{symbols}  {sansup}{OMS}{jkp}  {m}{n}

\DeclareMathVersion{boldsans}
\SetSymbolFont{operators}{boldsans}{OT1}{mdbch}{bx}{n}
\SetSymbolFont{letters}  {boldsans}{OML}{jkpss}{bx}{it}
\SetSymbolFont{symbols}  {boldsans}{OMS}{jkp}  {bx}{n}
\SetMathAlphabet{\mathit} {boldsans}{T1}{\savesffamily}{\savesfbfseries}{it}
\SetMathAlphabet{\mathtt} {boldsans}{T1}{cmtl}{b}{n}
\SetMathAlphabet{\mathcal}{boldsans}{OMS}{ntxsy}{b}{n}
\SetSymbolFont{largesymbols}{boldsans}{OMX}{mdbch}{bx}{n}

\DeclareMathVersion{boldsansup}
\SetSymbolFont{letters}{boldsansup}{OML}{jkpss}{bx}{it}
\SetSymbolFont{symbols}{boldsansup}{OMS}{jkp}  {bx}{n}

\DeclareSymbolFont{uprightglyphs}{T1}{\savermfamily}{\savermmdseries}{n}
\SetSymbolFont{uprightglyphs}{normal}    {T1}{\savermfamily}{\savermmdseries}{n}
\SetSymbolFont{uprightglyphs}{normalup}  {T1}{\savermfamily}{\savermmdseries}{n}
\SetSymbolFont{uprightglyphs}{bold}      {T1}{\savermfamily}{\savermbfseries}{n}
\SetSymbolFont{uprightglyphs}{boldup}    {T1}{\savermfamily}{\savermbfseries}{n}
\SetSymbolFont{uprightglyphs}{sans}      {T1}{\savesffamily}{\savesfmdseries}{n}
\SetSymbolFont{uprightglyphs}{sansup}    {T1}{\savesffamily}{\savesfmdseries}{n}
\SetSymbolFont{uprightglyphs}{boldsans}  {T1}{\savesffamily}{\savesfbfseries}{n}
\SetSymbolFont{uprightglyphs}{boldsansup}{T1}{\savesffamily}{\savesfbfseries}{n}
\DeclareSymbolFont{italicglyphs} {T1}{\savermfamily}{\savermmdseries}{it}
\SetSymbolFont{italicglyphs} {normal}    {T1}{\savermfamily}{\savermmdseries}{it}
\SetSymbolFont{italicglyphs} {normalup}  {T1}{\savermfamily}{\savermmdseries}{n}
\SetSymbolFont{italicglyphs} {bold}      {T1}{\savermfamily}{\savermbfseries}{it}
\SetSymbolFont{italicglyphs} {boldup}    {T1}{\savermfamily}{\savermbfseries}{n}
\SetSymbolFont{italicglyphs} {sans}      {T1}{\savesffamily}{\savesfmdseries}{it}
\SetSymbolFont{italicglyphs} {sansup}    {T1}{\savesffamily}{\savesfmdseries}{n}
\SetSymbolFont{italicglyphs} {boldsans}  {T1}{\savesffamily}{\savesfbfseries}{it}
\SetSymbolFont{italicglyphs} {boldsansup}{T1}{\savesffamily}{\savesfbfseries}{n}

\DeclareMathSymbol{0}{\mathalpha}{uprightglyphs}{`0}
\DeclareMathSymbol{1}{\mathalpha}{uprightglyphs}{`1}
\DeclareMathSymbol{2}{\mathalpha}{uprightglyphs}{`2}
\DeclareMathSymbol{3}{\mathalpha}{uprightglyphs}{`3}
\DeclareMathSymbol{4}{\mathalpha}{uprightglyphs}{`4}
\DeclareMathSymbol{5}{\mathalpha}{uprightglyphs}{`5}
\DeclareMathSymbol{6}{\mathalpha}{uprightglyphs}{`6}
\DeclareMathSymbol{7}{\mathalpha}{uprightglyphs}{`7}
\DeclareMathSymbol{8}{\mathalpha}{uprightglyphs}{`8}
\DeclareMathSymbol{9}{\mathalpha}{uprightglyphs}{`9}
\DeclareMathSymbol{+}{\mathbin}  {operators}    {`+}
\DeclareMathSymbol{=}{\mathrel}  {operators}    {`=}
\DeclareMathSymbol{.}{\mathord}  {uprightglyphs}{`.}
\DeclareMathSymbol{,}{\mathpunct}{uprightglyphs}{`,}
\DeclareMathSymbol{;}{\mathpunct}{uprightglyphs}{`;}
\DeclareMathSymbol{/}{\mathord}  {uprightglyphs}{`/}
\DeclareMathSymbol{\prime}{\mathord}{extrasymbols}{"30}
\DeclareMathDelimiter{(}      {\mathopen} {uprightglyphs}{`(} {largesymbols}{"00}
\DeclareMathDelimiter{)}      {\mathclose}{uprightglyphs}{`)} {largesymbols}{"01}
\DeclareMathDelimiter{[}      {\mathopen} {uprightglyphs}{`[} {largesymbols}{"02}
\DeclareMathDelimiter{]}      {\mathclose}{uprightglyphs}{`]} {largesymbols}{"03}
\DeclareMathDelimiter{\lbrace}{\mathopen} {uprightglyphs}{`\{}{largesymbols}{"08}
\DeclareMathDelimiter{\rbrace}{\mathclose}{uprightglyphs}{`\}}{largesymbols}{"09}
\DeclareMathSymbol{A}{\mathalpha}{italicglyphs}{`A}
\DeclareMathSymbol{B}{\mathalpha}{italicglyphs}{`B}
\DeclareMathSymbol{C}{\mathalpha}{italicglyphs}{`C}
\DeclareMathSymbol{D}{\mathalpha}{italicglyphs}{`D}
\DeclareMathSymbol{E}{\mathalpha}{italicglyphs}{`E}
\DeclareMathSymbol{F}{\mathalpha}{italicglyphs}{`F}
\DeclareMathSymbol{G}{\mathalpha}{italicglyphs}{`G}
\DeclareMathSymbol{H}{\mathalpha}{italicglyphs}{`H}
\DeclareMathSymbol{I}{\mathalpha}{italicglyphs}{`I}
\DeclareMathSymbol{J}{\mathalpha}{italicglyphs}{`J}
\DeclareMathSymbol{K}{\mathalpha}{italicglyphs}{`K}
\DeclareMathSymbol{L}{\mathalpha}{italicglyphs}{`L}
\DeclareMathSymbol{M}{\mathalpha}{italicglyphs}{`M}
\DeclareMathSymbol{N}{\mathalpha}{italicglyphs}{`N}
\DeclareMathSymbol{O}{\mathalpha}{italicglyphs}{`O}
\DeclareMathSymbol{P}{\mathalpha}{italicglyphs}{`P}
\DeclareMathSymbol{Q}{\mathalpha}{italicglyphs}{`Q}
\DeclareMathSymbol{R}{\mathalpha}{italicglyphs}{`R}
\DeclareMathSymbol{S}{\mathalpha}{italicglyphs}{`S}
\DeclareMathSymbol{T}{\mathalpha}{italicglyphs}{`T}
\DeclareMathSymbol{U}{\mathalpha}{italicglyphs}{`U}
\DeclareMathSymbol{V}{\mathalpha}{italicglyphs}{`V}
\DeclareMathSymbol{W}{\mathalpha}{italicglyphs}{`W}
\DeclareMathSymbol{X}{\mathalpha}{italicglyphs}{`X}
\DeclareMathSymbol{Y}{\mathalpha}{italicglyphs}{`Y}
\DeclareMathSymbol{Z}{\mathalpha}{italicglyphs}{`Z}
\DeclareMathSymbol{a}{\mathalpha}{italicglyphs}{`a}
\DeclareMathSymbol{b}{\mathalpha}{italicglyphs}{`b}
\DeclareMathSymbol{c}{\mathalpha}{italicglyphs}{`c}
\DeclareMathSymbol{d}{\mathalpha}{italicglyphs}{`d}
\DeclareMathSymbol{e}{\mathalpha}{italicglyphs}{`e}
\DeclareMathSymbol{f}{\mathalpha}{italicglyphs}{`f}
\DeclareMathSymbol{g}{\mathalpha}{italicglyphs}{`g}
\DeclareMathSymbol{h}{\mathalpha}{italicglyphs}{`h}
\DeclareMathSymbol{i}{\mathalpha}{italicglyphs}{`i}
\DeclareMathSymbol{\imath}{\mathalpha}{italicglyphs}{"19}
\DeclareMathSymbol{j}{\mathalpha}{italicglyphs}{`j}
\DeclareMathSymbol{\jmath}{\mathalpha}{italicglyphs}{"1A}
\DeclareMathSymbol{k}{\mathalpha}{italicglyphs}{`k}
\DeclareMathSymbol{l}{\mathalpha}{italicglyphs}{`l}
\DeclareMathSymbol{m}{\mathalpha}{italicglyphs}{`m}
\DeclareMathSymbol{n}{\mathalpha}{italicglyphs}{`n}
\DeclareMathSymbol{o}{\mathalpha}{italicglyphs}{`o}
\DeclareMathSymbol{p}{\mathalpha}{italicglyphs}{`p}
\DeclareMathSymbol{q}{\mathalpha}{italicglyphs}{`q}
\DeclareMathSymbol{r}{\mathalpha}{italicglyphs}{`r}
\DeclareMathSymbol{s}{\mathalpha}{italicglyphs}{`s}
\DeclareMathSymbol{t}{\mathalpha}{italicglyphs}{`t}
\DeclareMathSymbol{u}{\mathalpha}{italicglyphs}{`u}
\DeclareMathSymbol{v}{\mathalpha}{italicglyphs}{`v}
\DeclareMathSymbol{w}{\mathalpha}{italicglyphs}{`w}
\DeclareMathSymbol{x}{\mathalpha}{italicglyphs}{`x}
\DeclareMathSymbol{y}{\mathalpha}{italicglyphs}{`y}
\DeclareMathSymbol{z}{\mathalpha}{italicglyphs}{`z}

\makeatletter
\@for\@tempa:=%
	alpha,beta,gamma,delta,epsilon,zeta,eta,theta,iota,kappa,lambda,mu,nu,xi,%
	omicron,pi,rho,sigma,varsigma,tau,upsilon,phi,chi,psi,omega,digamma,%
	Alpha,Beta,Gamma,Delta,Epsilon,Zeta,Eta,Theta,Iota,Kappa,Lambda,Mu,Nu,Xi,%
	Omicron,Pi,Rho,Sigma,Tau,Upsilon,Phi,Chi,Psi,Omega,Digamma%
	\do{%
		\expandafter\let\csname\@tempa orig\expandafter\endcsname\csname\@tempa\endcsname%
		\expandafter\let\csname\@tempa uporig\expandafter\endcsname\csname\@tempa up\endcsname%
	}%
\makeatother

\makeatletter

\@for\@tempa:=%
epsilon,theta,pi,rho,phi%
\do{%
	\expandafter\let\csname var\@tempa orig\expandafter\endcsname\csname var\@tempa\endcsname%
}

\newcommand*{\sansmath}{%
	\@for\@tempa:=%
		alpha,beta,gamma,delta,epsilon,zeta,eta,theta,iota,kappa,lambda,mu,nu,xi,%
		omicron,pi,rho,sigma,varsigma,tau,upsilon,phi,chi,psi,omega,digamma,%
		Alpha,Beta,Gamma,Delta,Epsilon,Zeta,Eta,Theta,Iota,Kappa,Lambda,Mu,Nu,Xi,%
		Omicron,Pi,Rho,Sigma,Tau,Upsilon,Phi,Chi,Psi,Omega,Digamma%
	\do{%
		\expandafter\let\csname\@tempa\expandafter\endcsname\csname\@tempa LGR\endcsname%
		\expandafter\let\csname\@tempa up\expandafter\endcsname\csname\@tempa upLGR\endcsname%
		\expandafter\let\csname up\@tempa\expandafter\endcsname\csname\@tempa upLGR\endcsname%
	}%
	\@for\@tempa:=%
		epsilon,theta,pi,rho,phi%
	\do{%
		\expandafter\let\csname var\@tempa\expandafter\endcsname\csname\@tempa\endcsname%
	}%
}

\newcommand*{\unsansmath}{%
	\@for\@tempa:=%
		alpha,beta,gamma,delta,epsilon,zeta,eta,theta,iota,kappa,lambda,mu,nu,xi,%
		omicron,pi,rho,sigma,varsigma,tau,upsilon,phi,chi,psi,omega,digamma,%
		Alpha,Beta,Gamma,Delta,Epsilon,Zeta,Eta,Theta,Iota,Kappa,Lambda,Mu,Nu,Xi,%
		Omicron,Pi,Rho,Sigma,Tau,Upsilon,Phi,Chi,Psi,Omega,Digamma%
	\do{%
		\expandafter\let\csname\@tempa\expandafter\endcsname\csname\@tempa orig\endcsname%
		\expandafter\let\csname\@tempa up\expandafter\endcsname\csname\@tempa uporig\endcsname%
		\expandafter\let\csname up\@tempa\expandafter\endcsname\csname\@tempa uporig\endcsname%
	}%
	\@for\@tempa:=%
		epsilon,theta,pi,rho,phi%
	\do{%
		\expandafter\let\csname var\@tempa\expandafter\endcsname\csname var\@tempa orig\endcsname%
	}%
}

\newcommand*{\upgreekletters}{%
	\@for\@tempa:=%
		alpha,beta,gamma,delta,epsilon,zeta,eta,theta,iota,kappa,lambda,mu,nu,xi,%
		omicron,pi,rho,sigma,varsigma,tau,upsilon,phi,chi,psi,omega,digamma,%
		Alpha,Beta,Gamma,Delta,Epsilon,Zeta,Eta,Theta,Iota,Kappa,Lambda,Mu,Nu,Xi,%
		Omicron,Pi,Rho,Sigma,Tau,Upsilon,Phi,Chi,Psi,Omega,Digamma%
		\do{%
			\expandafter\let\csname\@tempa\expandafter\endcsname\csname\@tempa up\endcsname%
		}%
}
\newcommand*{\itgreekletters}{%
	\@for\@tempa:=%
		alpha,beta,gamma,delta,epsilon,zeta,eta,theta,iota,kappa,lambda,mu,nu,xi,%
		omicron,pi,rho,sigma,varsigma,tau,upsilon,phi,chi,psi,omega,digamma,%
		Alpha,Beta,Gamma,Delta,Epsilon,Zeta,Eta,Theta,Iota,Kappa,Lambda,Mu,Nu,Xi,%
		Omicron,Pi,Rho,Sigma,Tau,Upsilon,Phi,Chi,Psi,Omega,Digamma%
		\do{%
			\expandafter\let\csname\@tempa\expandafter\endcsname\csname\@tempa orig\endcsname%
		}%
}

\makeatother

\let \bmorig \bm
\renewcommand{\bm}[1]{%
	\IfInSansMode%
		\textbf{\mathversion{boldsans}\(#1\)}%
	\else%
		\bmorig{#1}%
	\fi\relax%
}
\renewcommand{\mathbf}[1]{\bm{#1}}

\renewcommand{\mathcal}[1]{\mathscr{#1}}

\newif\IfInSansMode
\newif\IfInBoldMode
\newif\IfInUpMode
\let \oldsf \sffamily
\renewcommand*{\sffamily}{%
	\oldsf\sansmath\InSansModetrue%
	\IfInBoldMode\mathversion{boldsans}\else\mathversion{sans}\fi\relax%
}
\let \oldbf \bfseries
\renewcommand*{\bfseries}{%
	\oldbf\InBoldModetrue%
	\IfInSansMode\sansmath\mathversion{boldsans}\else\mathversion{bold}\fi\relax%
}
\let \oldmd \mdseries
\renewcommand*{\mdseries}{%
	\oldmd\InBoldModefalse%
	\IfInSansMode\sansmath\mathversion{sans}\else\mathversion{normal}\fi\relax%
}
\let \oldnorm \normalfont
\renewcommand*{\normalfont}{%
	\oldnorm\InSansModefalse\InBoldModefalse\mathversion{normal}%
	\unsansmath%
}
\let \oldrm \rmfamily
\renewcommand*{\rmfamily}{%
	\oldrm\InSansModefalse%
	\IfInBoldMode\mathversion{bold}\else\mathversion{normal}\fi\relax%
	\unsansmath%
}

\let \mathnormalorig \mathnormal
\renewcommand{\mathnormal}[1]{%
	\IfInSansMode%
		\IfInBoldMode%
			\mathversion{boldsans}%
			{\textbf{\(#1\)}}%
		\else%
			\mathversion{sans}%
			{\textmd{\(#1\)}}%
		\fi\relax%
	\else%
		\mathnormalorig{#1}%
	\fi\relax%
}

\let \mathrmorig \mathrm
\renewcommand{\mathrm}[1]{%
	\IfInSansMode%
		{\textrm{%
			\IfInBoldMode%
				\mathversion{bold}%
				\(\mathrmorig{#1}\)%
			\else%
				\mathversion{normal}%
				\(\mathrmorig{#1}\)%
			\fi\relax%
		}}%
	\else%
		\mathrmorig{#1}%
	\fi\relax%
}

\newcommand{\mathup}[1]{%
	\IfInSansMode%
		{\textup{%
			\InUpModetrue%
			\IfInBoldMode%
				\mathversion{boldsansup}%
				\(#1\)%
			\else%
				\mathversion{sansup}%
				\(#1\)%
			\fi\relax%
		}}%
	\else%
		{\upgreekletters\mathrm{#1}\itgreekletters}%
	\fi\relax%
}

\let \operatornameorig \operatorname
\renewcommand{\operatorname}[1]{%
	\operatornameorig{\mathup{#1}}%
}
\makeatletter
\@for\@tempa:=%
	arccos,arccot,arccsc,arcsec,arcsin,arctan,arg,cos,cosh,cot,coth,csc,%
	deg,det,dim,exp,gcd,hom,inf,ker,lg,lim,liminf,limsup,ln,log,max,min,%
	Pr,sec,sin,sinh,sup,tan,tanh%
	\do{%
		\expandafter\let\csname\@tempa\endcsname\relax%
	}%
\makeatother

\DeclareMathOperator*{\lim}   {\mathup{lim}}

\usepackage[backend=biber, natbib=true, bibencoding=inputenc, bibstyle=authoryear, citestyle=authoryear-comp, mincitenames=1, maxcitenames=3, minbibnames=99, maxbibnames=99, uniquename=false, uniquelist=true, backref=true, backrefstyle=three, doi=true, isbn=false, dashed=false, sorting=ynt, sortcites=true, mergedate=true, dateabbrev=false, abbreviate=false, citetracker=true]{biblatex}
\DeclareBibliographyAlias{newspaper}{article}

\xpatchbibmacro{cite}
	{\printnames{labelname}}
	{\ifciteseen
		{\printnames{labelname}}
		{\printnames[][1-5]{labelname}}%
	}
	{}
	{}
\xpatchbibmacro{textcite}
	{\printnames{labelname}}
	{\ifciteseen
		{\printnames{labelname}}
		{\printnames[][1-5]{labelname}}%
	}
	{}
	{}

\renewcommand{\cite}{\citet}

\defbibheading{subbibliography}[\refname]{%
	\section*{#1}%
	\sectionmark{#1}%
	\addcontentsline{toc}{section}{#1}%
}

\DefineBibliographyStrings{english}{%
	andothers = {et~al\adddot},
	volume = {vol\adddot}
}

\xpatchbibmacro{date+extradate}{%
	\printtext[parens]%
}{%
	\setunit*{\mkbibbold{\addperiod\space}}%
	\mdseries\selectfont\printtext%
}{}{}
\xpatchbibmacro{author}{\printnames{author}}{\mkbibbold{\printnames{author}\addperiod}}{}{}
\xpretobibmacro{editor+others}{\mkbibbold\bgroup}{}{}%
\xapptobibmacro{editor+others}{\egroup\clearname{editor}}{}{}%
\xpretobibmacro{translator+others}{\mkbibbold\bgroup}{}{}%
\xapptobibmacro{translator+others}{\egroup\clearname{translator}}{}{}
\DeclareFieldFormat{pages}{\mkcomprange{#1}}
\DeclareFieldFormat[article, inbook, incollection, inproceedings, patent, thesis, unpublished, newspaper]{pages}{{\nopp\mkcomprange{#1}}} 
\AtEveryBibitem{

}
\xpretobibmacro{byeditor+others}  
	{%
	}
	{}
	{\DoNotContinue}

\renewcommand*{\bibnamedash}{---{\kern -2.25pt}---{\kern -2.25pt}---\addcomma\space}

\renewcommand*{\intitlepunct}{\space}
\renewbibmacro{in:}{%
	\ifentrytype{article}%
		{}%
		{\ifentrytype{newspaper}%
			{}%
			{\printtext{\bibstring{in}\intitlepunct}}%
		}%
}

\renewbibmacro*{doi+eprint+url}
{%
	\iffieldundef{doi}
	{}
	{\printfield{doi}}
	\newunit\newblock
	\iftoggle{bbx:eprint}
	{\usebibmacro{eprint}}
	{}%
	\newunit\newblock
	\iffieldundef{doi}  
	{\usebibmacro{url+urldate}}
	{}
}
\DeclareFieldFormat{url}{%
	\Urlmuskip = 0mu\relax%
	\mathchardef\UrlBigBreakPenalty=1\relax%
	\mathchardef\UrlBreakPenalty=2\relax%
	\caps{URL}\addcolon\space\url{#1}%
}
\DeclareFieldFormat{doi}{%
	\Urlmuskip = 0mu\relax
	\mathchardef\UrlBigBreakPenalty=1\relax%
	\mathchardef\UrlBreakPenalty=2\relax%
	\caps{DOI}\addcolon\space\href{https://doi.org/#1}{\nolinkurl{#1}}%
}

\DeclareFieldFormat[article]{journaltitle}{%
	\mkbibemph{#1}%
	\iffieldundef{volume}{\addcomma}{}%
}

\DeclareFieldFormat[newspaper]{journaltitle}{%
	\mkbibemph{#1} \mkbibparens{\thefield{edition}}\addcomma\addspace%
	\iflanguage{ngerman}{%
		\addnbspace\thefield{day}\addperiod\addnbspace\mkbibmonth{\thefield{month}}\addspace\thefield{year}%
	}{
		\mkbibmonth{\thefield{month}}\addnbspace\thefield{day}\addcomma\addspace\thefield{year}%
	}%
	\iffieldundef{volume}{\addcolon}{\addcomma}%
}
\DeclareFieldFormat[newspaper]{date}{%
	\thefield{year}%
}

\DeclareFieldFormat{month}{\mkbibmonth{#1}}

\DeclareFieldFormat[thesis, unpublished, report, misc, newspaper]{title}{\mkbibquote{#1}}

\DeclareFieldFormat[thesis, unpublished, report, misc]{type}{\MakeCapital{#1}}

\DeclareFieldFormat[book, inbook, incollection, inproceedings]{volume}{\bibstring{volume}\addnbspace#1\addcomma}
\DeclareFieldFormat[article]{volume}{%
	\iffieldundef{volume}{}{%
		\addspace#1%
		\iffieldundef{number}{\addcolon}{\nopunct}%
	}
}
\DeclareFieldFormat[article]{number}{\addnbthinspace\mkbibparens{#1}\addcolon}
\DeclareFieldFormat[article]{date}{#1}

\DeclareSourcemap{
    \maps[datatype=bibtex]{
        \map{
            \pertype{article}
			\step[notfield=number, final]
			\step[fieldsource=month, fieldtarget=number]
            \step[fieldset=month, null]
            \step[fieldset=month_numeric, null]
        }
    }
}

\DefineBibliographyStrings{english}{%
	backrefpage  = {},	
	backrefpages = {}	
}
\DefineBibliographyStrings{ngerman}{%
	backrefpage  = {},	
	backrefpages = {}	
}

\renewbibmacro*{pageref}{%
	\addperiod
	\iflistundef{pageref}
	{}
	{\printtext[brackets]{%
			\ifnumgreater{\value{pageref}}{1}
				{\bibstring{backrefpages}}
				{\bibstring{backrefpage}}%
			\printlist[pageref][-\value{listtotal}]{pageref}%
		}%
	}%
}

\DeclareSourcemap{
	\maps[datatype=bibtex]{
		\map[overwrite]{
		 	\step[fieldsource=annote]
		 	\step[fieldset=note, origfieldval, append]
			\step[fieldset=annote, null]
		}
		\map{
			\step[fieldsource=note, final]
			\step[fieldset=addendum, origfieldval, final]
			\step[fieldset=note, null]
		}
	}
}

\AtEveryBibitem{%
	\ifentrytype{newspaper}%
		{\clearfield{addendum}}
		{\clearfield{month}}
}

\DeclareSourcemap{
	\maps[datatype=bibtex]{
		\map{
			\pertype{techreport}
			\step[fieldset=type, fieldvalue={Working paper}]
		}
	}
}

\DeclareSourcemap{
	\maps[datatype=bibtex]{
		\map{
			\pertype{phdthesis}
			\step[fieldset=type, fieldvalue={Doctoral dissertation}]
		}
	}
}

\DeclareSourcemap{ 
	\maps[datatype=bibtex]{
		\map{
			\step[fieldsource=journal, match={\regexp{^The\s}}, replace={}]
		}
	}      
}

\DeclareSourcemap{ 
	\maps[datatype=bibtex]{
		\map{
			\step[fieldsource=number, match={\regexp{^jan$}}, replace=\regexp{\\mkbibmonth\{1\}}]
			\step[fieldsource=number, match={\regexp{^feb$}}, replace=\regexp{\\mkbibmonth\{2\}}]
			\step[fieldsource=number, match={\regexp{^mar$}}, replace=\regexp{\\mkbibmonth\{3\}}]
			\step[fieldsource=number, match={\regexp{^apr$}}, replace=\regexp{\\mkbibmonth\{4\}}]
			\step[fieldsource=number, match={\regexp{^may$}}, replace=\regexp{\\mkbibmonth\{5\}}]
			\step[fieldsource=number, match={\regexp{^jun$}}, replace=\regexp{\\mkbibmonth\{6\}}]
			\step[fieldsource=number, match={\regexp{^jul$}}, replace=\regexp{\\mkbibmonth\{7\}}]
			\step[fieldsource=number, match={\regexp{^aug$}}, replace=\regexp{\\mkbibmonth\{8\}}]
			\step[fieldsource=number, match={\regexp{^sep$}}, replace=\regexp{\\mkbibmonth\{9\}}]
			\step[fieldsource=number, match={\regexp{^oct$}}, replace=\regexp{\\mkbibmonth\{10\}}]
			\step[fieldsource=number, match={\regexp{^nov$}}, replace=\regexp{\\mkbibmonth\{11\}}]
			\step[fieldsource=number, match={\regexp{^dec$}}, replace=\regexp{\\mkbibmonth\{12\}}]
		}
	}      
}

\let\comment\undefined  

\usepackage[todonotes={textsize=scriptsize}, authormarkuptext=name, authormarkup=none]{changes}

\makeatletter
\tikzstyle{notestyleraw} = [
	draw=\@todonotes@currentbordercolor,
	fill=\@todonotes@currentbordercolor,
	text=white,
	line width = 0.35pt,
	text width = \@todonotes@textwidth - 1.5ex - 1pt,
	inner sep = 0.75ex,
	rounded corners = 0pt
]
\colorlet{authorcolor}{magenta}
\@namedef{Changes@AuthorColor}{magenta}
\colorlet{Changes@Color}{magenta}
\newcommand{\changesfontsettings}{\sffamily\scriptsize\baselineskip=2.25ex}
\renewcommand{\@todonotes@useSizeCommand}{\changesfontsettings}
\xpatchcmd{\@todonotes@drawLineToLeftMargin} {connectstyle}{connectstyle, line width=0.6pt}{}{}
\xpatchcmd{\@todonotes@drawLineToRightMargin}{connectstyle}{connectstyle, line width=0.6pt}{}{}
\makeatother

\definecolor{electricultramarine}{rgb}{0.25, 0.0, 1.0}
\definecolor{alizarin}{rgb}{0.82, 0.1, 0.26}
\definecolor{dartmouthgreen}{rgb}{0.05, 0.5, 0.06}
\definecolor{goldenpoppy}{HTML}{FCC200}
\definecolor{internationalorange}{rgb}{1.0, 0.31, 0.0}
\definechangesauthor[name={Holger}, color=alizarin]           {HG}
\definechangesauthor[name={Lou~E.}, color=internationalorange]{LV}
\definechangesauthor[name={U.~R.},  color=electricultramarine]{UR}

\LetLtxMacro{\todoorig}{\todo}
\makeatletter
\renewcommand{\todo}[2][]{%
	\todoorig[color=Changes@Color, bordercolor=Changes@Color, #1, tickmarkheight=0.1cm, linecolor=authorcolor]{#2}%
}
\makeatother

\LetLtxMacro{\commentorig}{\comment}
\newsavebox{\commentbox}
\newlength{\commentboxwidth}
\makeatletter
\newcommand{\commentpatched}[2][]{%
	\savebox{\commentbox}[\commentboxwidth][t]{%
		\parbox{\commentboxwidth}{%
			\changesfontsettings\color{white}\RaggedRight%
			\hrule\smallskip%
			#2%
		}%
	}%
	\commentorig[#1]{%
		\strut\newline%
		\usebox{\commentbox}%
	}%
}
\makeatother
\renewcommand{\comment}[2][]{%
	\commentpatched[#1]{#2}%
}

\LetLtxMacro{\addedorig}{\added}
\renewcommand{\added}[2][]{%
	\addedorig[#1]{#2}%
	\IfSubStr{#1}{comment}{}{\commentpatched[#1]{\textit{Added text.}}}%
}
\LetLtxMacro{\replacedorig}{\replaced}
\renewcommand{\replaced}[3][]{%
	\replacedorig[#1]{#2}{}%
	\IfSubStr{#1}{comment}{}{\commentpatched[#1]{\textit{Replaced:} ``#3''}}%
}
\setdeletedmarkup{}
\LetLtxMacro{\deletedorig}{\deleted}
\renewcommand{\deleted}[2][]{%
	\deletedorig[#1]{#2}%
	\IfSubStr{#1}{comment}{}{\commentpatched[#1]{\textit{Removed:} ``#2''}}%
}
\newcommand{\coloruline}[2]{%
    \newcommand{\tempuline}{%
    	\bgroup\markoverwith{\textcolor{#1}{\rule[-1ex]{1.5pt}{0.25ex}}}%
    	\ULon%
    }%
    \tempuline{#2}%
}
\sethighlightmarkup{%
	\colorlet{authorcoloraux}{authorcolor!33}%
	\IfIsColored{%
		\sethlcolor{authorcoloraux}%
		\hl{#1}%
	}{}%
}
\LetLtxMacro{\highlightorig}{\highlight}
\renewcommand{\highlight}[2][]{%
	\highlightorig[#1]{#2}%
	\IfSubStr{#1}{comment}{}{\commentorig[#1]{~}}%
}

\usepackage{pdfcomment}
	{\end{pdfsidelinecomment}%
}

\newcommand{\sigstar}{\raisebox{0.66ex}{\scalebox{0.95}{$\star$}}}

\newcommand{\balCL}[1][1]{\mbox{\caps{BAL}$_{\mathup{CL}}$}\xspace}
\newcommand{\unbalCLA}[1][1]{\mbox{\caps{UNBAL}$^{\mathup{I}}_{\mathup{CL}}$}\xspace}
\newcommand{\unbalCLB}[1][1]{\mbox{\caps{UNBAL}$^{\mathup{II}}_{\mathup{CL}}$}\xspace}

\bibliography{references.bib}

\definechangesauthor[name={Tobias}, color=alizarin]{T}
\definechangesauthor[name={Janos}, color=internationalorange]{J}
\definechangesauthor[name={Klara}, color=dartmouthgreen]{K}
\definechangesauthor[name={HM}, color=blue]{HM}

\begin{document}

\selectlanguage{USenglish}
\frenchspacing


\title{\sffamily\bfseries%
	\papertitle%
	\thanks{\protect The authors are grateful for support by the Deutsche Forschungsgemeinschaft (DFG, German
Research Foundation) under Germany´s Excellence Strategy – EXC 2126/1– 390838866 – and
through CRC-TR 224 (Projects A02 and C01), by the IZA Institute of Labor Economics, and
by the Google Cloud Covid-19 research credits program.

}%
}

\author[a, b]{Janoś Gabler}
\author[c]{Tobias Raabe}
\author[a]{Klara Röhrl}
\author[b,d]{Hans-Martin von Gaudecker}

\affil[a]{Bonn Graduate School of Economics}
\affil[b]{IZA Institute of Labor Economics}
\affil[c]{Private sector}
\affil[d]{Rheinische Friedrich-Wilhelms-Universität Bonn}

\date{%
	\today%
}

\maketitle

\vfill



\begin{abstract}%
	\noindent
In order to slow the spread of the CoViD-19 pandemic, governments around the world have
enacted a wide set of policies limiting the transmission of the disease. Initially,
these focused on non-pharmaceutical interventions; more recently, vaccinations and
large-scale rapid testing have started to play a major role. The objective of this study
is to explain the quantitative effects of these policies on determining the course of
the pandemic, allowing for factors like seasonality or virus strains with different
transmission profiles. To do so, the study develops an agent-based simulation model,
which is estimated using data for the second and the third wave of the CoViD-19 pandemic
in Germany. The paper finds that during a period where vaccination rates rose from 5\%
to 40\%, rapid testing had the largest effect on reducing infection numbers. Frequent
large-scale rapid testing should remain part of strategies to contain CoViD-19; it can
substitute for many non-pharmaceutical interventions that come at a much larger cost to
individuals, society, and the economy.

\vspace{1cm}
\noindent \textbf{JEL Classification:} C63, I18

\noindent \textbf{Keywords:} CoViD-19, agent based simulation model, rapid testing,
non-pharmaceutical interventions%
	\comment[id=T]{
		\textbf{Commenting is on!} \\
		To~switch it off, activate
		\mbox{\texttt{\textbackslash PassOptionsToPackage}}\\
		\mbox{\texttt{\{final\}\{changes\}}} \\
		in preamble.tex.%
	}
\end{abstract}

\vfill

\clearpage

\maketitle
\pagenumbering{arabic}
\setcounter{page}{1}

\newpage

Since early 2020,\comment[id=J]{Put stronger focus on time dependent rapid test
sensitivity (30\% before day of onset of infectiousness)} the CoViD-19 pandemic has
presented an enormous challenge to humanity on many dimensions. The development of
highly effective vaccines holds the promise of containment in the medium term. However,
most countries find themselves many months---and often years---away from reaching
vaccination-induced herd immunity \citep{Swaminathan2021}.\comment[id=HM]{Need to find
some more scientific citation before submission} In the meantime, it is of utmost
importance to employ an effective mix of strategies for containing the virus. The most
frequent initial response was a set of non-pharmaceutical interventions (NPIs) to reduce
contacts between individuals. While this has allowed some countries to sustain
equilibria with very low infection numbers,\footnote{See \citet{Contreras2021} for a
theoretical equilibrium at low case numbers which is sustained with
test-trace-and-isolate policies.} most have seen large fluctuations of infection rates
over time. Containment measures have become increasingly diverse and now include rapid
testing, more nuanced NPIs, and contact tracing. Neither these policies' effect nor the
influence of seasonal patterns or of more infectious virus strains are well understood
in quantitative terms.

This paper develops a quantitative model incorporating these factors simultaneously. The
framework allows to combine a wide variety of data and mechanisms in a timely fashion,
making it useful to predict the effects of various interventions. We apply the model to
Germany, where new infections fell by almost 80\% during the month of May 2021. Our
analysis shows that, aside from seasonality, frequent and large-scale rapid testing
caused the bulk of this decrease, which is in line with prior predictions
\citep{Mina2021}. We conclude that it should have a large role for at least as long as
vaccinations have not been offered to an entire population.

\begin{figure}   
    \centering

    \begin{subfigure}[b]{0.425\textwidth}
        \centering
        \includegraphics[width=\textwidth]{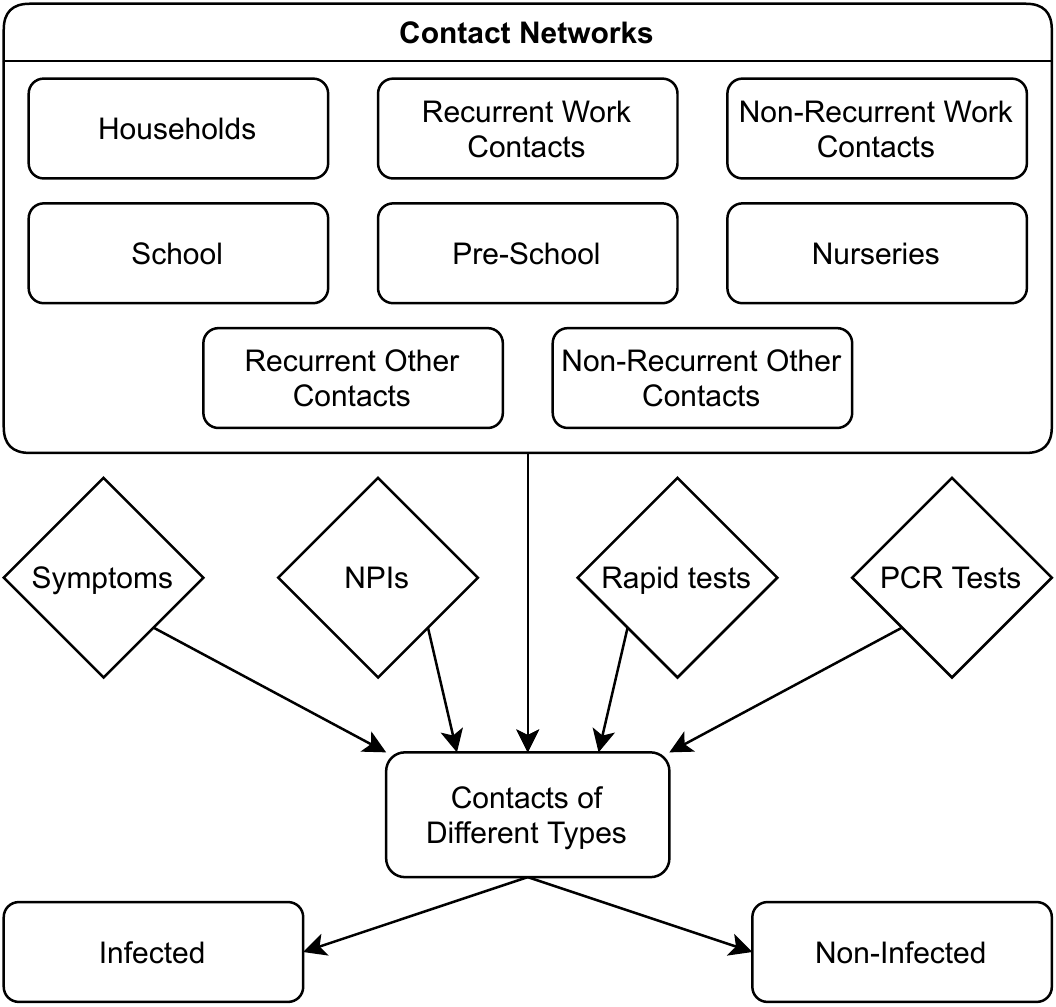}
        \caption{{Model for contacts and infections}}
        \label{fig:model_contacts_infections}
    \end{subfigure}
    \hfill
    \begin{subfigure}[b]{0.425\textwidth}
        \centering
        \includegraphics[width=\textwidth]{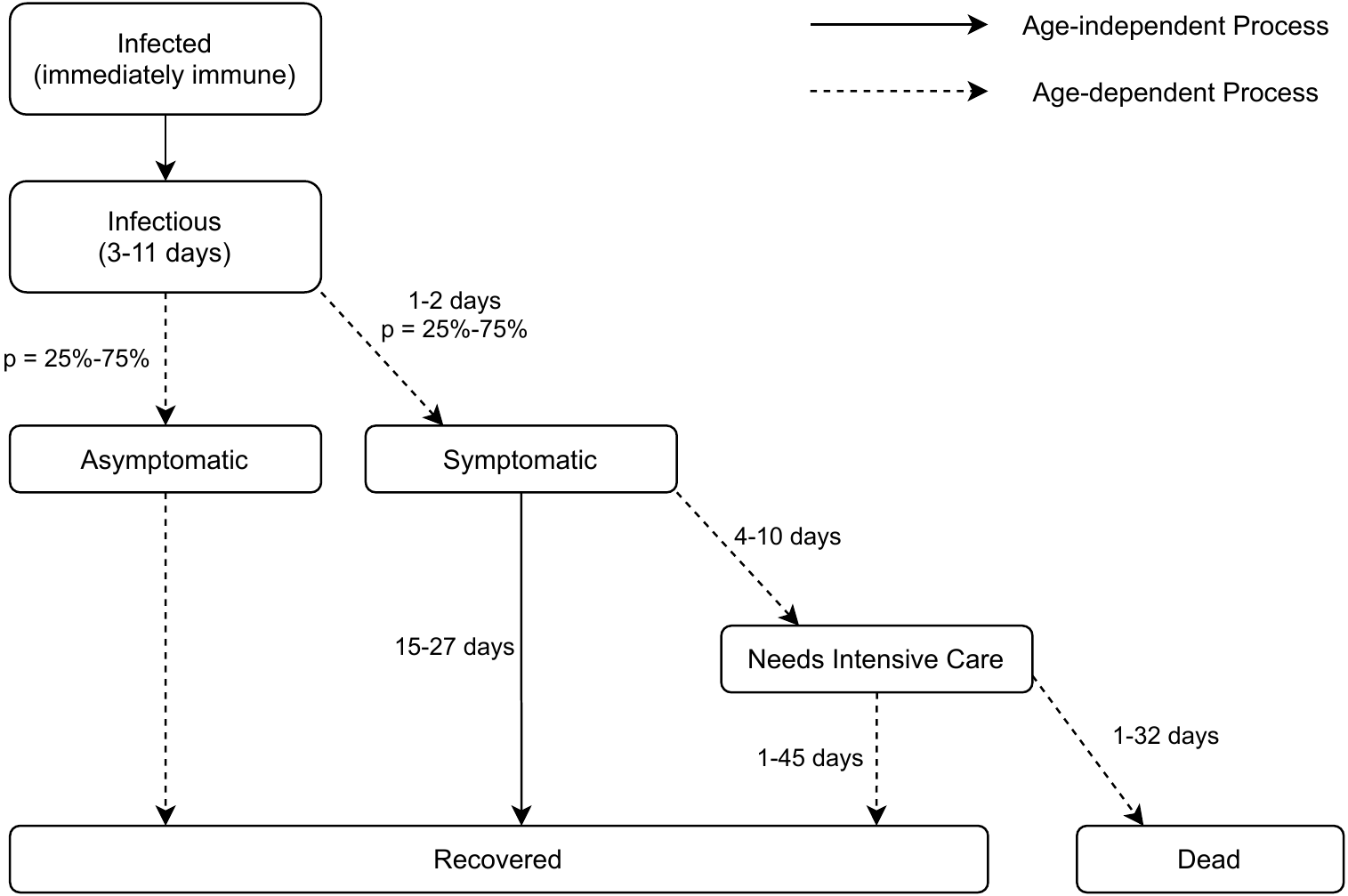}
        \vskip5ex

        \caption{Disease progression}
        \label{fig:model_disease_progression}
    \end{subfigure}
    \vskip3ex
    \begin{subfigure}[b]{0.425\textwidth}
        \centering

        \includegraphics[width=\textwidth]{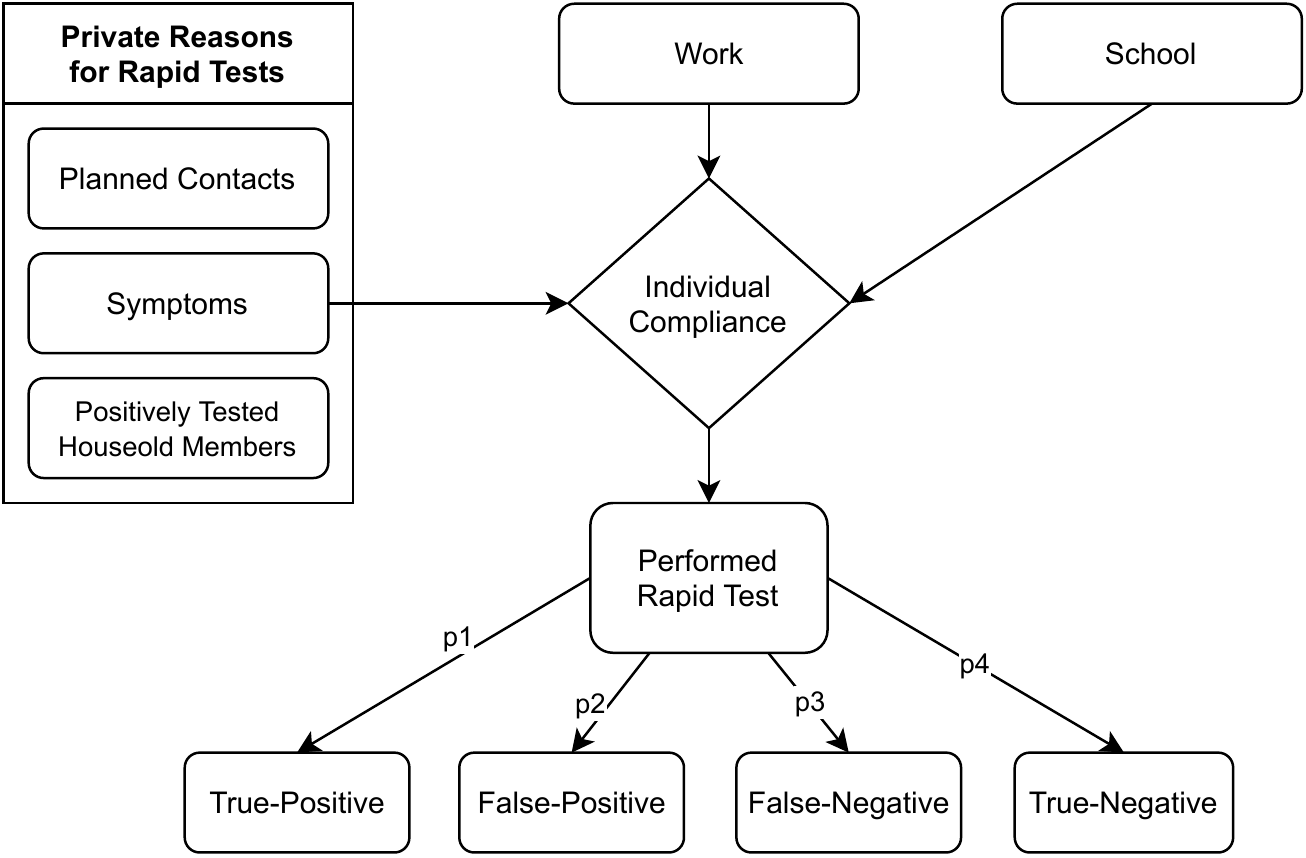}
        \caption{{Model for rapid tests}}
        \label{fig:model_rapid_tests}
    \end{subfigure}
    \hfill
    \begin{subfigure}[b]{0.425\textwidth}
        \centering
        \includegraphics[width=\textwidth]{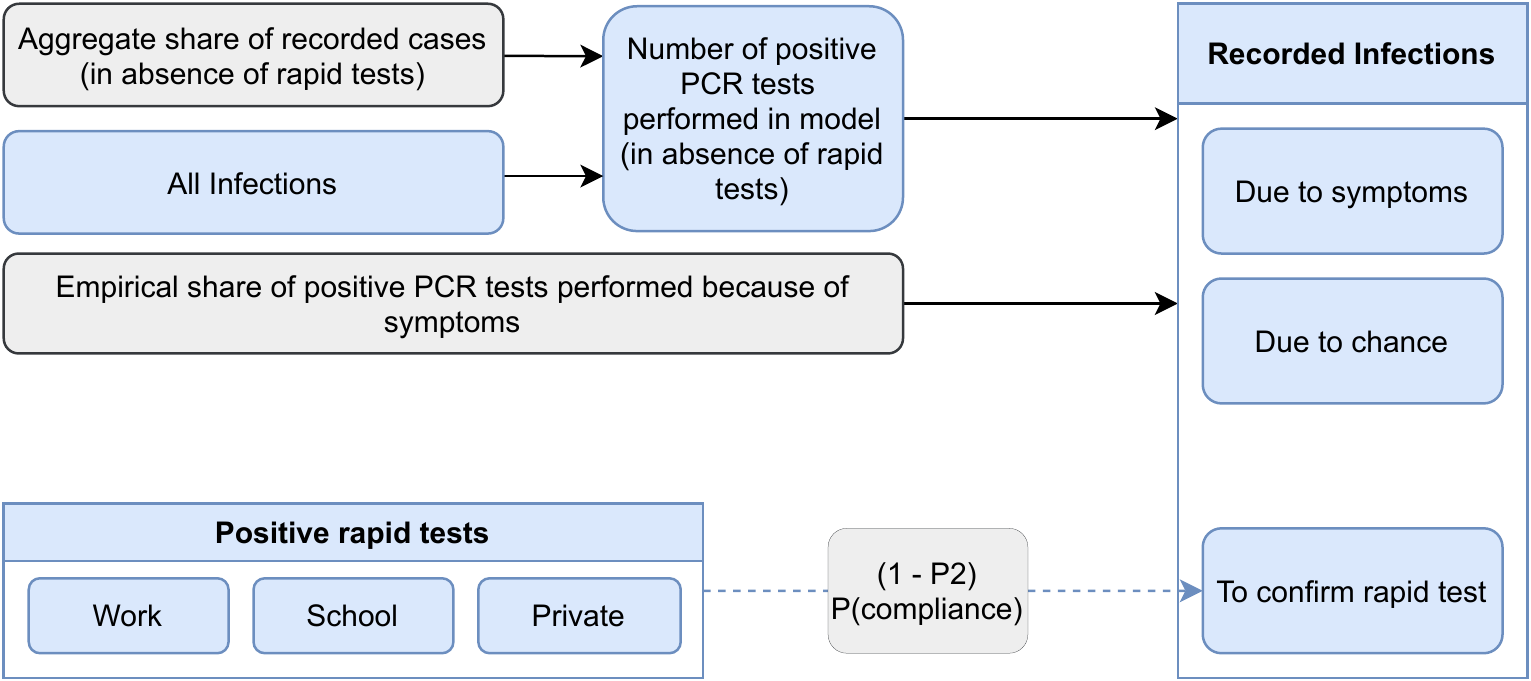}
        \vskip5ex

        \caption{{Translating all infections to recorded ones}}
        \label{fig:model_official_cases}
    \end{subfigure}

    \caption{Model description}
    \label{fig:model_description}

    \floatfoot{\noindent \textit{Note:} A
        description of the model can be found in Supplementary Material~\ref{sec:model}.
        Figure~\ref{fig:model_contacts_infections} shows the influence on an agent's
        contacts to other agents. Demographic characteristics set the baseline number of
        contacts in different networks. She may reduce the number of contacts due to
        NPIs, showing symptoms, or testing positively for SARS-CoV-2. Infections may
        occur when a susceptible agent meets an infectious agent; the probability depends
        on the type of contact, on seasonality, and on NPIs. If infected, the infection
        progresses as depicted in Figure~\ref{fig:model_disease_progression}. If rapid
        tests are available, agents' demand is modeled as in
        Figure~\ref{fig:model_rapid_tests}. All reasons trigger a test only for a
        fraction of individuals depending on an individual compliance parameter; the
        thresholds for triggering test demand differ across reasons and they may depend
        on calendar time. Figure~\ref{fig:model_official_cases} shows the model of
        translating all infections in the simulated data to age-specific recorded
        infections. The model uses data on the aggregate share of recorded cases, on the
        share of positive PCR tests triggered by symptoms, and on the false positive rate
        of rapid tests. The lower part of the graph is relevant only for periods where
        rapid tests are available.}
    \end{figure}

At the core of our agent-based model are physical contacts between heterogeneous agents
(Figure~\ref{fig:model_contacts_infections}).\footnote{A detailed comparison with other
approaches  is relegated to Supplementary Material~\ref{sec:literature_review}. The
model most closely related to ours is described in \citet{Hinch2020}.} Each contact
between an infectious individual and somebody susceptible to the disease bears the risk
of transmitting the virus. Contacts occur in up to four networks: Within the household,
at work, at school, or in other settings (leisure activities, grocery shopping, medical
appointments, etc.). Some contacts recur regularly, others occur at random. Empirical
applications can take the population and household structure from census data and the
network-specific frequencies of contacts from diary data measuring contacts before the
pandemic \citep[e.g.][]{Mossong2008,Hoang2019}. Within each network, meeting frequencies
depend on age and geographical location (see Supplementary Material
Section~\ref{sub:number_of_contacts}).

The four contact networks are chosen so that the most common NPIs can be modeled in
great detail. NPIs affect the number of contacts or the risk of transmitting the disease
upon having physical contact. The effect of different NPIs will generally vary across
contact types. For example, a mandate to work from home will reduce the number of work
contacts to zero for a fraction of the working population. Schools and daycare can be
closed entirely, operate at reduced capacity---including an alternating schedule---, or
implement mitigation measures like masking requirements or air filters
\citep{Lessler2021}. Curfews may reduce the number of contacts in
non-work/non-school/non-work settings. In any setting, measures like masking
requirements would reduce the probability of infection associated with a contact
\citep{Cheng2021}.

In the model, susceptibility to contracting the SARS-CoV-2 virus is dependent on age. A
possible infection progresses as shown in Figure~\ref{fig:model_disease_progression}. We
differentiate between an initial period of infection without being infectious or showing
symptoms, being infectious (presymptomatic or asymptomatic), showing symptoms, requiring
intensive care, and recovery or death \citep[similar
to][]{Grimm2021}.\comment[id=HM]{Need to look for non-economists} The probabilities of
transitioning between these states depend on age; their duration is random within
intervals calibrated to medical literature (for a detailed description see
Section~\ref{sec:medical_params}). Conditional on the type of contact, infectiousness is
independent of age \citep{Jones2021}.

The model includes several other features, which are crucial to describe the evolution
of the pandemic in 2020-2021. New virus strains with different profiles regarding
infectiousness can be introduced. Agents may receive a vaccination. With a probability
of 75\% \citep{Hunter2021}, vaccinated agents become immune and they do not transmit the
virus \citep{Petter2021, LevineTiefenbrun2021, Pritchard2021}.\footnote{75\% is lower
than what is usually reported for after the second dose of the Biontech/Pfizer vaccine,
which is most commonly used in Germany. We choose it because our model neither includes
booster shots, nor does it allow vaccinated individuals who became immune to transmit
the disease\citep{Petter2021, LevineTiefenbrun2021, Pritchard2021}. If anything, these
assumptions would overstate the effect of vaccines for our study period. This would be
different if a large fraction of vaccinated individuals had received a second dose
already.} During the vaccine roll-out, priority may depend on age and occupation.

We include two types of tests. Polymerase chain reaction (PCR) tests reveal whether an
individual is infected or not; there is no uncertainty to the result. PCR tests require
some days to be processed and there are aggregate capacity constraints throughout. In
contrast, rapid antigen tests yield immediate results; after a phase-in period, all tests
that are demanded will be performed. Specificity and sensitivity of these tests is set
according to data analyzed in \cite{Bruemmer2021, Smith2021}; sensitivity depends on the
timing of the test relative to the onset of infectiousness.
Figure~\ref{fig:model_rapid_tests} shows our model for rapid test demand. Schools may
require staff and students to be tested regularly. Rapid tests may be offered by
employers to on-site workers. Individuals may demand tests for private reasons, which
include having plans to meet other people, showing symptoms of CoViD-19, and because a
household member tested positively for the virus. We endow each agent with an individual
compliance parameter. This parameter determines whether she takes up rapid
tests.\footnote{A positive test result (of either kind), as well as symptoms leads most
individuals to reduce their contacts; this is why tests impact the actual contacts in
Figure~\ref{fig:model_description}.}

Modelling a population of agents according to actual demographic characteristics means
that we can use a wide array of data to identify and estimate the model's many
parameters.\footnote{See section~\ref{sec:data_and_parameters} of the supplementary
materials for an overview.} Contact diaries yield pre-pandemic distributions of contacts
for different contact types and their assortativity by age group. Mobility data is used
to model the evolution of work contacts. School and daycare policies can be incorporated
directly from official directives. Administrative records on the number of tests,
vaccinations by age and region, and the prevalence of virus strains are generally
available. Surveys may ask about test offers, propensities to take them up, and past
tests. Other studies' estimates of the seasonality of infections can be incorporated
directly. The remaining parameters---most notably, these include infection probabilities
by contact network and the effects of NPIs, see Supplementary
Material~\ref{subsec:estimated_params}---will be chosen numerically so that the model
matches features of the data \citep[see][for the general method, also described in
Supplementary Material \ref{sec:model}]{McFadden1989}.\comment[id=HM]{Make reference
more specific once Appendix has new structure} In our application, we keep the number of
free parameters low in order to avoid overfitting. The data features to be matched
include official case numbers for each age group and region, deaths, and the share of
the B.1.1.7 strain.

The main issue with official case numbers is that they will contain only a fraction of
all infections. We thus model official cases as depicted in
Figure~\ref{fig:model_official_cases}. We take aggregate estimates of the share of
detected cases and use data on \replaced[id=K]{the share of PCR tests administered to
people with CoViD-19 symptoms}{whether CoViD-19 symptoms led to a PCR test}. As the share
of asymptomatic individuals varies by age group, this gives us age-specific shares (see
Figure~\ref{fig:share_known_cases_by_age_group} for the share of known cases by age group
over time in our model). Our estimates suggest that---in the absence of rapid
testing---the detection rate is 80\% higher on average for individuals above age 80
compared to school age children. Once rapid test become available, confirmation of a
positive result is another reason leading to positive PCR tests.

\begin{figure}[!tp]   
    \centering

    \begin{subfigure}[b]{0.475\textwidth}
        \centering
        \includegraphics[width=\textwidth]{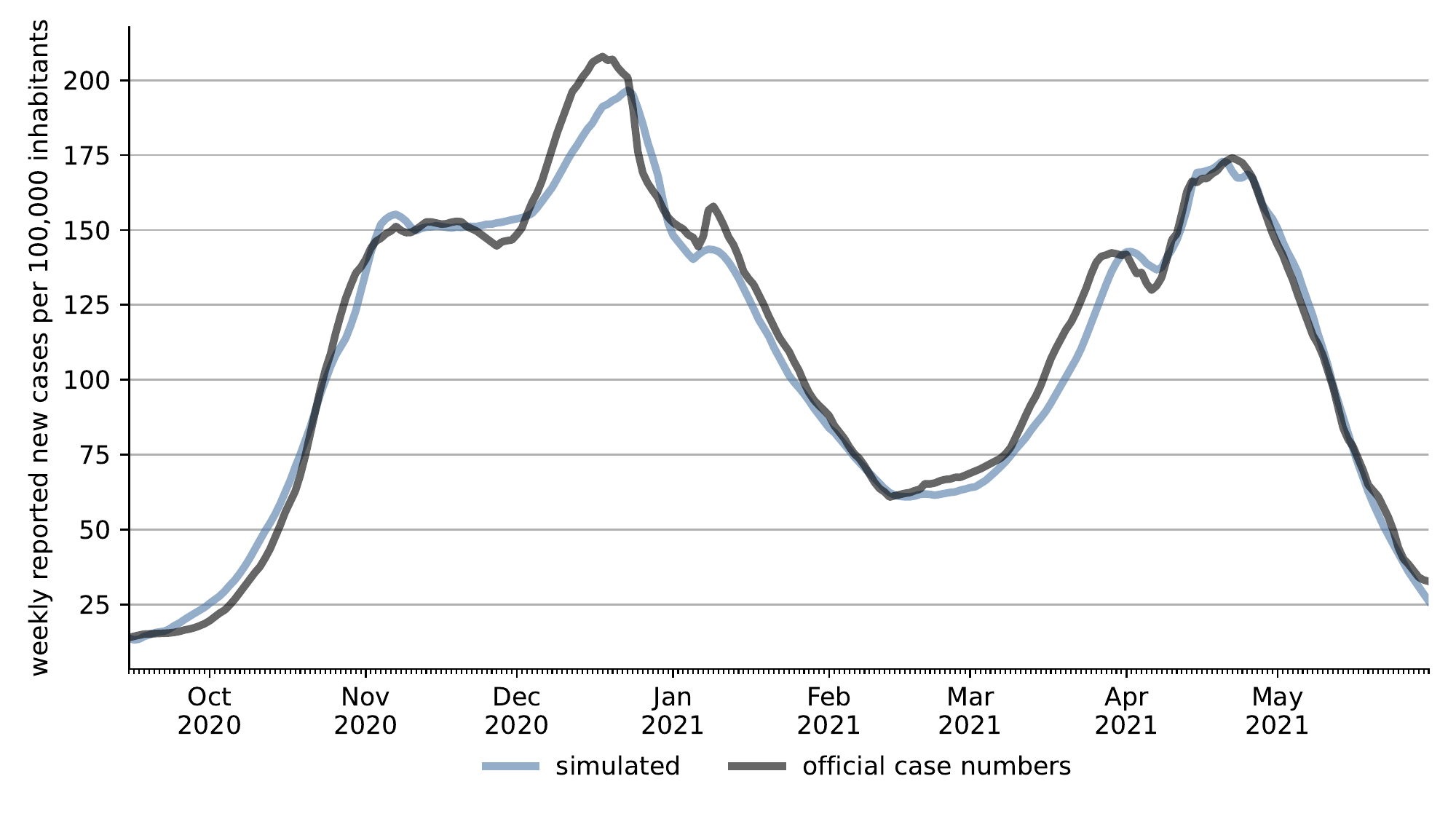}
        \caption{{Recorded cases: Empirical and simulated}}
        \label{fig:aggregated_fit}
    \end{subfigure}
    \hfill
    \begin{subfigure}[b]{0.475\textwidth}
        \centering
        \includegraphics[width=\textwidth]{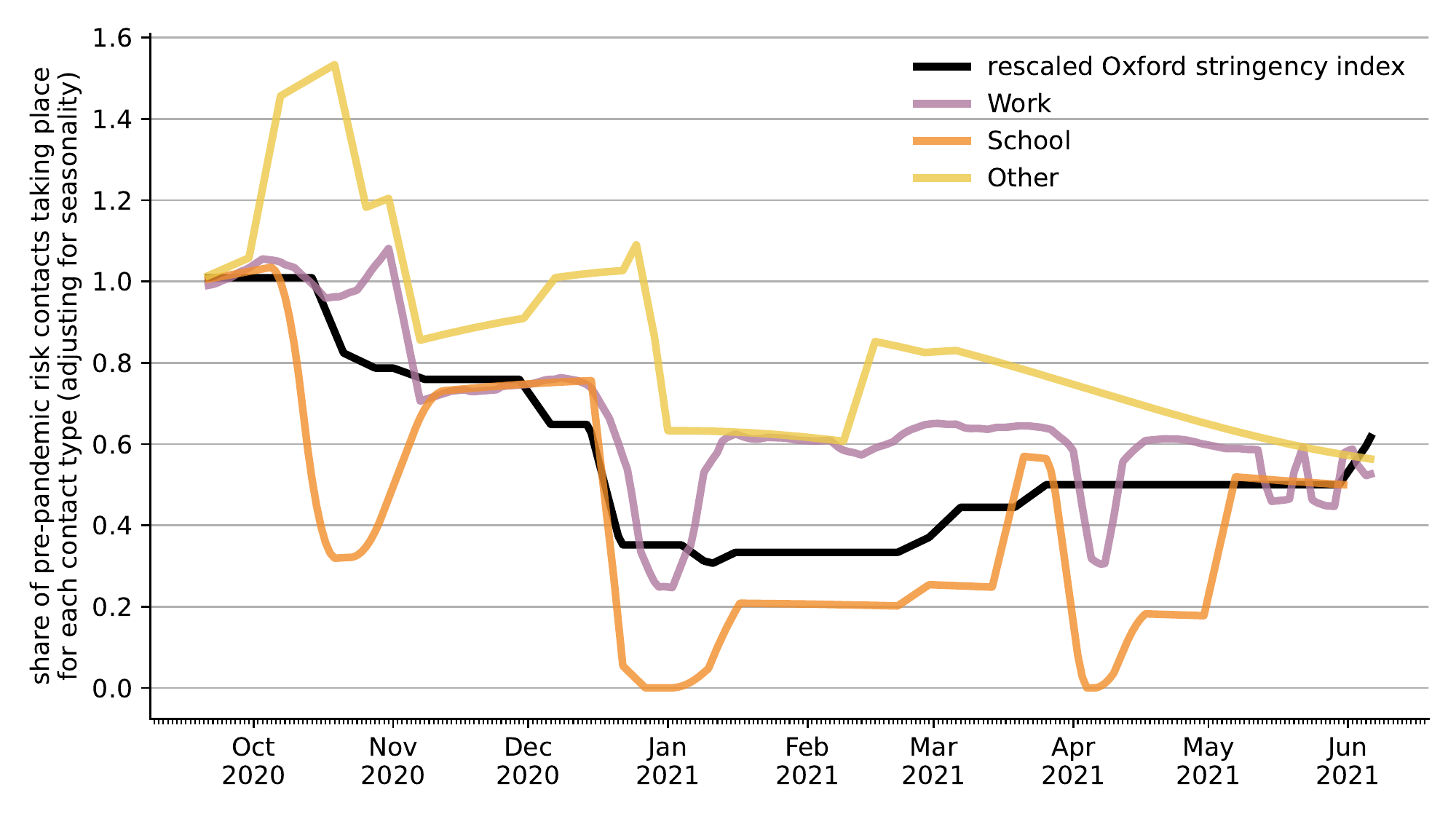}

        \caption{{Stringency of NPIs and infectious contacts}}
        \label{fig:stringency_infectious_contacts}
    \end{subfigure}

    \vskip3ex

    \begin{subfigure}[b]{0.475\textwidth}
        \centering

        \includegraphics[width=\textwidth]{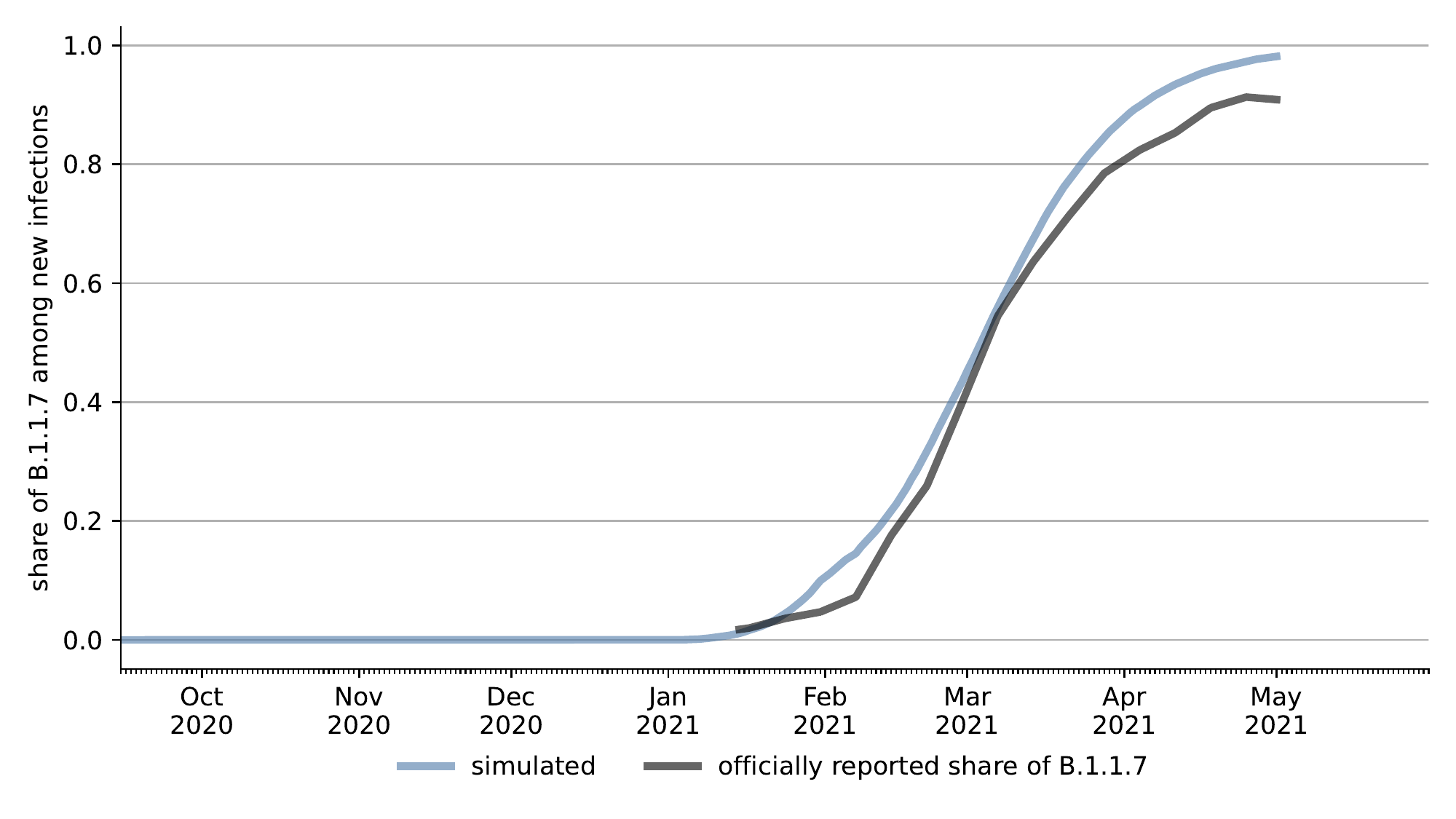}

        \caption{Fraction of B.1.1.7 strain}
        \label{fig:share_b117}
    \end{subfigure}
    \hfill
    \begin{subfigure}[b]{0.475\textwidth}
        \centering

        \includegraphics[width=\textwidth]{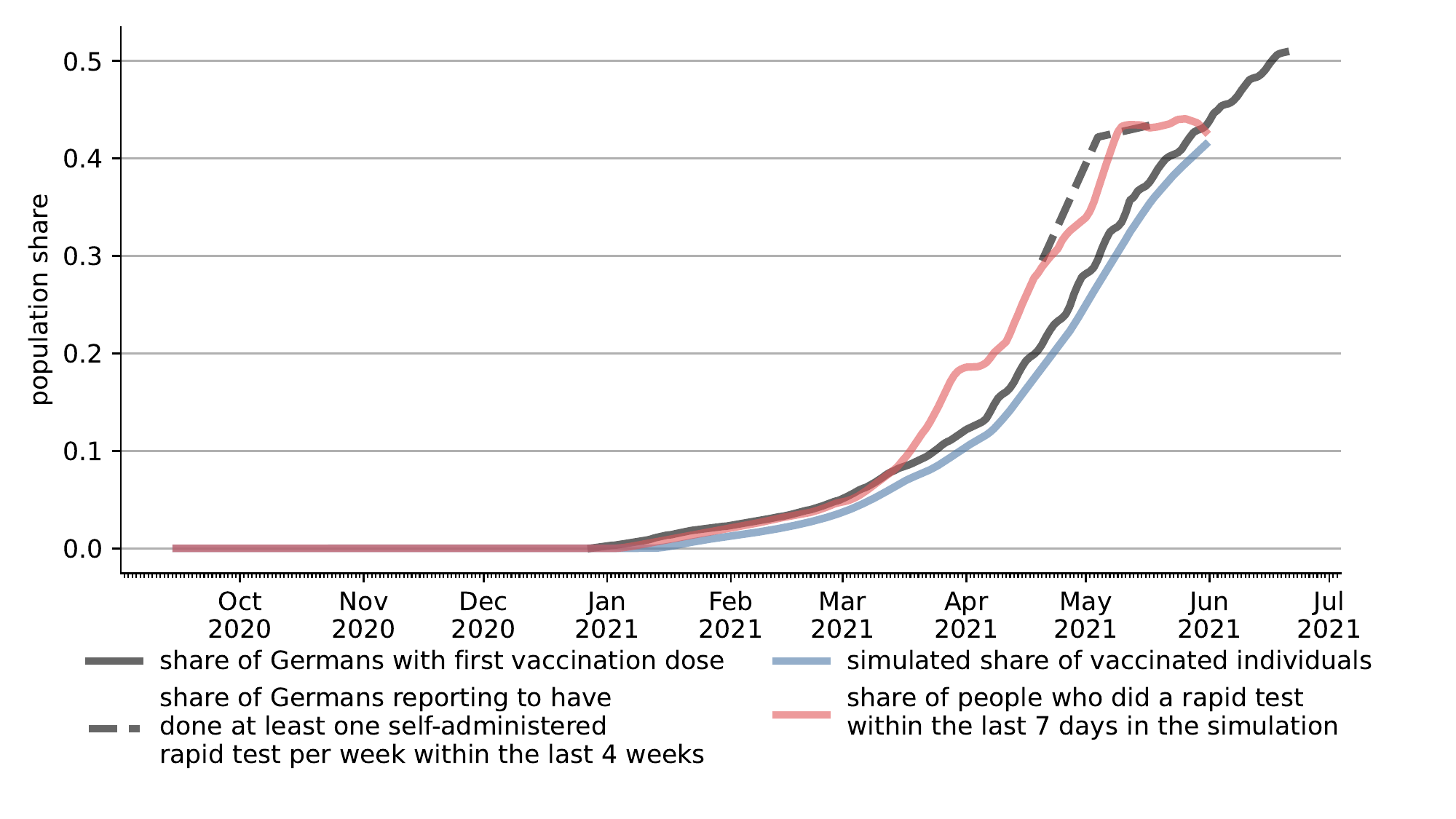}

        \caption{{Tests and vaccinations}}
        \label{fig:antigen_tests_vaccinations}
    \end{subfigure}

    \caption{Evolution of the pandemic, its drivers, and model fit, September 2020 to May 2021}
    \label{fig:pandemic_drivers_model_fit}

    \floatfoot{\noindent \textit{Note:} Data sources are described in Supplementary
        Material~\ref{sec:data_and_parameters}. Age- and region-specific analogues to
        Figure~\ref{fig:aggregated_fit} can be found in Supplementary Material
        \ref{subsec:fit_results}. For legibility reasons, all lines in
        Figure~\ref{fig:stringency_infectious_contacts} are rolling 7-day averages. The
        Oxford Response Stringency Index is scaled as $2 \cdot (1 -  x / 100)$, so that
        a value of 1 refers to the situation at the start of our sample period and 0
        means that all NPIs included in the index are turned on. The other lines in
        Figure~\ref{fig:stringency_infectious_contacts} show the product of the effect
        of contact reductions, increased hygiene regulations, and seasonality. See
        Appendix~\ref{subsec:policies_seasonality} for separate plots of the three
        factors by contact type.}
\end{figure}

The model is applied to the second and third wave of the CoViD-19 pandemic in Germany,
covering the period mid-September 2020 to the end of May 2021.
Figure~\ref{fig:pandemic_drivers_model_fit} describes the evolution of the pandemic and
of its drivers. The black line in Figure~\ref{fig:aggregated_fit} shows officially
recorded cases; the black line in Figure~\ref{fig:stringency_infectious_contacts} the
Oxford Response Stringency Index \citep{Hale2020}, which tracks the tightness of
non-pharmaceutical interventions. The index is shown for illustration of the NPIs, we
never use it directly. For legibility reasons, we transform the index so that lower
values represent higher levels of restrictions. A value of zero means all measures
incorporated in the index are turned on. The value 1 represents the situation in
mid-September, with restrictions on gatherings and public events, masking requirements,
but open schools and workplaces. In the seven weeks between mid September and early
November, cases increased by a factor of 10. Restrictions were somewhat tightened in
mid-October and again in early November. New infections remained constant throughout
November before rising again in December, prompting the most stringent lockdown to this
date. Schools and daycare centers were closed again, so were customer-facing businesses
except for grocery and drug stores. From the peak of the second wave just before
Christmas until the trough in mid-February, newly detected cases decreased by almost
three quarters. The third wave in the spring of 2021 is associated with the B.1.1.7
strain, which became dominant in March (Figure~\ref{fig:share_b117}). In early March,
some NPIs were being relaxed; e.g., hairdressers and home improvement stores were
allowed to open again to the public. There were many changes in details of regulations
afterwards, but they did not change the stringency index.

By this time, the set of policy instruments had become much more diverse. Around the
turn of the year, the first people were vaccinated with a focus on older age groups and
medical personnel (Figure~\ref{fig:antigen_tests_vaccinations}). By the end of May, just
over 40\% had received at least one dose of a vaccine. Around the same time, rapid tests
started to replace regular PCR tests for staff in many medical and nursing facilities.
These had to be administered by medical doctors or in pharmacies. At-home tests approved
by authorities became available in mid-March, rapid test centers were opened, and one
test per person and week was made available free of charge. In several states, customers
were only allowed to enter certain stores with a recent negative rapid test result.
These developments are characteristic of many countries: The initial focus on NPIs to
slow the spread of the disease has been accompanied by vaccines and a growing acceptance
and use of rapid tests. At broadly similar points in time, novel strains of the virus
have started to pose additional challenges.

We draw simulated samples of agents from the population structure in September 2020 and
use the model to predict recorded infection rates until the end of May 2021. See
Supplementary Material~\ref{sub:initial_conditions} for a detailed description of this
procedure. The blue line in Figure~\ref{fig:aggregated_fit} shows that our model's
predictions are very close to officially recorded cases in the aggregate. This is also
true for infections by age and geographical region, which are shown in the supplementary
materials (Figures~\ref{fig:age_group_fit} and \ref{fig:state_fit}, respectively).

The effects of various mechanisms can be disentangled due to the distinct temporal
variation in the drivers of the pandemic. Next to the stringency index, the three lines
in Figure~\ref{fig:stringency_infectious_contacts} summarize how contact reductions,
increased hygiene regulations, and seasonality evolved since early September for each of
the three broad contact networks. For example, a value of 0.75 for the work multiplier
means that if the environment was the same as in September (levels of infection rates, no
rapid tests or vaccinations, only the wildtype virus present), infections at the
workplace would be reduced by 25\%. The lines show the product of the effect of contact
reductions, increased hygiene regulations, and seasonality. Two aspects are particularly
interesting. First, all lines broadly follow the stringency index and they would do so
even more if we left out seasonality and school vacations (roughly the last two weeks of
October, two weeks each around Christmas and Easter, and some days in late May). Second,
the most stringent regulations are associated with the period of strong decreases in new
infections between late December 2020 and mid-February 2021. The reversal of the trend is
associated with the spread of the B.1.1.7 variant. The steep drop in recorded cases
during May 2021 is associated with at least weekly rapid tests for 42\% of the
population, a vaccination rate that rose from 28\% to 43\%, and \deleted[id=K]{mostly}
seasonality \replaced[id=K]{further lowering}{impacting a fall in} the relative
infectiousness of contacts \deleted[id=K]{outside of work and school.}

\begin{figure}[!tp]
    \centering

    \begin{subfigure}[b]{0.475\textwidth}
        \centering
        \includegraphics[width=\textwidth]{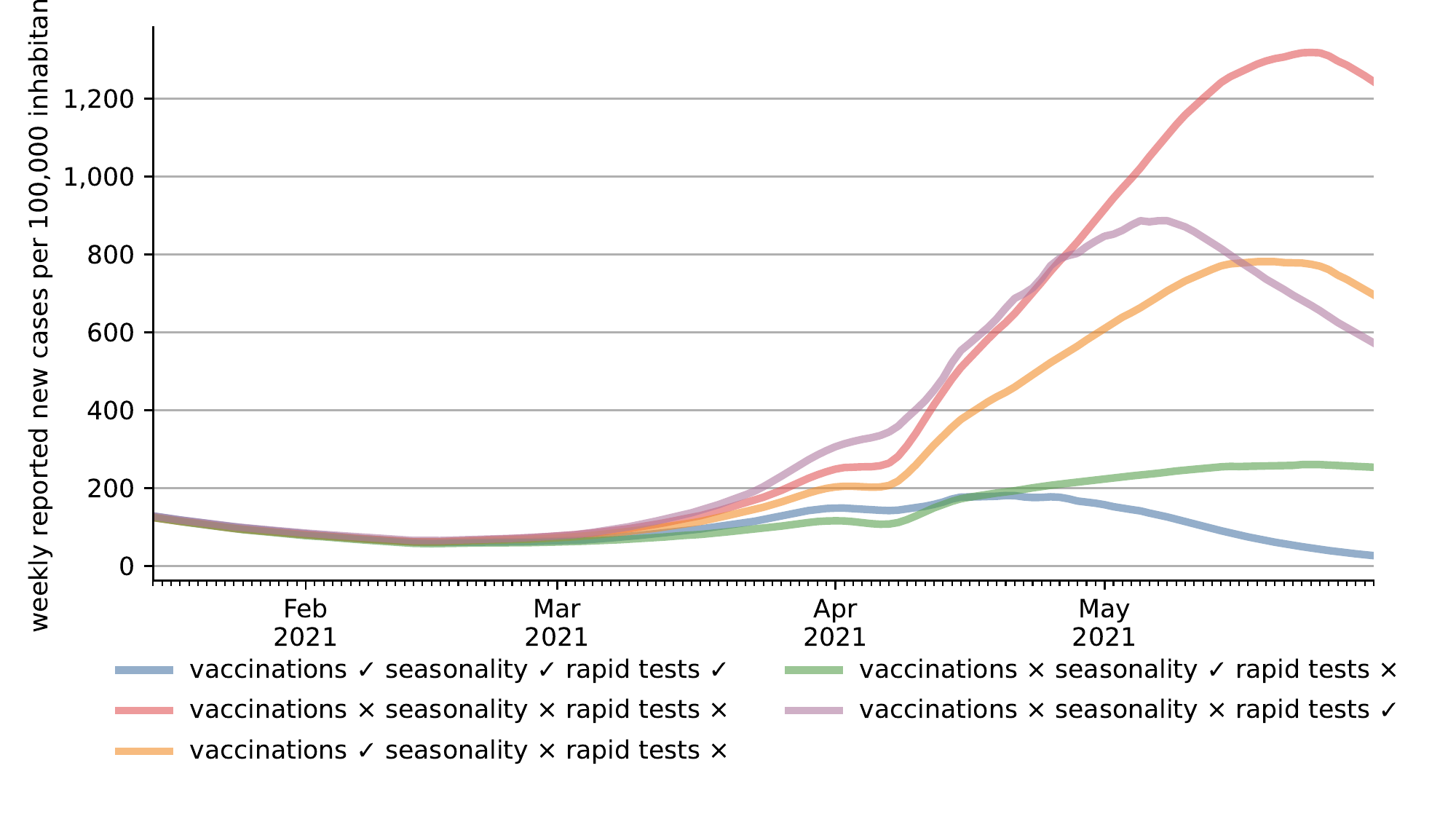}
        \caption{{Recorded cases: 2021 scenarios}}
        \label{fig:2021_scenarios_recorded}
    \end{subfigure}
    \hfill
    \begin{subfigure}[b]{0.475\textwidth}
        \centering
        \includegraphics[width=\textwidth]{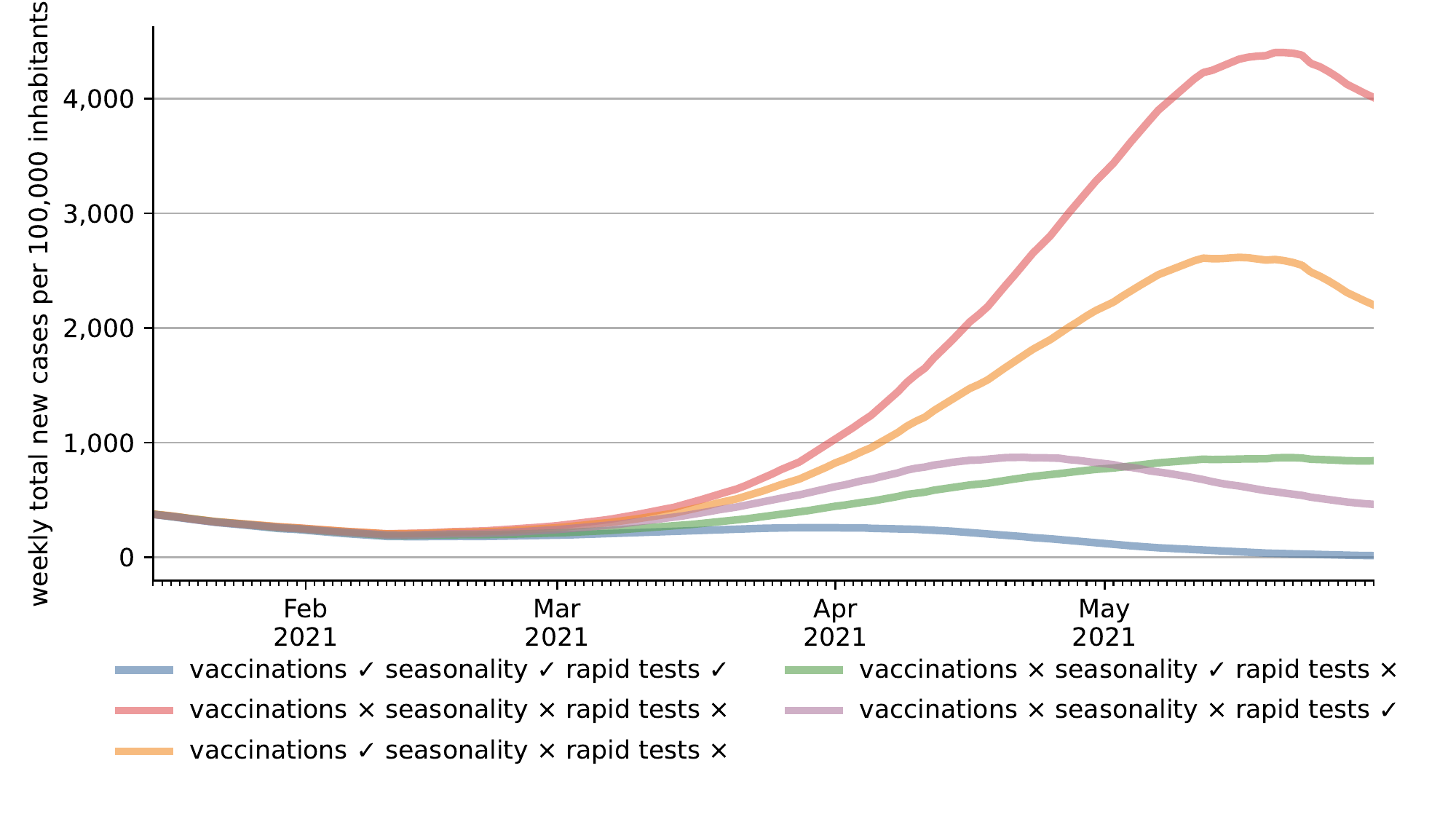}
        \caption{{Total cases: 2021 scenarios}}
        \label{fig:2021_scenarios_newly_infected}
    \end{subfigure}

    \begin{subfigure}[b]{0.475\textwidth}
        \centering
        \includegraphics[width=\textwidth]{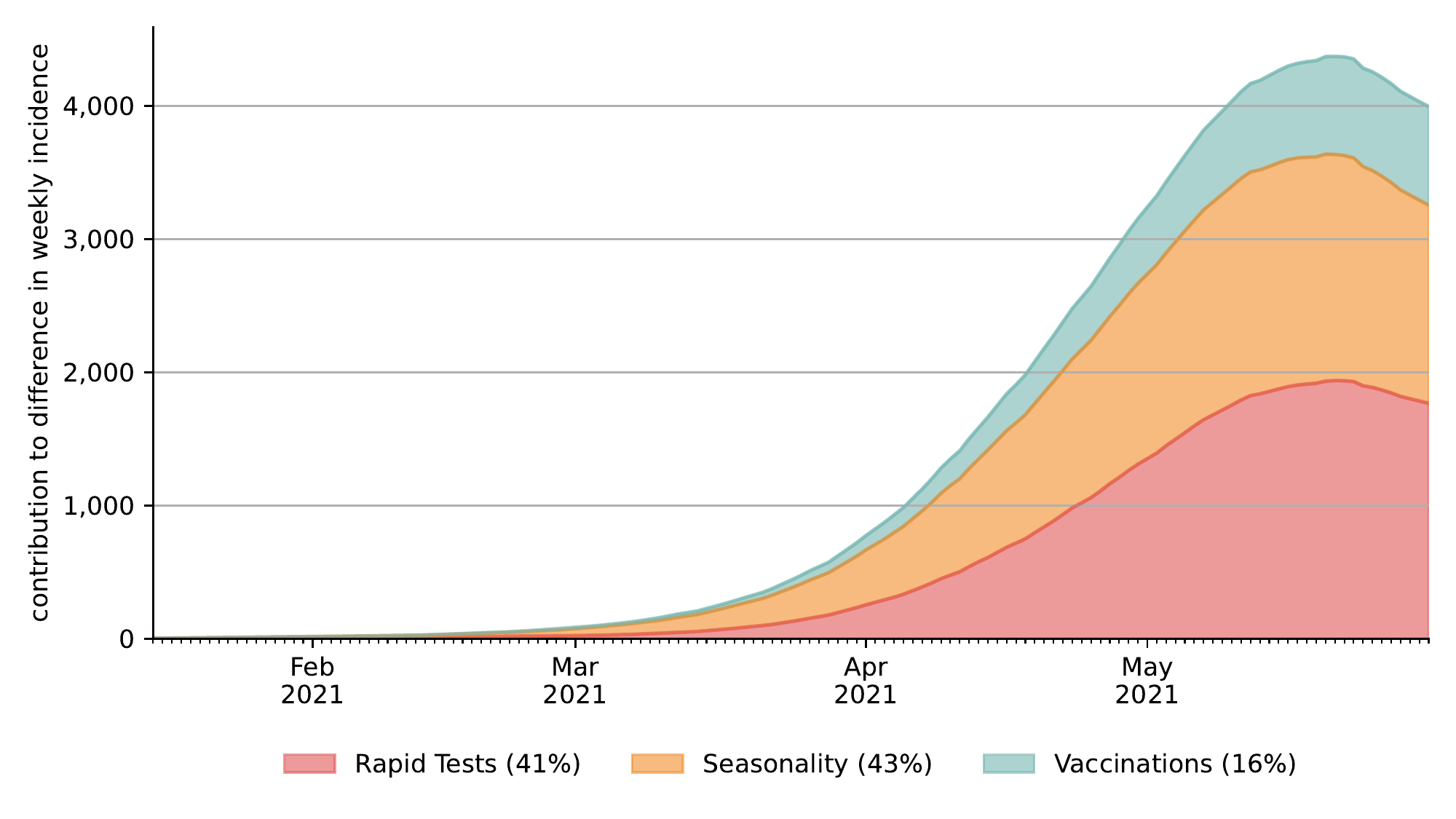}
        \caption{Decomposition of the difference between the scenario without any of the
            three factors and the main scenario in
            Figure~\ref{fig:2021_scenarios_newly_infected}.}
        \label{fig:2021_scenarios_decomposition}
    \end{subfigure}
    \hfill
    \begin{subfigure}[b]{0.475\textwidth}
        \centering
        \includegraphics[width=\textwidth]{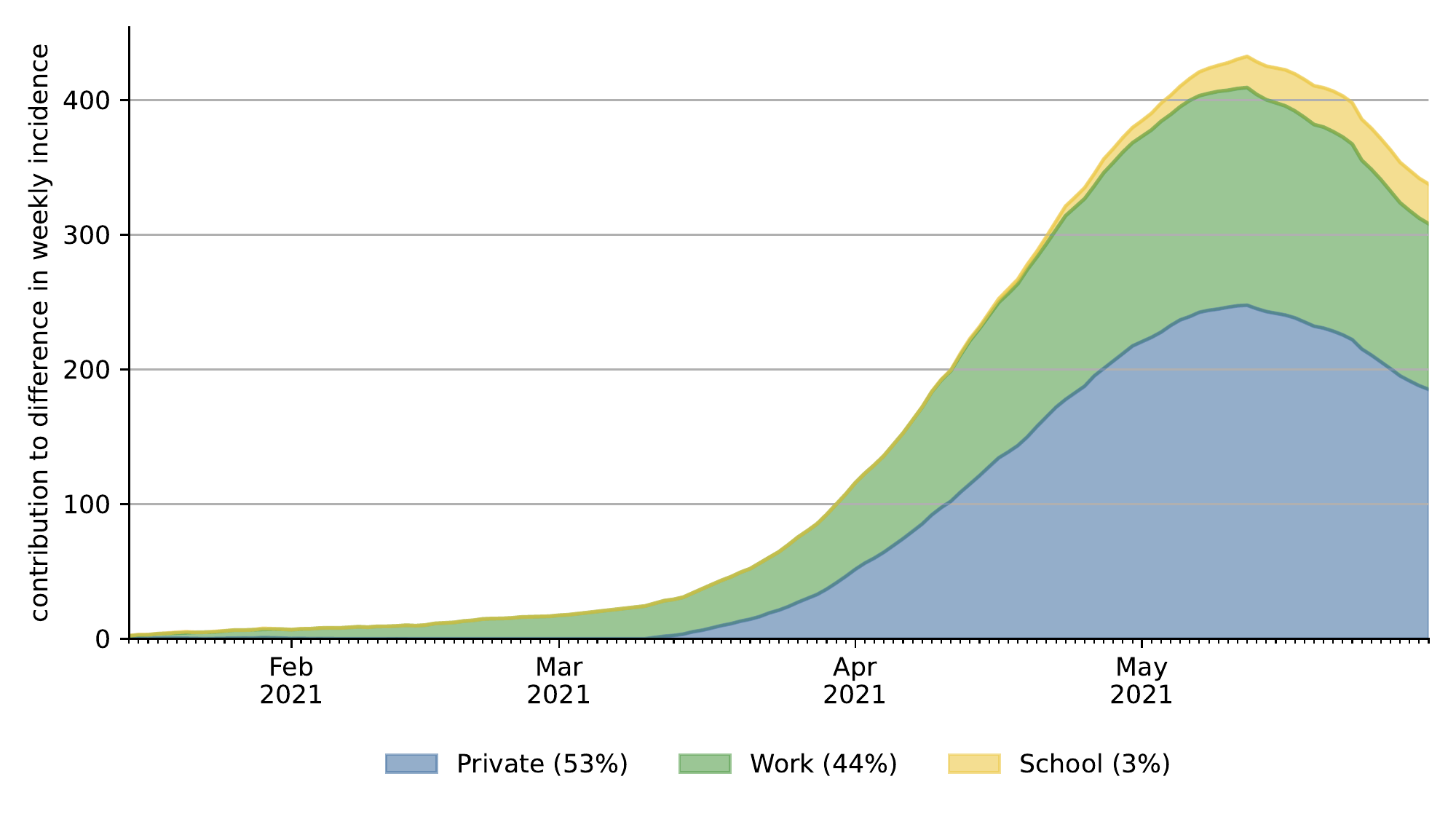}
        \caption{Decomposition of the difference between the scenario without rapid
            tests and the main scenario in
            Figure~\ref{fig:2021_scenarios_newly_infected}.}
        \label{fig:2021_scenarios_decomposition_tests}
    \end{subfigure}

    \caption{The effect of different interventions on recorded and actual infections}
    \label{fig:2021_scenarios_broad}

    \floatfoot{\noindent \textit{Note:} The blue line in
        Figure~\ref{fig:2021_scenarios_recorded} is the same as in
        Figure~\ref{fig:aggregated_fit} and refers to our baseline scenario, so does the
        blue line in Figure~\ref{fig:2021_scenarios_newly_infected}. The red lines refer
        to a situation where NPIs evolve as in the baseline scenario and the B.1.1.7
        variant is introduced in the same way; vaccinations, rapid tests, and
        seasonality remain at their January levels. The other scenarios turn these three
        factors on one-by-one. The decompositions in
        Figures~\ref{fig:2021_scenarios_decomposition} and
        \ref{fig:2021_scenarios_decomposition_tests} are based on Shapley values, which
        are explained more thoroughly in Appendix~\ref{sub:shapley_value}}
\end{figure}

In order to better understand the contributions  of rapid tests, vaccinations, and
seasonality on the evolution of infections in 2021, Figure~\ref{fig:2021_scenarios_broad}
considers various scenarios. NPIs are always the same as in the baseline scenario.
Figure~\ref{fig:2021_scenarios_recorded} shows the model fit (the blue line, same as in
Figure~\ref{fig:aggregated_fit}), a scenario without any of the three factors (red line),
and three scenarios turning \replaced[id=K]{each factor on individually}{these factors
off one by one}. Figure~\ref{fig:2021_scenarios_newly_infected} does the same for total
infections in the model. Figure~\ref{fig:2021_scenarios_decomposition} employs Shapley
values \citep{Shapley2016} to decompose the difference in total infections between the
scenario without any of the three factors and our main specification.

Until mid-March, there is no visible difference between the different scenarios.
Seasonality hardly changes, and only few vaccinations and rapid tests were administered.
Even thereafter, the effect of the vaccination campaign is surprisingly small at first
sight. Whether considering recorded or total infections with only one channel active,
the final level is always the highest in case of the vaccination campaign (orange
lines). The Shapley value decomposition shows that vaccinations contribute 16\% to the
cumulative difference between scenarios. Reasons for this are the slow start---it took
until March~24th until 10\% of the population had received their first vaccination, the
20\% mark was reached on April 19th---and the focus on older individuals. These groups
contribute less to the spread of the disease than others due to a lower number of
contacts\replaced[id=K]{as they do not work, do not go to school and tend to live in small households}{, see Supplementary Material~\ref{subsec:contacts_by_age}}.
By the end of our study period, when first-dose vaccination rates reached around 40\% of
the population, the numbers of new cases would have started to decline. It is important
to note that the initial focus of the campaign was to prevent deaths and severe disease.
Indeed, the case fatality was rate considerably lower during the third wave when compared
to the second (4.4\% between October and February and 1.4\% between March and the end of
May).\comment[id=K]{This needs a source.}


Seasonality has a large effect in slowing the spread of SARS-CoV-2. By May 31, both
observed and recorded cases would be reduced by a factor of four if only seasonality
mattered. However, in this period, cases would have kept on rising throughout, just at a
much lower pace \citep[this is in line with results in][, which our seasonality measure
is based on]{Gavenciak2021}. Nevertheless, we estimate seasonality to be a
quantitatively important factor determining the evolution of the pandemic, explaining
most of the early changes and 43\% of the cumulative difference by the end of May.

A similar-sized effect---41\% in the decompositions---comes from rapid testing. Here, it
is crucial to differentiate between recorded cases and actual cases. Additional testing
means that additional infections will be recorded which would otherwise remain
undetected. Figure~\ref{fig:2021_scenarios_recorded} shows that this effect is large.
Until late April, recorded cases are higher in the scenario with rapid testing alone
when compared to the setting where none of the three mechanisms are turned on. The
effect on total cases, however, is visible immediately in
Figure~\ref{fig:2021_scenarios_newly_infected}. Despite the fact that only 10\%  of the
population performed weekly rapid tests in March on average, new infections on April 1
would be reduced by 53\% relative to the scenario without vaccinations, rapid tests, or
seasonality.

So why is rapid testing so effective? In order to shed more light on this question,
Figure~\ref{fig:2021_scenarios_decomposition_tests} decomposes the difference in the
scenario without rapid tests only (purple line in
Figure~\ref{fig:2021_scenarios_newly_infected}) and the main specification into the three
channels for rapid tests. Tests at schools have the smallest effect, which is largely
explained by schools not operating at full capacity during our period of study and the
relatively small number of students.\footnote{18\% of our population are in the education
sector (pupils, teachers, etc.); 46\% are workers outside the education sector.} Almost
40\% come from tests at the workplace. Despite the fact that rapid tests for private
reasons are phased in only late, they make up for more than half of the total effect. The
reason lies in the fact that a substantial share of these tests is driven by an elevated
probability to carry the virus, i.e., showing symptoms of CoViD-19 or following up on a
positive test of a household member. The latter is essentially a form of contact tracing,
which has been shown to be very effective \citep{Contreras2021,
Kretzschmar2020,FetzerGraeber2021}.

\begin{figure}[!tp]
    \centering

    \begin{subfigure}[b]{0.475\textwidth}
        \centering
        \includegraphics[width=0.9 \textwidth]{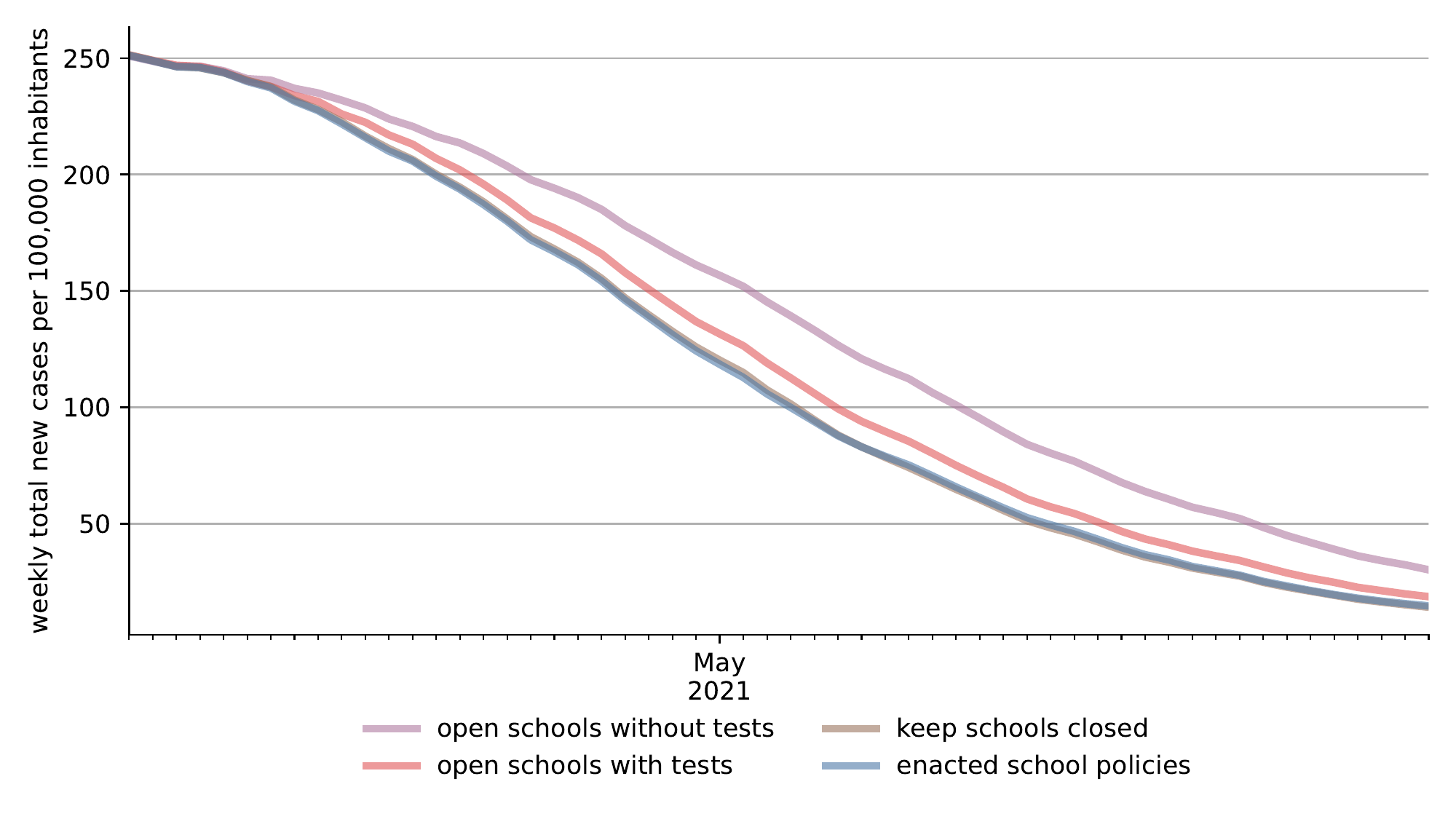}
        \caption{{Effects of different schooling scenarios}}
        \label{fig:school_scenarios}
    \end{subfigure}
    \hfill
    \begin{subfigure}[b]{0.475\textwidth}
        \centering
        \includegraphics[width=0.9 \textwidth]{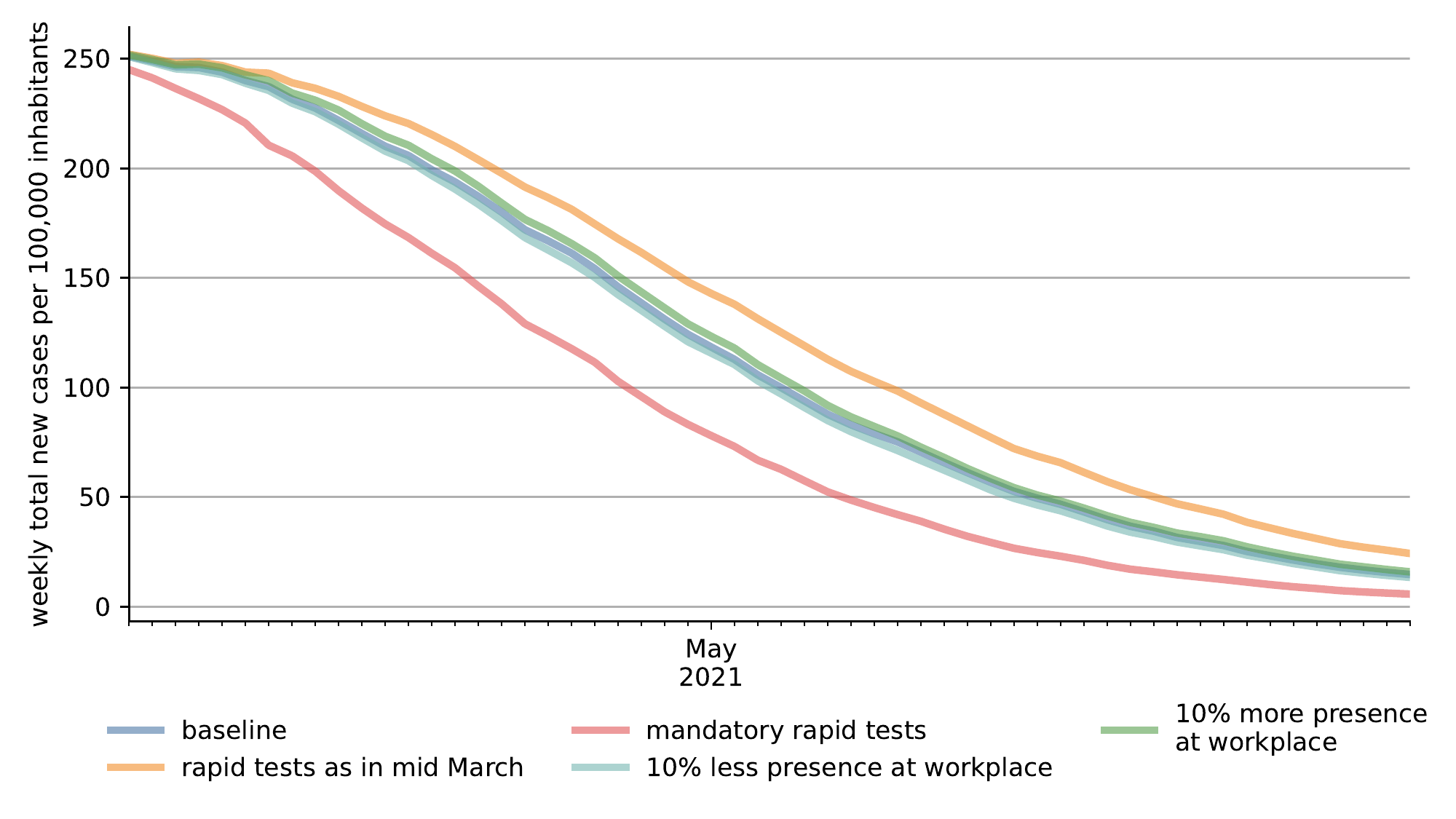}
        \caption{{Effects of different work scenarios}}
        \label{fig:workplace_scenarios}
    \end{subfigure}
    \vskip3ex

    \caption{Effects of different scenarios for policies regarding schools and workplaces.}
    \label{fig:school_workplace_scenarios}

    \floatfoot{\noindent \textit{Note:} Blue lines in both figures refer to our baseline
        scenario; they are the same as in
        Figure~\ref{fig:2021_scenarios_newly_infected}. Interventions start at Easter
        because there were no capacity constraints for rapid tests afterwards.}

\end{figure}

Two of the most contentious NPIs concern schools and mandates to work from home. In many
countries, schools switched to remote instruction during the first wave, so did Germany.
After the summer break, they were operating at full capacity with increased hygiene
measures, before being closed again from mid-December onwards. Some states started
opening them gradually in late February, but operation at normal capacity did not resume
until the beginning of June. Figure~\ref{fig:school_scenarios} shows the effects of
different policies regarding school starting at Easter, at which point rapid tests had
become widely available. We estimate the realized scenario to have essentially the same
effect as a situation with closed schools. Under fully opened schools with mandatory
tests, total infections would have been 6\% higher; this number rises to 20\% without
tests. These effect sizes are broadly in line with empirical studies
\citep[e.g.,][]{Vlachos2021}. \comment[id=HM]{Need a back-of the envelope calculation} To use
another metric, the effective weekly reproduction number differs by $0.018$ and $0.052$,
respectively. In light of the large negative effects school closures have on children
and parents \citep{Luijten2021, Melegari2021}---and in particular on those with low
socio-economic status---these results in conjunction with hindsight bias suggest that
opening schools combined with a testing strategy would have been beneficial. In other
situations, and particular when rapid test are not available at scale, trade-offs may
well be different.

Figure~\ref{fig:workplace_scenarios} shows that with a large fraction of workers
receiving tests, testing at the workplace has larger effects than mandating employees to
work from home. Whether the share of workers working at the usual workplace is reduced
or increased by ten percent changes infection rates by 2.5\% or less in either
direction. Making testing mandatory twice a week---assuming independent compliance by
employers and workers of 95\% each---would have reduced infections by 23\%. Reducing
rapid tests offers by employers to the level of March would have increased infections by
13\%.

Our analysis has shown that during the transition to high levels of vaccination and
possibly thereafter, large-scale rapid testing can substitute for some NPIs. This comes
at a fraction of the cost. A week of the fairly strict lockdown in early 2021 is
estimated to have cost around 20~Euros per capita \citep{Wollmershauser2021}; retail
prices for rapid tests were below one Euro in early June 2021. Despite these large
effects, the results on testing likely understate the benefits. Disadvantaged groups are
less likely to be reached by testing campaigns relying on voluntary participation
\citep[e.g.][]{StillmanTonin2021}; at the same time, these groups have a higher risk to
contract CoViD-19 \citep{KochInstitut2021a}. Mandatory tests at school and at the
workplace will extend more into these groups. The same goes for individuals who exhibit
a low level of compliance with CoViD-19-related regulations. Compared to vaccinations,
rapid testing programmes allow a much quicker roll-out, making it arguably the most
effective tool to contain the pandemic in the short run.

\phantomsection 

\begin{refcontext}[sorting=nyt]  
\printbibliography[heading=bibintoc]
\end{refcontext}

\clearpage

\begin{center}
    {
        \Large 
        Supplementary Material for\\[2ex]
    
        \textbf{\papertitle}
    }
    \vskip4ex
    {
        \large
        Janoś Gabler,
        
        Tobias Raabe,
        
        Klara Röhrl,
        
        Hans-Martin von Gaudecker

    }
\end{center}

\clearpage

\begin{appendices}
    \begin{refsection}

    \section{Data and Parameters}
\label{sec:data_and_parameters}

The model is described by a large number of parameters that govern the number of
contacts a person has, the likelihood of becoming infected on each contact, the
likelihood of developing light or strong symptoms or even dying from the disease as well
as the duration each stage of the disease takes.

Most of these parameters can be calibrated from existing datasets or the medical
literature or calibrated from surveys and empirical datasets.

\subsection{Medical Parameters}
\label{sec:medical_params}

This section discusses the medical parameters used in the model, their sources and how we
arrived at the distributions used in the model.\footnotemark See
Figure~\ref{fig:model_disease_progression} for a summary of our disease progression model.

\footnotetext{Additional information can be found in the
\href{https://sid-dev.readthedocs.io/en/latest/reference_guides/epi_params.html}{online
documentation}.}


The first medical parameter we need is the length of the period between infection and the
start of infectiousness, the so called latent period. We infer it from two other measures
that are more common in the medical literature: Firstly, the time between infection and
the onset of symptoms, the incubation period. Secondly, the time between the start of
infectiousness and the onset of symptoms. We assume that the latency period is the same
for symptomatic and asymptomatic individuals.

Once individuals become infectious a share of them goes on to develop symptoms while
others remain asymptomatic. We rely on data by \cite{Davies2020} for the age-dependent
probability to develop symptoms. It varies from 25\% for children and young adults to
nearly 70\% for the elderly.

The incubation period is usually estimated to be two to twelve days. A meta analysis by
\citet{McAloon2020} comes to the conclusion that ``The incubation period distribution may
be modeled with a lognormal distribution with pooled $\mu$ and $\sigma$ parameters (95\%
CIs) of 1.63 (95\% CI 1.51 to 1.75) and 0.50 (95\% CI 0.46 to 0.55), respectively.'' For
simplicity we discretize this distribution into four bins.

The \href{https://bit.ly/3pi18Ax}{European Centre for Disease Prevention and Control}
reports that people become infectious between one and two days before symptoms
start.\footnotemark

\footnotetext{This is similar to \citet{He2020} and in line with \citet{Peak2020}.}

Taking these estimates together, we arrive at a latent period of one to five days.


We assume that the duration of infectiousness is the same for both symptomatic and
asymptomatic individuals as evidence suggests little differences in the transmission
rates between symptomatic and asymptomatic patients (\citet{Yin2020}) and that the viral
load between symptomatic and asymptomatic individuals are similar (\citet{Zou2020},
\citet{Byrne2020}, \citet{Singanayagam2020}). Our distribution of the duration of
infectiousness is based on \citet{Byrne2020}. For symptomatic cases they arrive at zero
to five days before symptom onset (see their figure 2) and three to eight days of
infectiousness afterwards.\footnote{Viral loads may be detected much later but eight days
seems to be the time after which most people are culture negative, as also reported by
\citet{Singanayagam2020}.} Thus, we arrive at 0 to 13 days as the range for
infectiousness among individuals who become symptomatic (see also figure 5).


We use the duration to recovery of mild and moderate cases reported by \cite[Figure~S3,
Panel~2]{Bi2020} for the duration of symptoms for non-ICU requiring symptomatic cases. We
only disaggregate by age how likely individuals are to require intensive
care.\footnote{The length of symptoms is not very important in our model given that
individuals mostly stop being infectious before their symptoms cease.}


For the time from symptom onset until need for intensive care we rely on data by the US
CDC (\citet{Stokes2020}) and \href{https://bit.ly/3yKXFyu}{the OpenABM-Project}.
%
%
%
%

For those who will require intensive care we follow \citet{Chen2020} who estimate the
time from symptom onset to ICU admission as 8.5 $\pm$ 4 days. This aligns well with
numbers reported for the time from first symptoms to hospitalization:
\citet{Gaythorpe2020} report a mean of 5.76 with a standard deviation of 4. This is also
in line with the duration estimates collected by \href{https://bit.ly/3gkGtaU}{the
Robert-Koch-Institut}. We assume that the time between symptom onset and ICU takes 4, 6,
8 or 10 days with equal probabilities. As we do not model nursing homes, do not focus on
matching deaths and do not use the number of individuals in intensive care to estimate
our parameters, these numbers are not important for our empirical results.


We take the survival probabilities and time to death and time until recovery from
intensive care from the \href{https://tinyurl.com/y5owhyts}{OpenABM Project}. They report
time until death to have a mean of 11.74 days and a standard deviation of 8.79 days. To
match this approximately we discretize that 41\% of individuals who will die from
Covid-19 do so after one day in intensive care, 22\% day after 12 days, 29\% after 20
days and 7\% after 32 days. Again, we rescale this for every age group among those that
will not survive. For survivors the \href{https://tinyurl.com/y5owhyts}{OpenABM Project}
reports a mean duration of 18.8 days until recovery and a standard deviation of 12.21
days. We discretize this such that of those who recover in intensive care, 22\% do so
after one day, 30\% after 15 days, 28\% after 25 days and 18\% after 45 days.

\FloatBarrier

\subsection{The Synthetic Population \comment[id=K]{To be written}}

We build a synthetic population based on the German microcensus \citep{FDSAeDBUDL2018}.
We only use private households, i.e. exclude living arrangements such as nursing homes as
non-private households vary widely in size and it is very difficult to know which
contacts take place in such households.

We sample households to build our synthetic population of over one million households
keeping for each individual their age, gender, occupation and whether they work on
Saturdays and Sundays.

\FloatBarrier

\subsection{Number of Contacts}
\label{sub:number_of_contacts}

We calibrate the parameters for the predicted numbers of contacts from contact diaries
of over 2000 individuals from Germany, Belgium, the Netherlands and Luxembourg
\citep{Mossong2008}. Each contact diary contains all contacts an individual had
throughout one day, including information on the other person (such as age and gender)
and information on the contact. Importantly, for each contact individuals entered of
which type the contact (school, leisure, work etc.) was and how frequent the contact
with the other person is.

Simplifying the number of contacts, we arrive at the following distributions of the
numbers of contacts by contact type.

\begin{figure}
    \centering

    \begin{subfigure}[b]{0.25\textwidth}
        \centering

        \includegraphics[width=\textwidth]{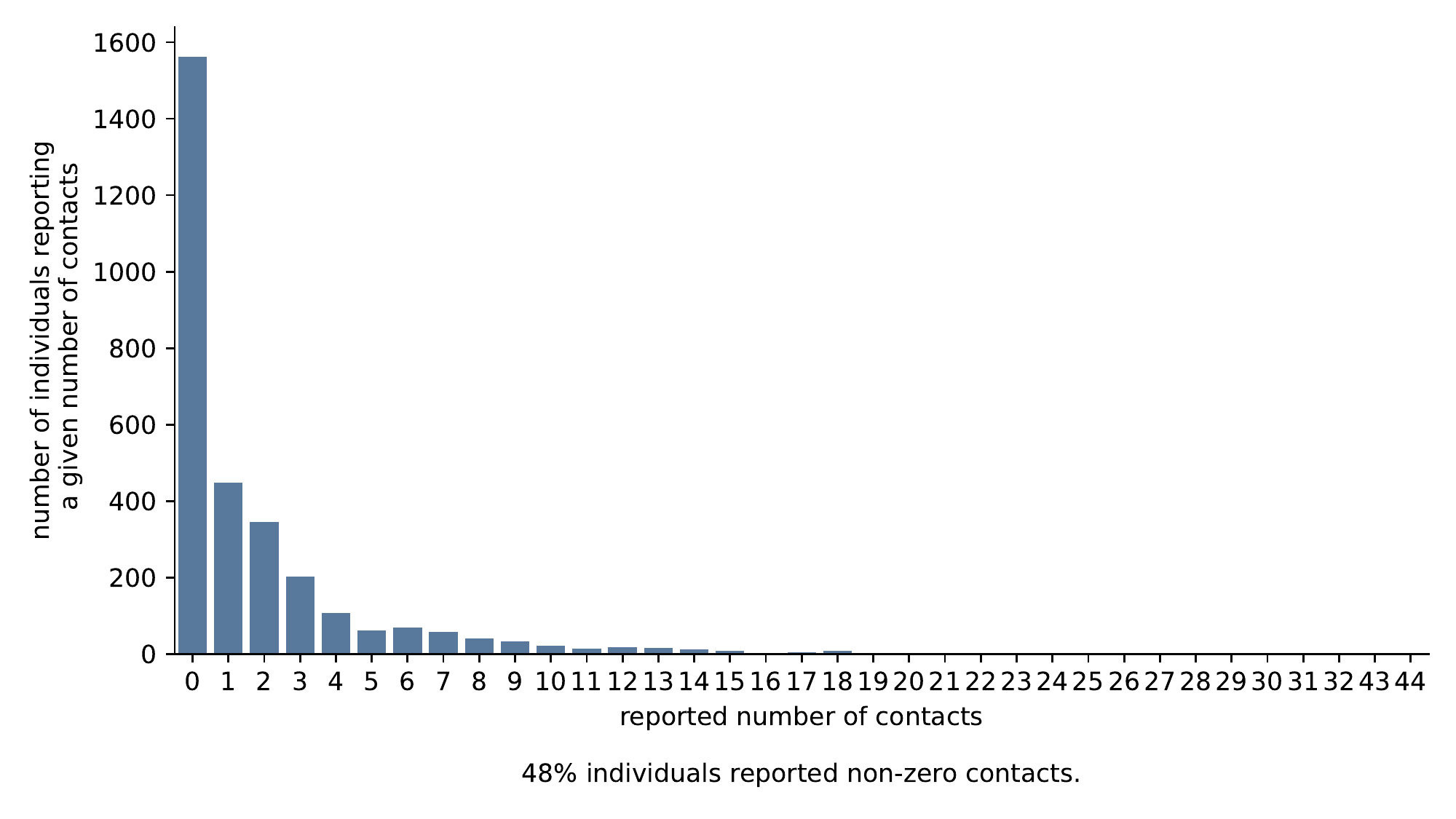}
        \caption{Number of Non Recurrent Other Contacts}
        \label{n_contacts_other_non_recurrent}
    \end{subfigure}

    \hfill

    \begin{subfigure}[b]{0.25\textwidth}
        \centering
        \includegraphics[width=\textwidth]{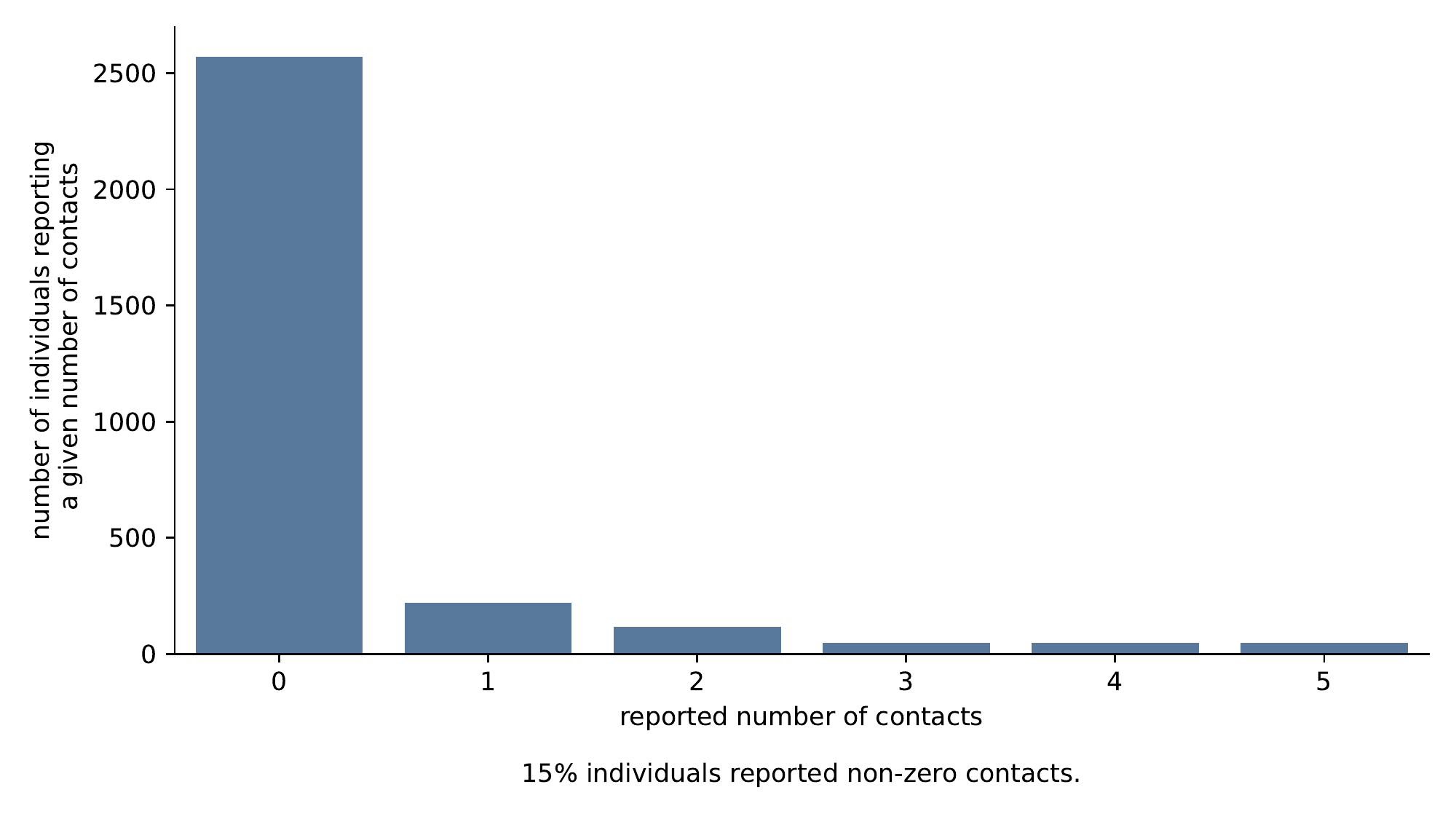}
        \caption{Number of Daily Recurrent Other Contacts}
        \label{n_contacts_other_daily_recurrent}
    \end{subfigure}

    \hfill

    \begin{subfigure}[b]{0.25\textwidth}
        \centering
        \includegraphics[width=\textwidth]{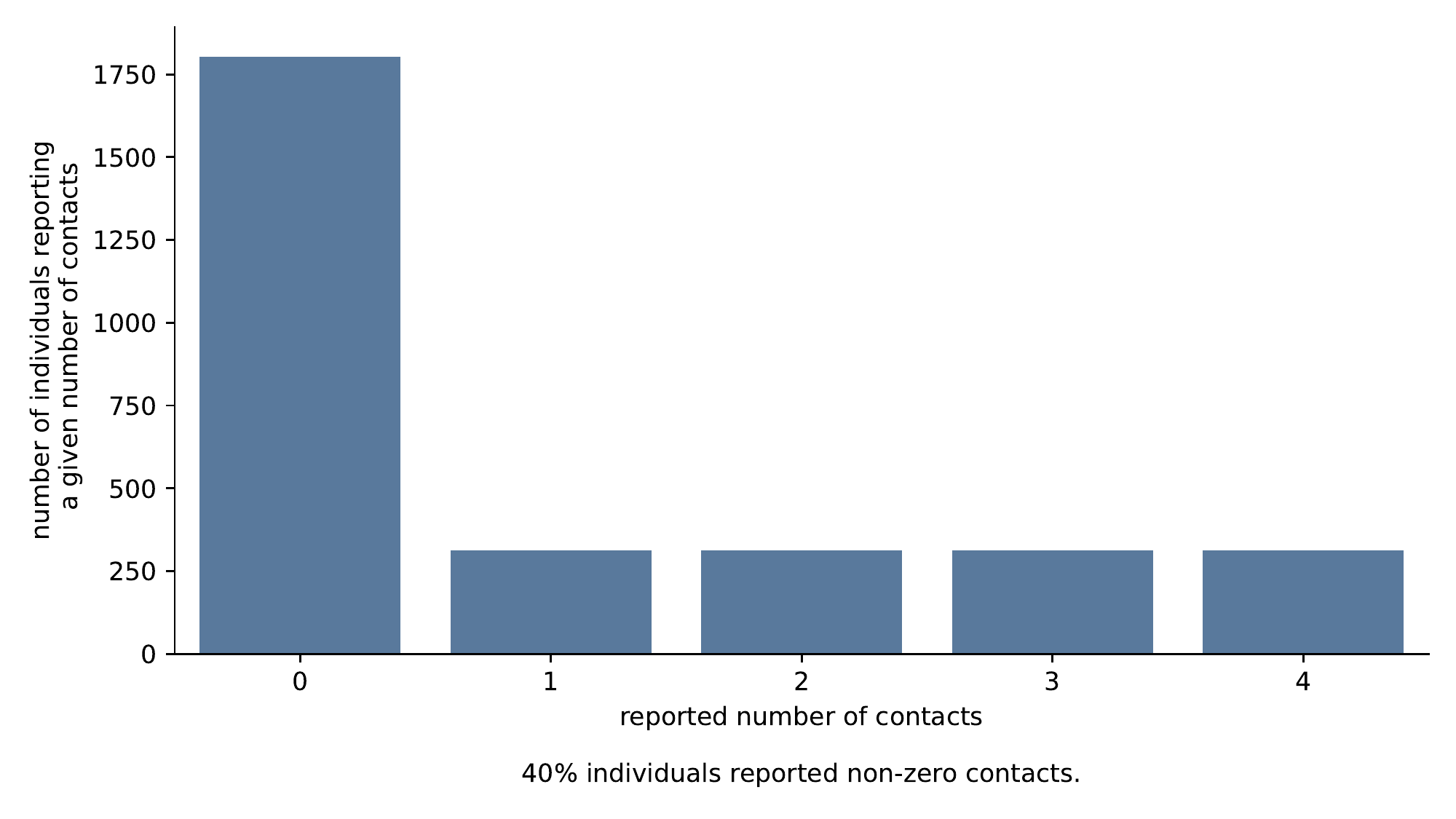}
        \caption{Number of Weekly Recurrent Other Contacts}
        \label{n_contacts_other_weekly_recurrent}
    \end{subfigure}

    \caption{Number of Contacts of the Other Contact Type}
    \label{fig:n_contacts_other}

    \floatfoot{\noindent
        \textit{Note:} Other contacts include all contacts that are not household
        members, school contacts or work contacts, for example leisure contacts or
        contacts during grocery shopping. The planned number of contacts is reduced by
        policies, seasonality and individual responses to events such as receiving a
        positive rapid test to the number of actual contacts with transmission potential.        
        In the model it is sampled every day which of the numbers of non recurrent
        contacts a person is planned to have. Note that the contact diaries include such
        high values that super spreading events are well possible in our model through
        non recurrent other models. We assume that individuals in households with
        children or teachers or retired individuals have additional non recurrent
        contacts during school vacations to cover things like family visits or travel
        during vacations. We estimate this to be on average 0.5 additional contacts per
        vacation day.
        For the recurrent other contacts, individuals are assigned to groups that are
        time constant and that meet daily or weekly. The share of individuals who attend
        in a way that has transmission potential is reduced by policies, seasonality and
        individual responses to events such as receiving a positive rapid test. For
        weekly contacts, individuals are assigned to up to four groups that are time
        constant and that meet weekly. The day on which meetings take place varies
        between groups but stays the same for each group.
    }
\end{figure}


\begin{figure}
    \centering
    \includegraphics[width=\textwidth]{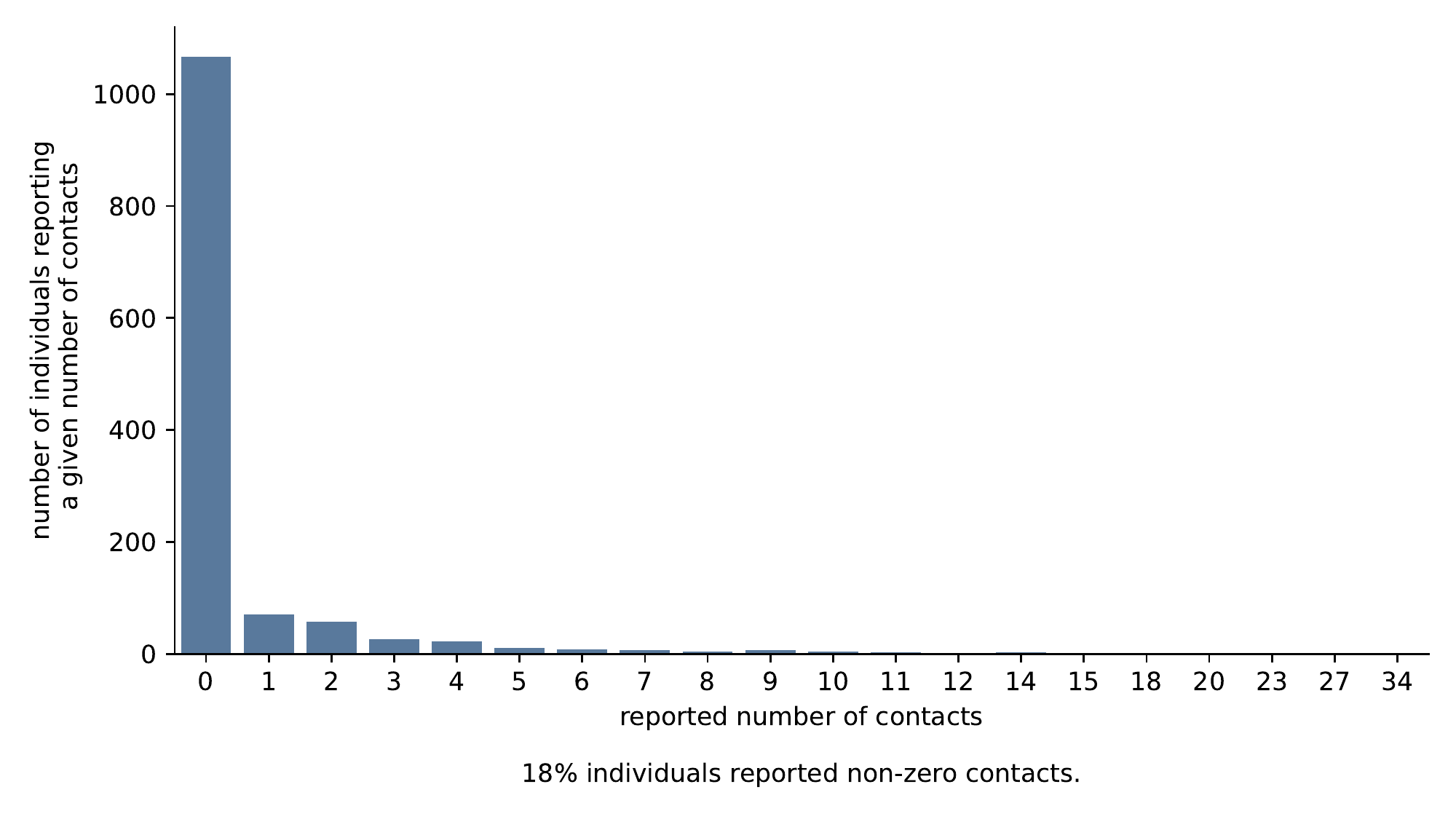}
    \caption{Number of Non Recurrent Work Contacts}
    \label{n_contacts_work_non_recurrent}
    \floatfoot{\noindent \textit{Note:} In the model it is sampled every day which of
        these numbers of contacts a working person is planned to have. Note that the
        contact diaries include such high values that super spreading events are well
        possible in our model. The planned number of contacts is reduced by policies,
        seasonality and individual responses to events such as receiving a positive rapid
        test to the number of actual contacts with transmission potential. Work contacts
        only take place between working individuals.}
\end{figure}

\begin{figure}
    \centering
    \includegraphics[width=\textwidth]{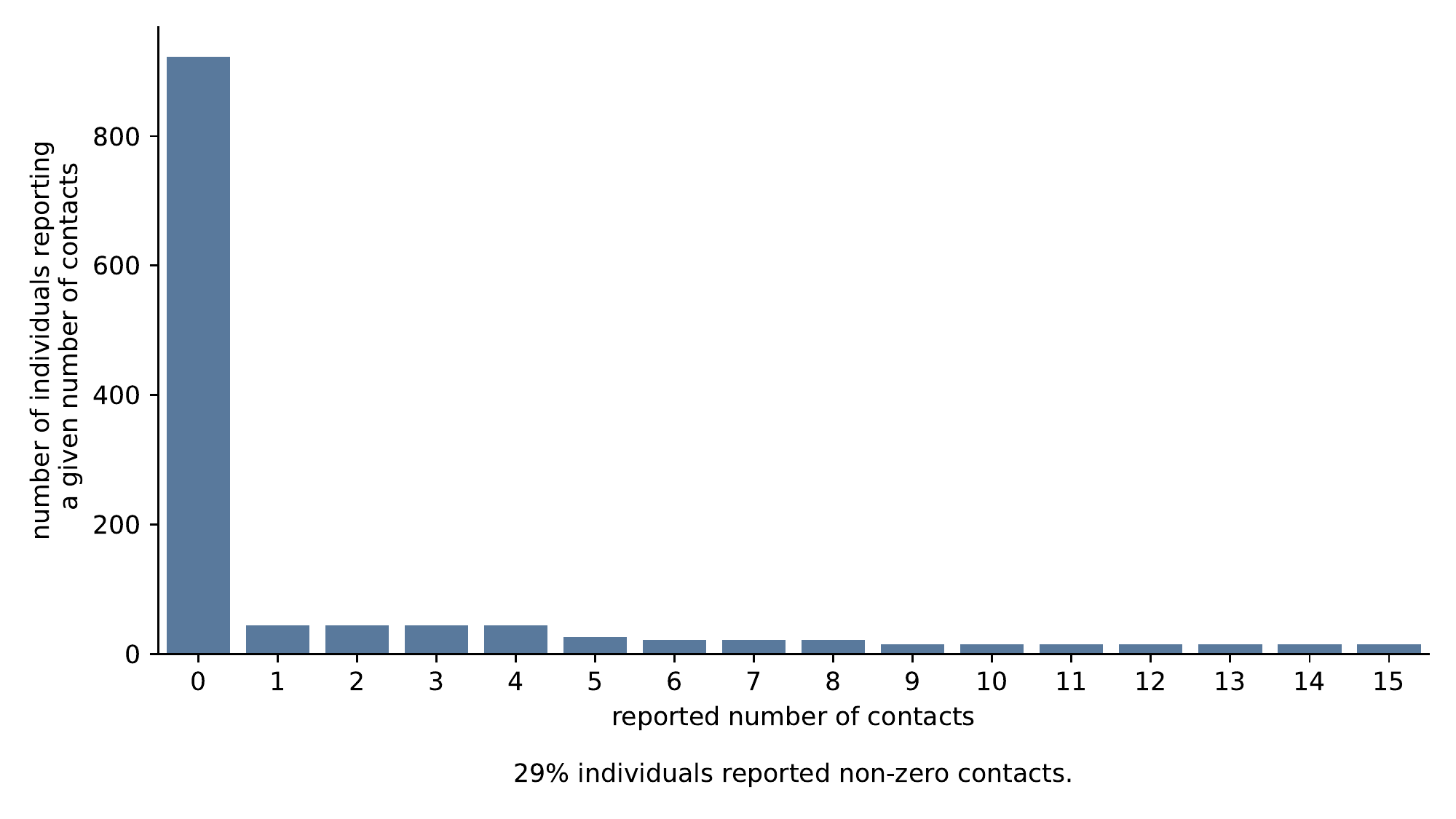}
    \caption{Number of Daily Recurrent Work Contacts}
    \label{n_contacts_work_daily_recurrent}
    \floatfoot{\noindent \textit{Note:} Working individuals are assigned to groups that
        are time constant and that meet daily to match the given distribution of daily
        work contacts. You can think of these as for example colleagues with which one
        shares an office space. The share of individuals who attend in a way that has
        transmission potential is reduced by policies (such as a work from home mandate),
        seasonality and individual responses to events such as receiving a positive rapid
        test. Work contacts only take place between working individuals.}
\end{figure}

\begin{figure}
    \centering
    \includegraphics[width=\textwidth]{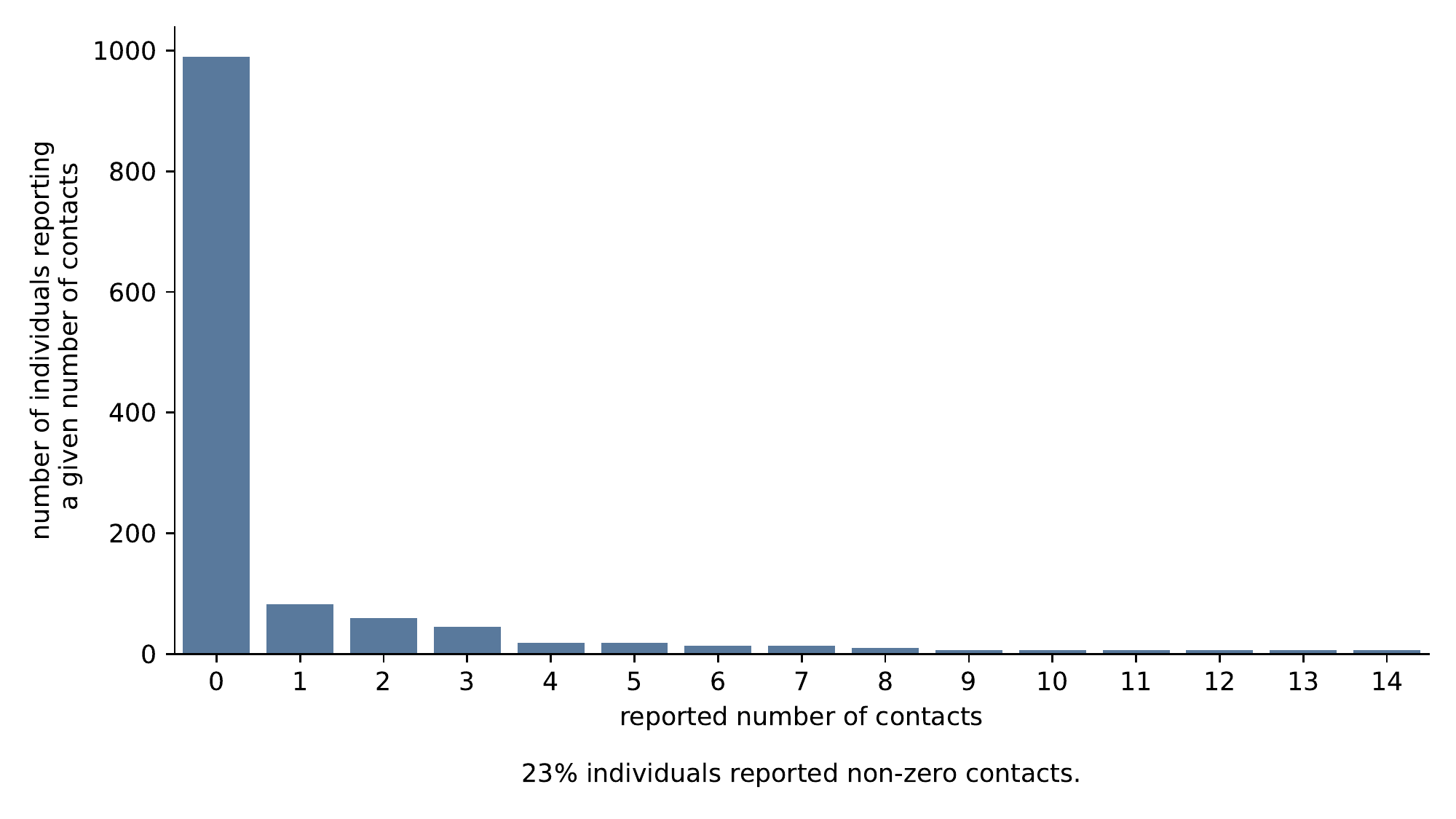}
    \caption{Number of Weekly Recurrent Work Contacts}
    \label{n_contacts_work_weekly_recurrent}
    \floatfoot{\noindent  \textit{Note:} Working individuals are assigned to up to 14
        groups that are time constant and meet weekly. Groups are scheduled to meet on
        separate days of the work week. These contact models cover weekly team meetings
        etc. The share of individuals that attend in a way that has transmission
        potential is reduced by policies, seasonality and individual responses to events
        such as receiving a positive rapid test. Work contacts only take place between
        working individuals.}
\end{figure}

An exception where we do not rely on the data by \cite{Mossong2008} are the household
contacts. Since household are included in the the German microcensus
\citep{FDSAeDBUDL2018} on which we build our synthetic population we simply assume for
the household contact model that individuals meet all other household members every day.

\begin{figure}
    \centering
    \includegraphics[width=\textwidth]{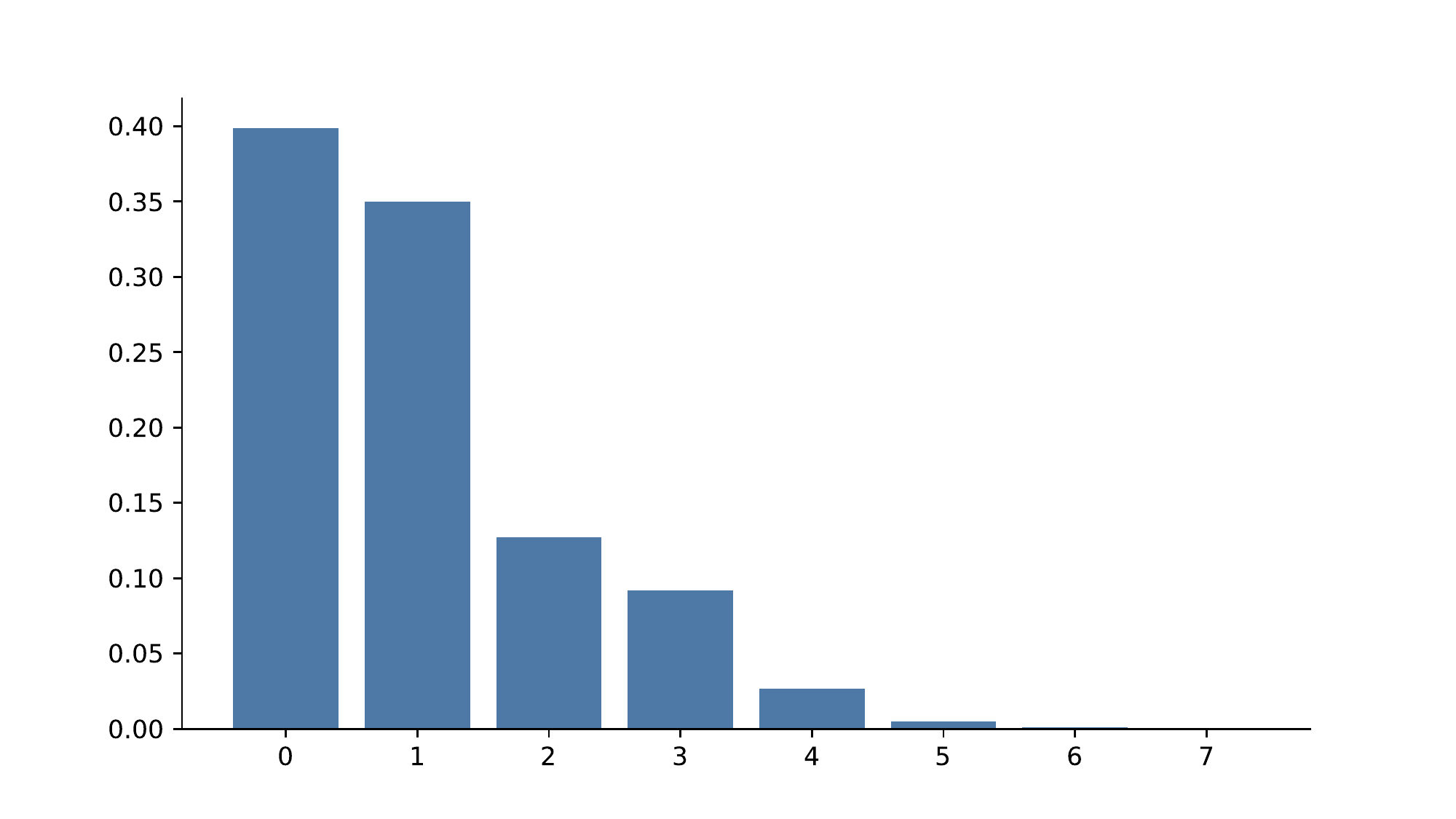}
    \caption{Number of Household Contacts}
    \label{n_contacts_hh}
    \floatfoot{\noindent
        \textit{Note:} Every individual meets all other household members every day. The
        German microcensus sampled full households such that our synthetic population
        automatically fits population characteristics such as size and age distribution.}
\end{figure}

\FloatBarrier

\subsection{Contacts by age}
\label{subsec:contacts_by_age}

\begin{figure}[ht]
    \centering
    \includegraphics[width=0.9 \textwidth]{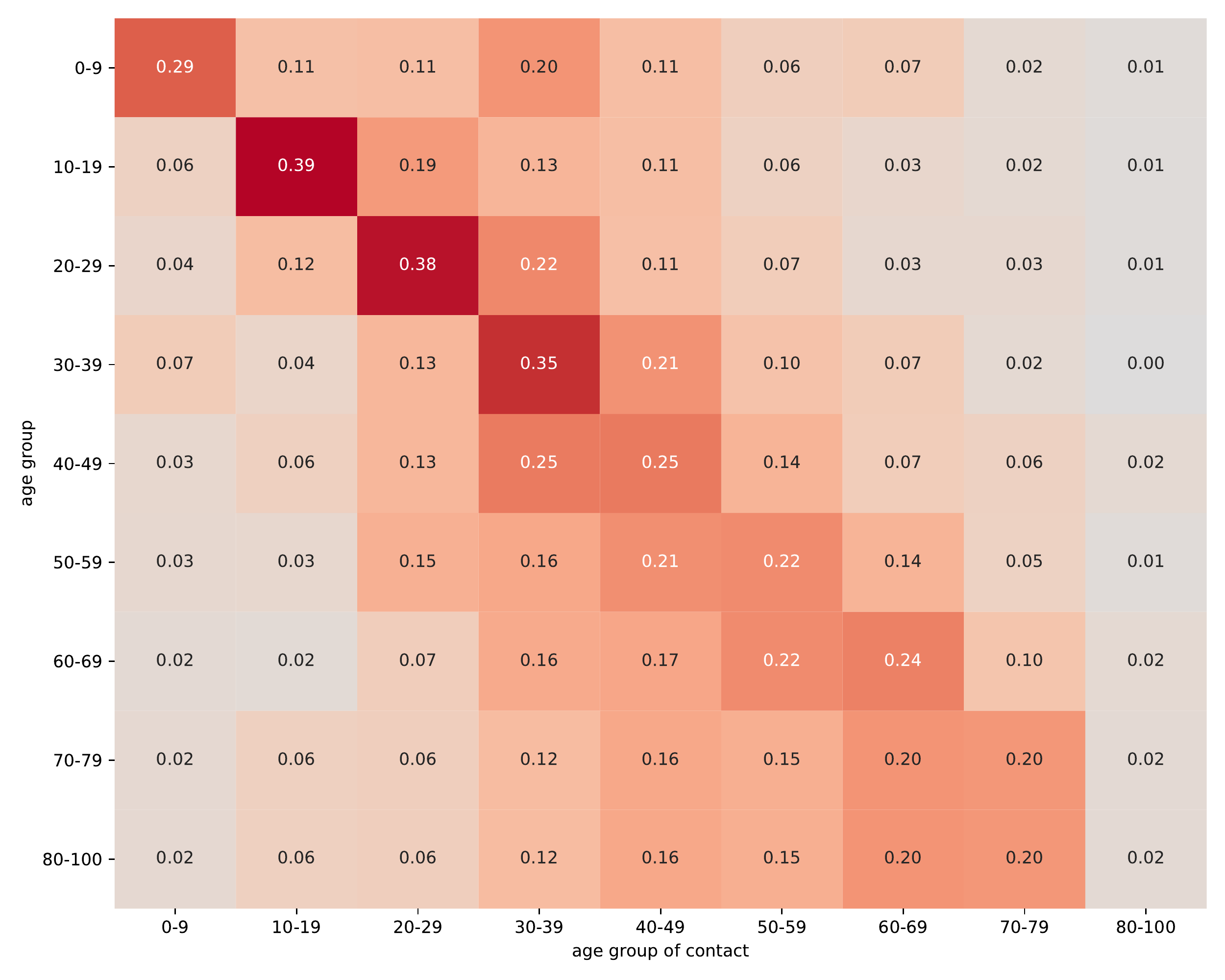}
    \caption{Distribution of Non Recurrent Other Contacts by Age Group}
    \label{fig:assortativity_other}
    \floatfoot{\noindent \textit{Note:} The figure shows the distribution of non
        recurrent other contacts by age group. A row shows the share of contacts a
        certain age group has with all other age groups. Higher values are colored in
        darker red tones. The diagonal represents the share of contacts with individuals
        from the same age group.}
\end{figure}

\begin{figure}[ht]
    \centering
    \includegraphics[width=0.9 \textwidth]{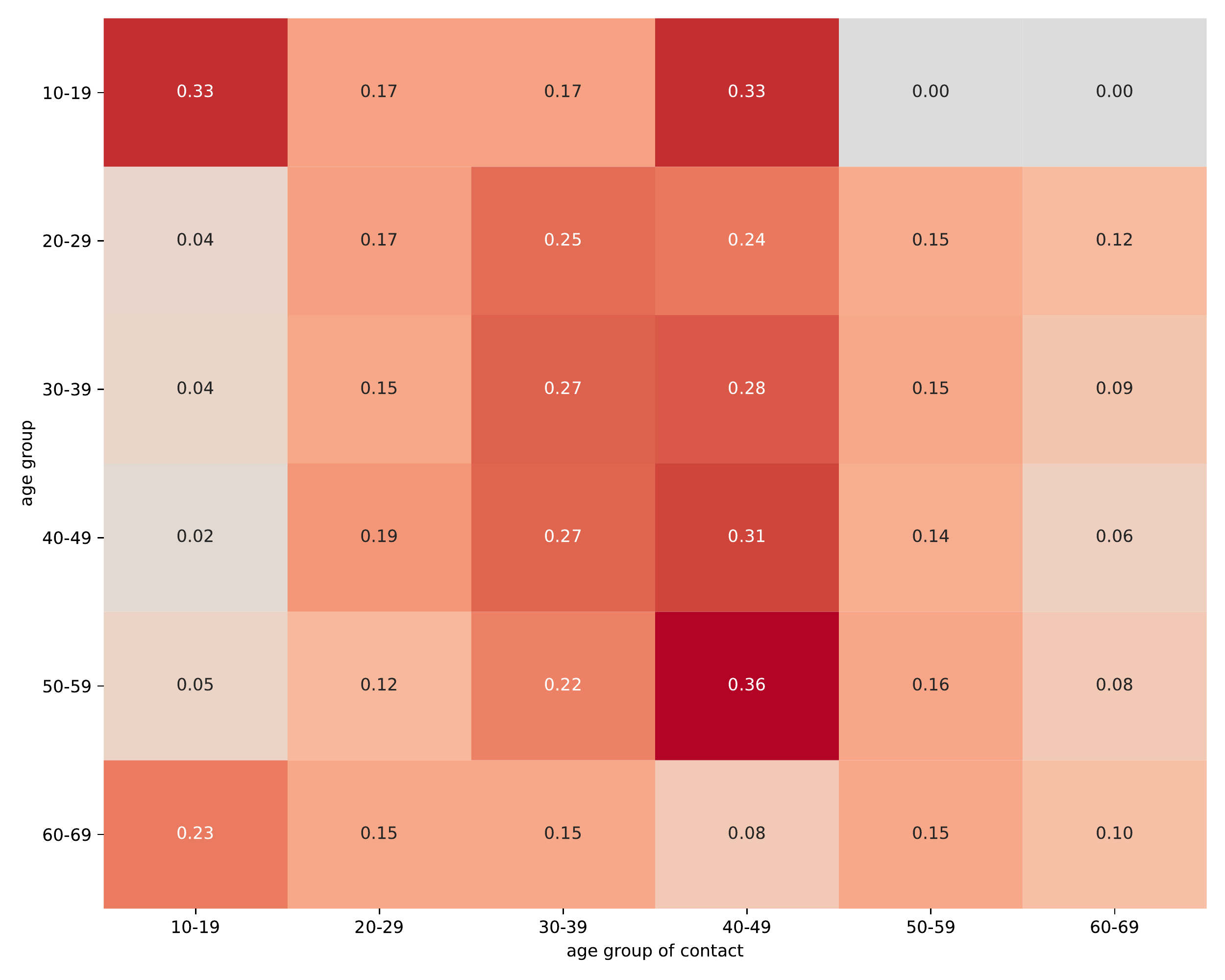}
    \caption{Distribution of Random Work Contacts by Age Group}
    \label{fig:assortativity_work}
    \floatfoot{\noindent \textit{Note:} The figure shows the distribution of non
        recurrent work contacts by age group. A row shows the share of contacts a certain
        age group has with all other age groups. Higher values are colored in darker red
        tones. The diagonal represents the share of contacts with individuals from the
        same age group. We only show age groups that have a significant fraction of
        working individuals.}
\end{figure}

As mentioned in section \ref{sec:matching}, the probability that two individuals are
matched can depend on background characteristics. In particular, we allow this
probability to depend on age and county of residence. While we do not have good data on
geographical assortativity and just roughly calibrate it such that 80\% of contacts are
within the same county, we can calibrate the assortative mixing by age from the same data
we use to calibrate the number of contacts.\comment[id=HM]{Redo
\ref{fig:assortativity_other} /\ref{fig:assortativity_work} with total number of contacts
or better add a similar figure showing total number of contacts by age in all networks}

Figure~\ref{fig:assortativity_other} shows that assortativity by age is especially strong
for children and younger adults. For older people, the pattern becomes more dispersed
around their own age group, but within-age-group contacts are still the most common
contacts.

Figure~\ref{fig:assortativity_work} shows that assortativity by age is also important
among work contacts.

Our two other types of contacts, households and schools, get their assortativity by
construction. Schools are groups where the same children of the mostly same age and
county meet with teachers every day. Household composition follows directly from the
German microcensus data we use to construct our synthetic population.

\subsection{Policies and Seasonality}
\label{subsec:policies_seasonality}

\FloatBarrier

\begin{figure}
    \centering
    \includegraphics[width=\textwidth]{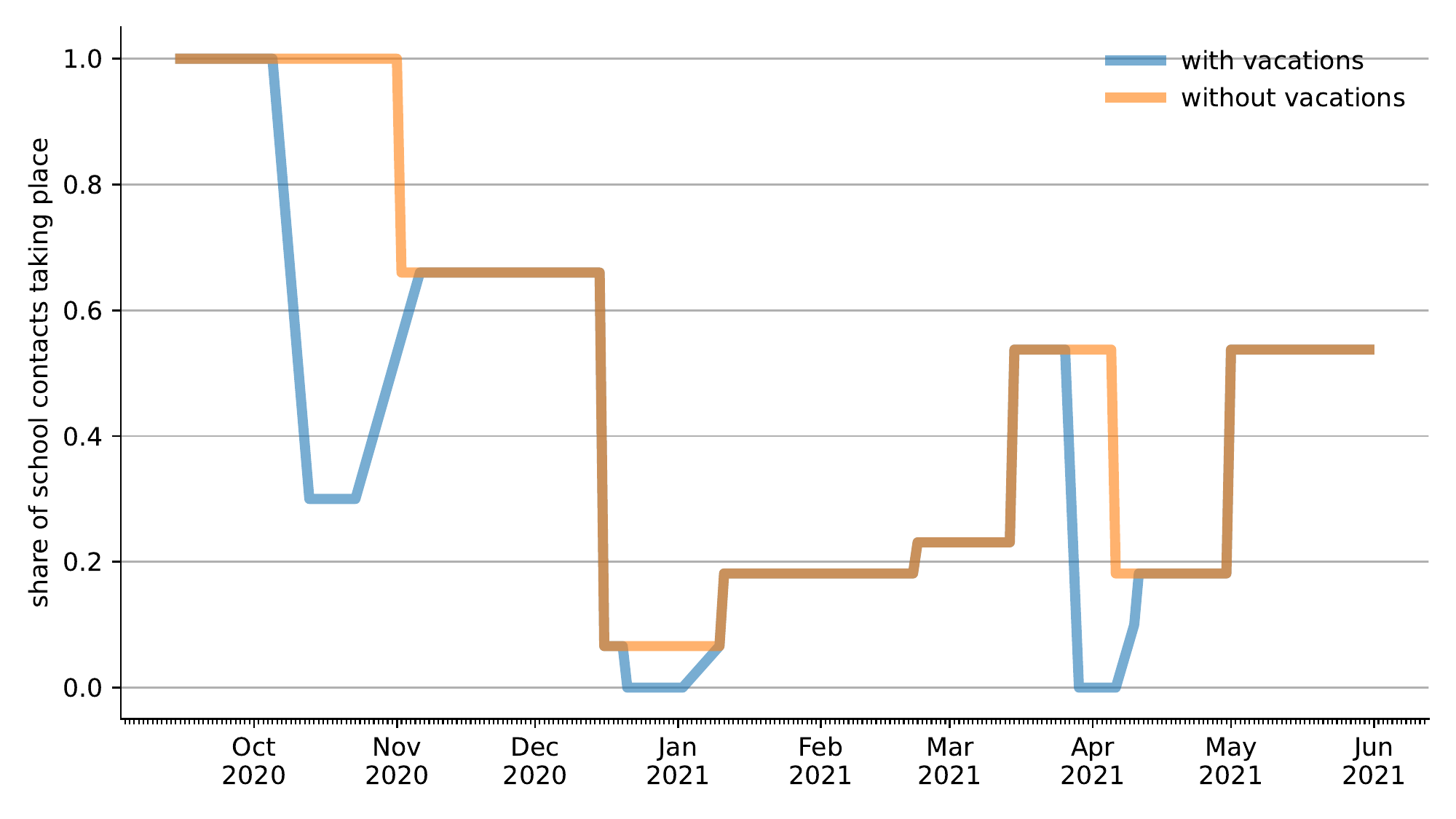}
    \caption{School Multiplier With and Without Vacations Factored In}
    \floatfoot{\noindent \textit{Note:} The dates on which schools have vacation are
    decided at the federal level. Vacations are directly implemented in our model with no
    school contacts taking place on weekends and during vacations (by federal state) just
    like the schooling mode (full operation, emergency care, rotating schemes with half
    class sizes etc.). The figure is thus only an illustration that roughly shows the
    share of contacts taking place compared to pre-pandemic level with and without
    vacations. The difference between the lines show when vacations take place and to
    what degree. For example all states have fall vacations but the timing varies
    strongly between states.}
    \label{fig:school_multiplier}
\end{figure}

In our empirical application we distinguish four groups of contact types: households,
education, work and other contacts.
For households we assume that the individuals'
contacts in their households do not change over our estimation period.
For nurseries, preschools and schools we implement vacations as announced by the German
federal states as well as school closures, emergency care and A / B schooling where only
one half of students attends every other week or day. For the moment we ignore that lack
of childcare leads working parents to stay home. An approximation of the share of contacts still taking place with the different school regulations can be found in Figure~\ref{fig:school_multiplier}.

%
%
%
%

\begin{figure}[ht]
    \centering
    \includegraphics[width=\textwidth]{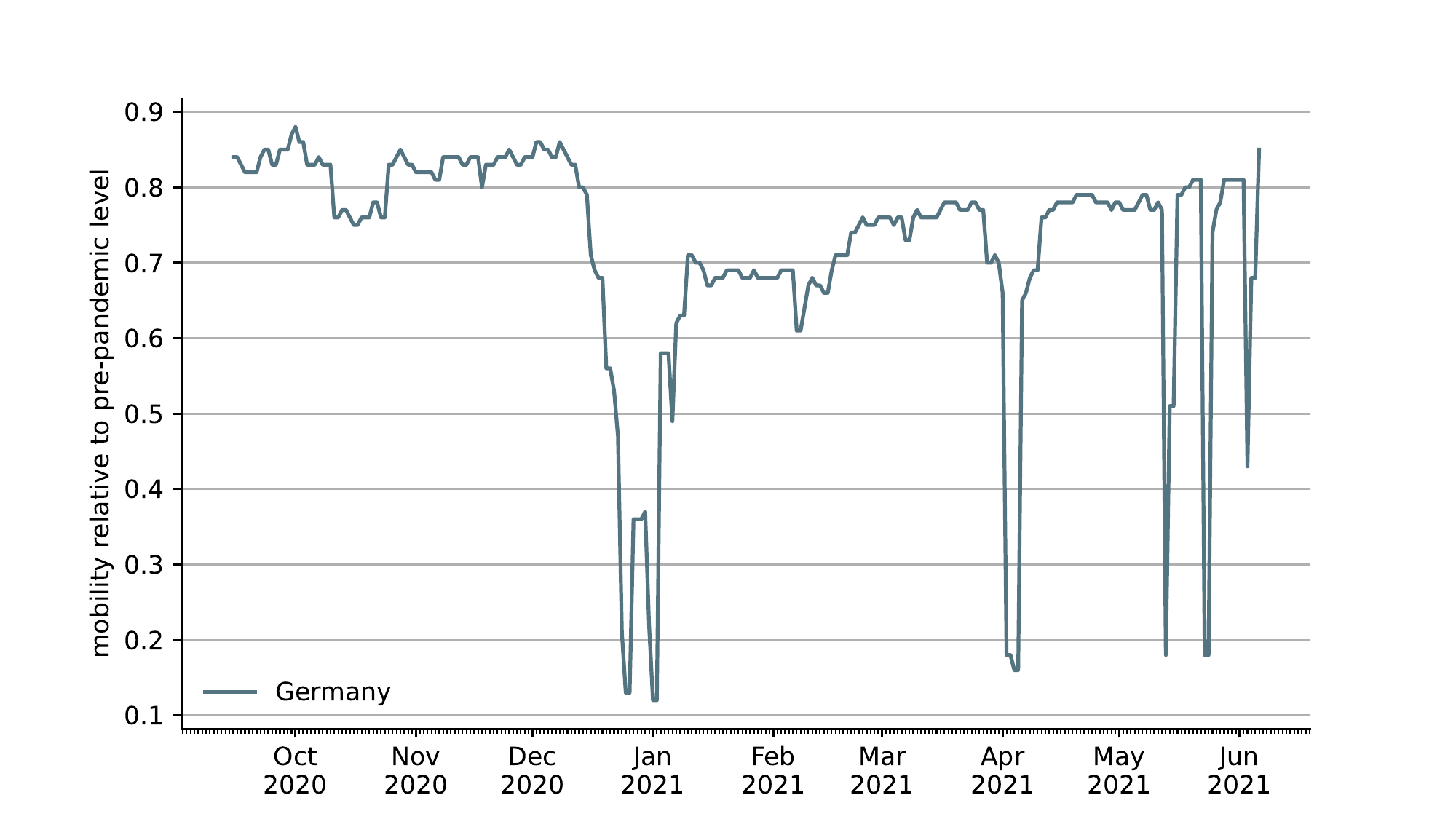}
    \caption{Share of Workers with Work Contacts}
    \label{fig:work_multiplier}
    \floatfoot{\noindent \textit{Note:} The figure shows the work mobility as reported by
        \cite{Google2021}. We take this as a proxy of the share of workers who are not in
        home office, i.e. who still have physical work contacts. The figure interpolates
        over weekends as we handle weekend effects through information on work on
        weekends in the German census data we use. The figure shows the share aggregated
        over Germany as a whole. To capture the effect that local policies, school
        vacations and public policies have on work contacts we use the data on the level
        of the federal states to determine which workers go to work depending on the
        state they live in.}
\end{figure}

For our work models\footnote{We distinguish non-recurrent work contacts, daily work
    contacts and weekly work contacts.} we use the reductions in work mobility reported in
the Google Mobility Data \citep{Google2021} to calibrate our work policies. Reductions in
work contacts are not random but governed through a work contact priority where the
policy changes the threshold below which workers stay home. Figure
\ref{fig:work_multiplier} shows the share of workers that go to work in our model over
time.

For both work and school contacts we assume that starting November with the lockdown
light in Germany, hygiene measures (such as masks, ventilation and hand washing) became
more strict and more conscientiously observed, leading to a reduction of 33\% in the
number of contacts with the potential to transmit Covid-19.

For the last group of contacts, which cover things like leisure activities, grocery
shopping, etc., we have no reliable data by how much policies reduce them. In addition,
they are likely to be affected by social and psychological factors such as pandemic
fatigue and vacations. Because of this we estimate them like the infection probabilities
to fit the time series data. We use very few change points and tie them to particular
events such as policy announcements or particular holidays. Because of the scarce data
situation we cannot distinguish between a hygiene factor (such as mask wearing) during
meetings and physical distancing (such as virtual meetings with
friends).\comment[id=K]{@Janos: Maybe make more concrete when the estimation is finished
which phases we have and why the switching points are where they are.}

\begin{figure}
    \centering
    \includegraphics[width=\textwidth]{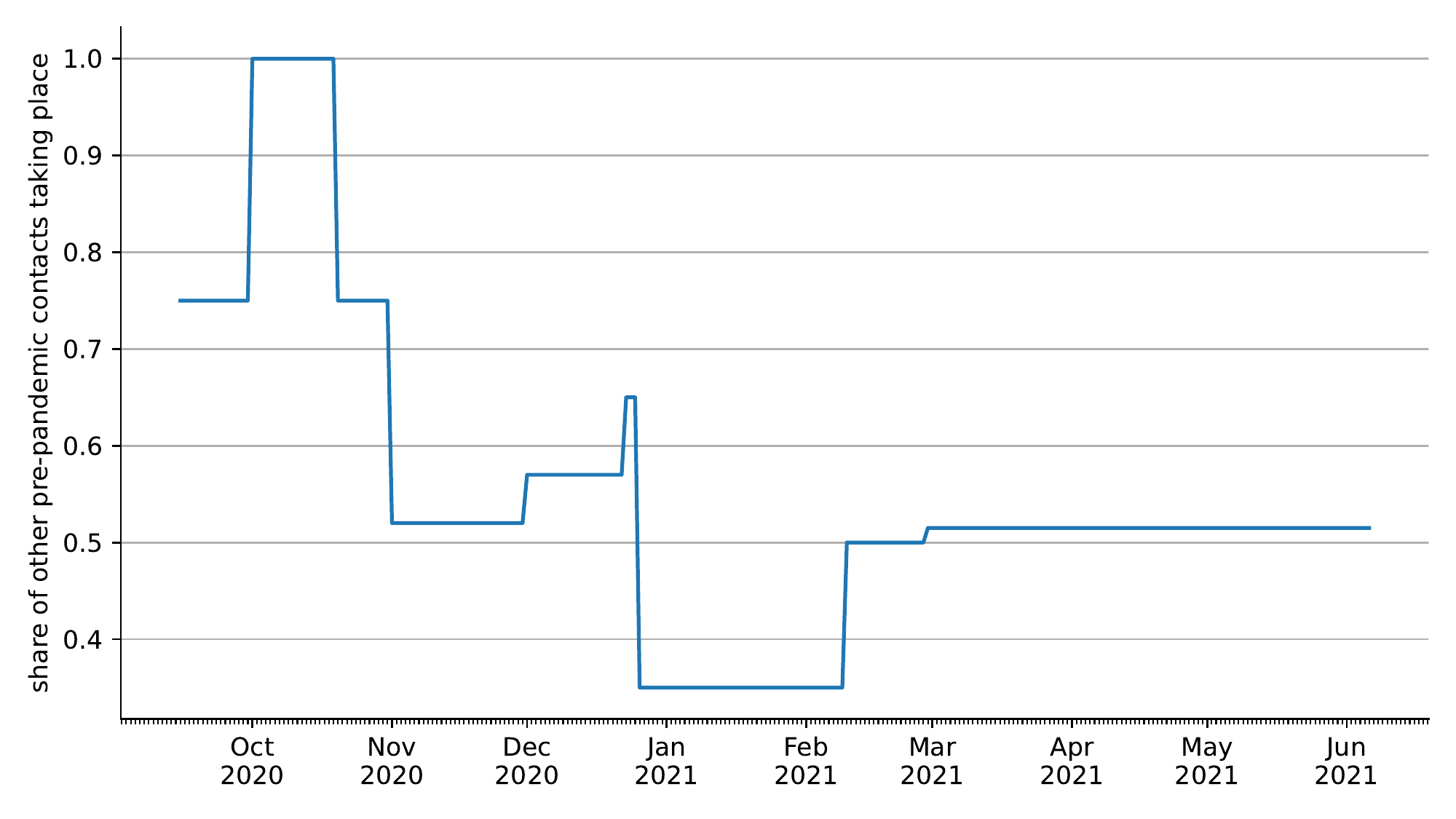}
    \caption{Share of Pre-Pandemic Other Contacts Taking Place with Infection Potential}
    \label{fig:other_multiplier}
    \floatfoot{\noindent \textit{Note:} All values are estimated. We try to use as little
    switching points as possible and tie them to political events (such as lockdown
    announcements) unless changes are used to capture anticipation or pandemic fatigue
    (for example we model an anticipation of the November lockdown and model lockdown
    fatigue in early March).}
\end{figure}

\begin{figure}
    \centering
    \includegraphics[width=\textwidth]{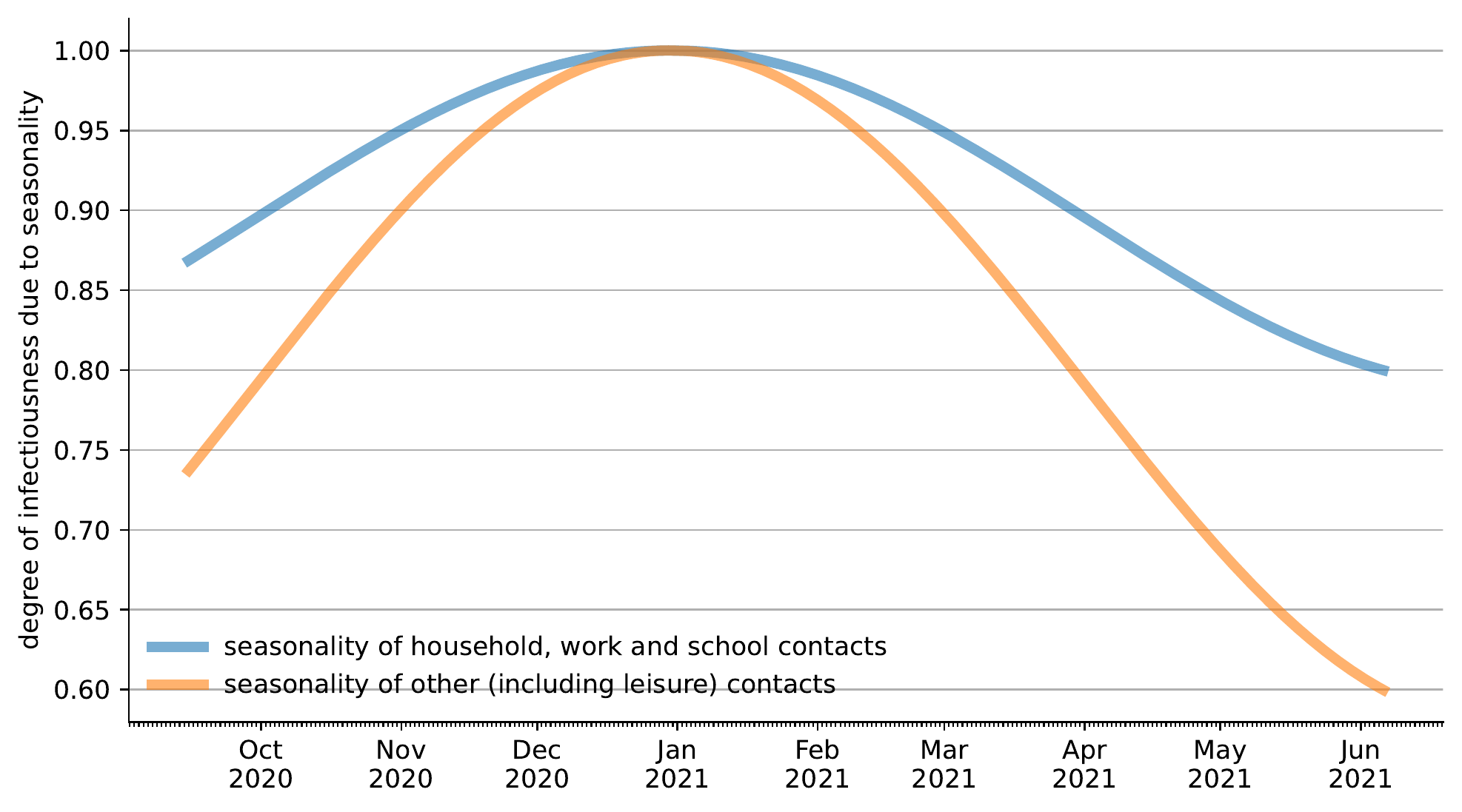}
    \caption{Seasonality by Type of Contact}
    \label{fig:seasonality}
    \floatfoot{\noindent \textit{Note:} We model seasonality as a factor that reduces the
    probability of infection of all encounters. The factor depends on the day and is
    calculated from a sinus shaped function with its maximum on January 1st. Since
    seasonality can affect the transmission both through physical conditions such as
    temperature and humidity as well as through the numbers of contacts that take place
    outside we assume two seasonality factors. One for other contacts which we expect to
    be strongly affected by fairer weather with a maximum reduction of 42\% in the
    infection probability. The other seasonality only makes contacts up to 21\% less
    infectious and is applied to household, work and school contacts.}
\end{figure}

Another potentially important factor for a contact to lead to an infection is the
seasonality \citep{Kuehn2020, Carlson2020} There are two channels through which
seasonality affects the infectiousness of contacts. One has to do with the physical
conditions like the temperature and the humidity. The other has to do with where people
meet. Especially leisure contacts are more likely to take place outdoors and individuals
are more likely to have windows open when the weather is nicer. To capture both channels
we allow for other contacts to have a higher seasonality than our other contact models.
Figure~\ref{fig:seasonality} shows our seasonality factors.

\FloatBarrier

\subsection{Rapid Test Demand}
\label{subsec:rapid_test_demand}

In our model, there are five reasons why rapid tests are done:\comment[id=J]{ Add a
    section on how we calibrate rapid test demand; Mainly describe the datapoints we have
    and say that we usually interpolate linearly in between data points. (Only exception
    to that is private rapid test demand, which we fit to data)
}

\begin{enumerate}
    \item someone plans to have work contacts
    \item someone is an employee of an educational facility or a school pupil
    \item a household member has tested positive or developed symptoms
    \item someone has developed symptoms but has not received a PCR test
    \item someone plans to participate in a weekly non-work meeting
\end{enumerate}


For work contacts, we know from the COSMO study (\cite{Betsch2021}, 20th/21st of April)
that 60\% of workers who receive a test offer by their employer regularly use it. We
assume this share to be time constant.

In addition, there are some surveys that allow us to trace the expansion of employers who
offer tests to their employees. Mid march, 20\% of employers offered tests to their
employees \citep{DIHK2021}. In the second half of March, 23\% of employees reported being
offered weekly rapid tests by their employer \citep{Ahlers2021}. This share increased to
60\% until the first days of April \cite{ZDF2021}.\comment[id=K]{ToDo: Find the survey
that the ZDF is citing here}

Until mid April 70\% of workers were expected to receive a weekly test offer
\citep{AerzteZeitung2021}. However, according to surveys conducted in mid April
\citep{Betsch2021}, less than two thirds of individuals with work contacts receive a test
offer. Starting on April 19th employers were required by law to provide two weekly tests
to their employees \citep{Bundesanzeiger2021}. We assume that compliance is incomplete
and only 80\% of employers actually offer tests.


We assume that employees in educational facilities start getting tested in 2021 and that
by March 1st 30\% of them are tested weekly. The share increases to 90\% for the week
before Easter. At that time both Bavaria \citep{BayrischerRundfunk2021} and
Baden-Württemberg \citep{MinisteriumKultus2021} were offering tests to teachers and
North-Rhine
Westphalia\footnote{\url{https://www.land.nrw/de/pressemitteilung/umfassende-informationen-fuer-die-schulen-zu-corona-selbsttests-fuer-schuelerinnen}}
\cite{DPA2021} and Lower Saxony \citep{SueddeutscheZeitung2021} were already testing
students and tests for students and teachers were already mandatory in Saxony
\citep{SueddeutscheZeitung2021a}. After Easter we assume that 95\% of teachers get tested
twice per week.

Tests for students started
later\footnote{\url{https://www.land.nrw/de/pressemitteilung/umfassende-informationen-fuer-die-schulen-zu-corona-selbsttests-fuer-schuelerinnen}}
\citep{MinisteriumKultus2021} so we assume that they only start in February and only 10\%
of students get tested by March 1st. Relying on the same sources as above we approximate
that by the week before Easter this share had increased to
40\%.\footnote{\url{https://www.land.nrw/de/pressemitteilung/umfassende-informationen-fuer-die-schulen-zu-corona-selbsttests-fuer-schuelerinnen},
}

After Easter the share of students receiving twice weekly tests is set to 75\%. This as
based on tests becoming mandatory becoming mandatory in Bavaria after Easter
break\footnote{Bavaria\footnote{\url{https://bit.ly/3nz5fXS}}, in North-Rhine Westphalia
on April
12th\footnote{https://www.schulministerium.nrw/ministerium/schulverwaltung/schulmail-archiv/14042021-schulbetrieb-im-wechselunterricht-ab-montag},
\url{https://bit.ly/2QHilX3}} and on the 19th in
Baden-Württemberg\footnote{https://bit.ly/3vuetaD, https://bit.ly/3vuetaD}.


To limit our degrees of freedom, we only have one parameter that governs how many
individuals do a rapid test because of any of the private demand reasons (own symptoms
but no PCR test, planned weekly leisure meeting or a symptomatic or positively tested
household member).

We assume that there is no private rapid test demand until March when both the citizens'
tests and rapid tests for lay people started to become available
\footnote{\url{https://bit.ly/3ehmGcj}, \url{https://bit.ly/3xJCIn8}} and other access to
rapid tests was very limited.

According to the COSMO study\footnote{\url{https://bit.ly/2QSFAgR}} 63\% would have been
willing to take a test in the round of 23rd of February 2021 when an acquaintance would
have tested positive. Since this is only asking for willingness not actual behavior and
the demand when meeting with friends is very likely lower, we take this as the upper
bound of private rapid test demand which is reached on May 4th. To cover that many people
are likely to have sought and done their first rapid test before the Easter holidays to
meet friends or family, we let the share of individuals doing rapid tests in that time
increase more rapidly than before and after. By end of March 25\% of individuals would do
a rapid test due to a private reason.

\begin{figure}
    \centering
    \includegraphics[width=\textwidth]{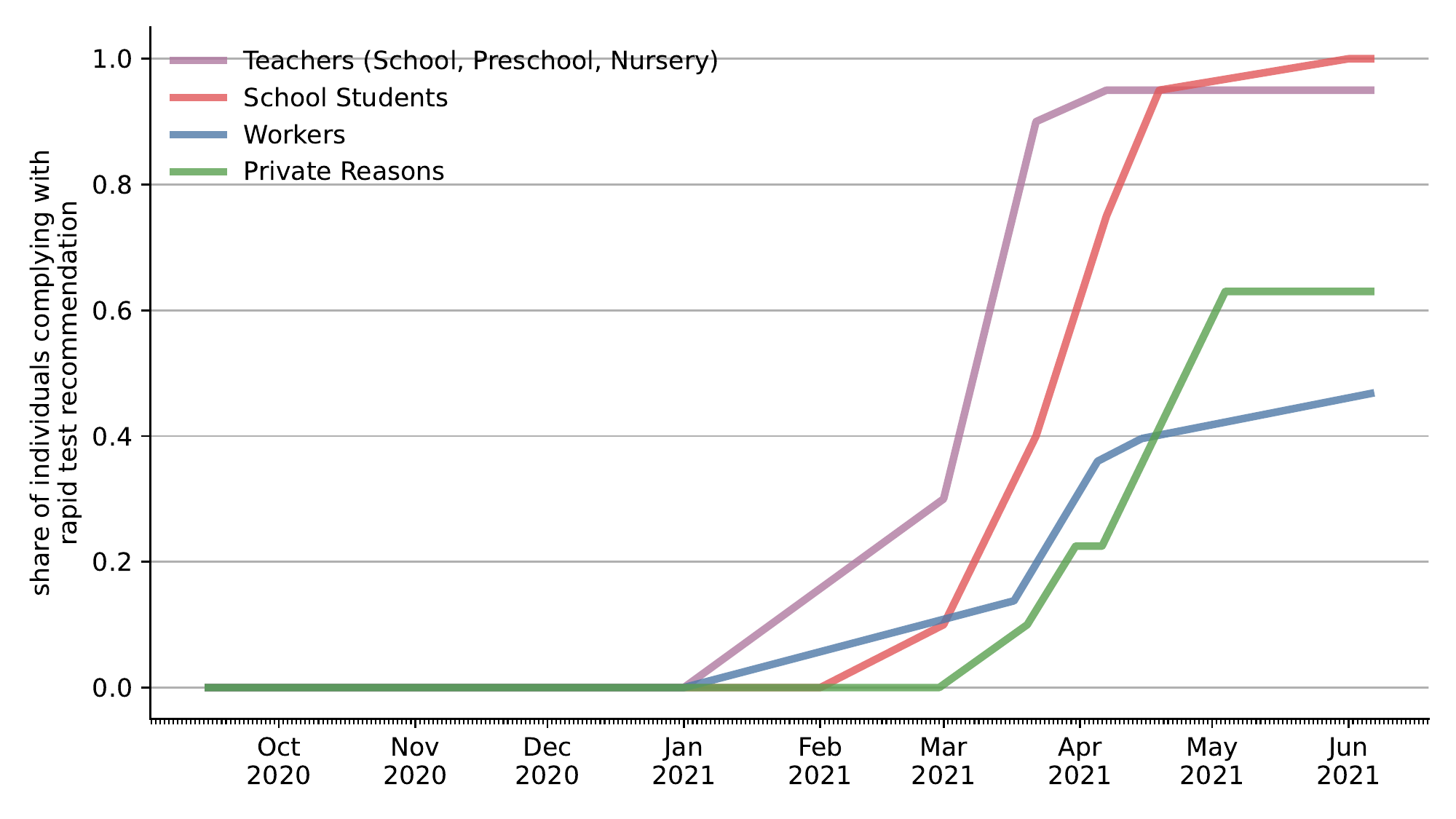}
    \caption{\textbf{Share of Individuals Doing a Rapid Test.}}
    \floatfoot{\noindent \textit{Note:} Rapid test demand can be triggered by individuals
    planning to have education contacts, work contacts, developing symptoms without
    access to a PCR test, having a household member with a positive test or symptoms. In
    each case whether a rapid test is done depends on how long it has been since the
    individual's last rapid test and her individual compliance parameters. As an example,
    take a worker in May. In that time workers are encouraged to test themselves twice
    weekly but there is no general requirement to test themselves. If the worker has not
    done a test within the last four days in our model she will demand a test if her
    (time-constant) compliance parameter belongs to the upper 60\% in the population.}
    \label{fig:rapid_test_demand}
\end{figure}

All shares of individuals who would take a rapid test if the conditions were met can be
seen in Figure~\ref{fig:rapid_test_demand}.\comment[id=J]{Talk about the interpretation
of each line.}

\begin{figure}[ht]
    \centering
    \caption{Share of Individuals With Rapid Tests}
    \label{fig:share_ever_rapid_test}
    \begin{subfigure}{.55\textwidth}
        \includegraphics[width=0.9 \textwidth]{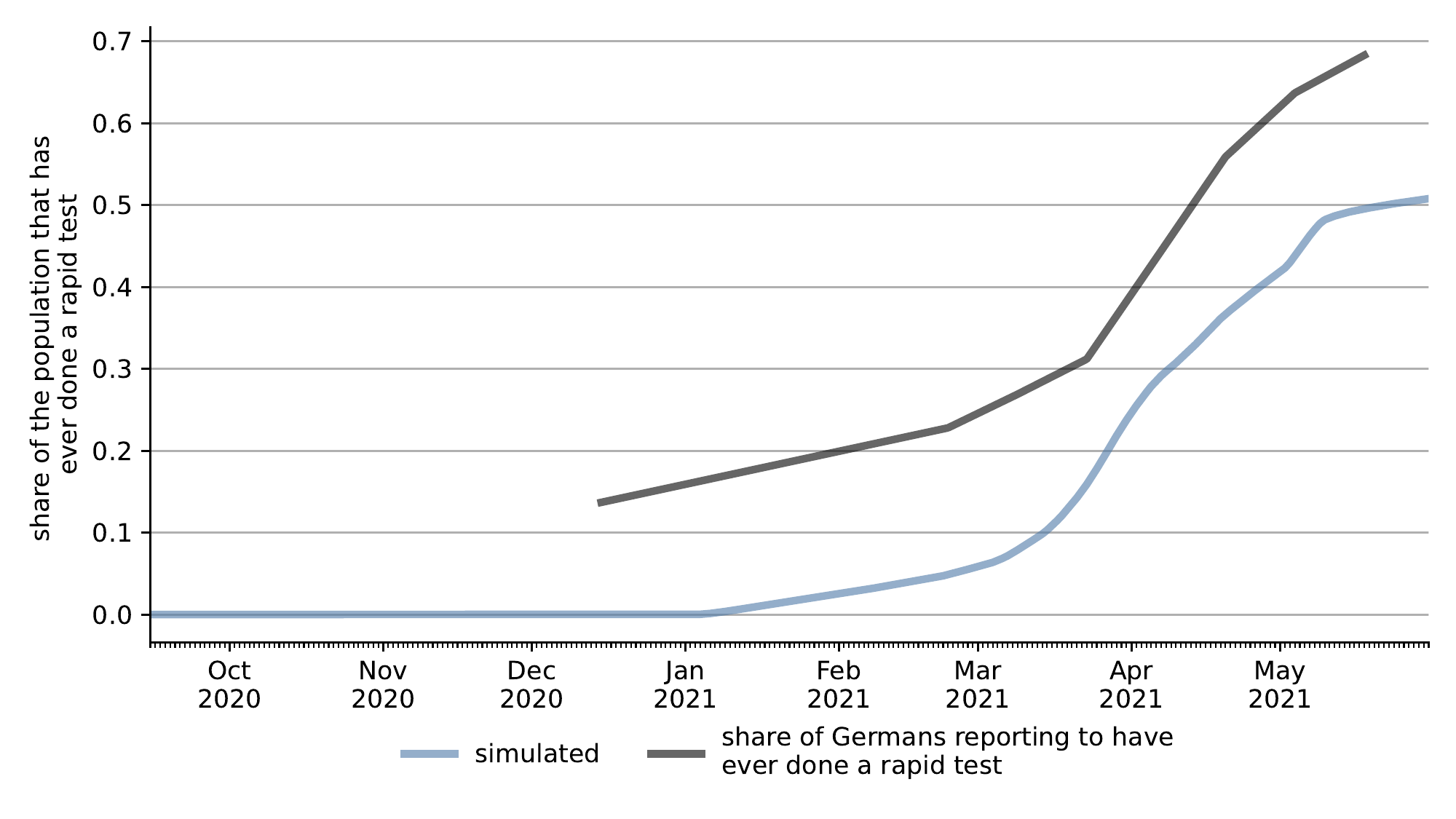}
    \end{subfigure}%
    \begin{subfigure}{.55\textwidth}
        \includegraphics[width=0.9 \textwidth]{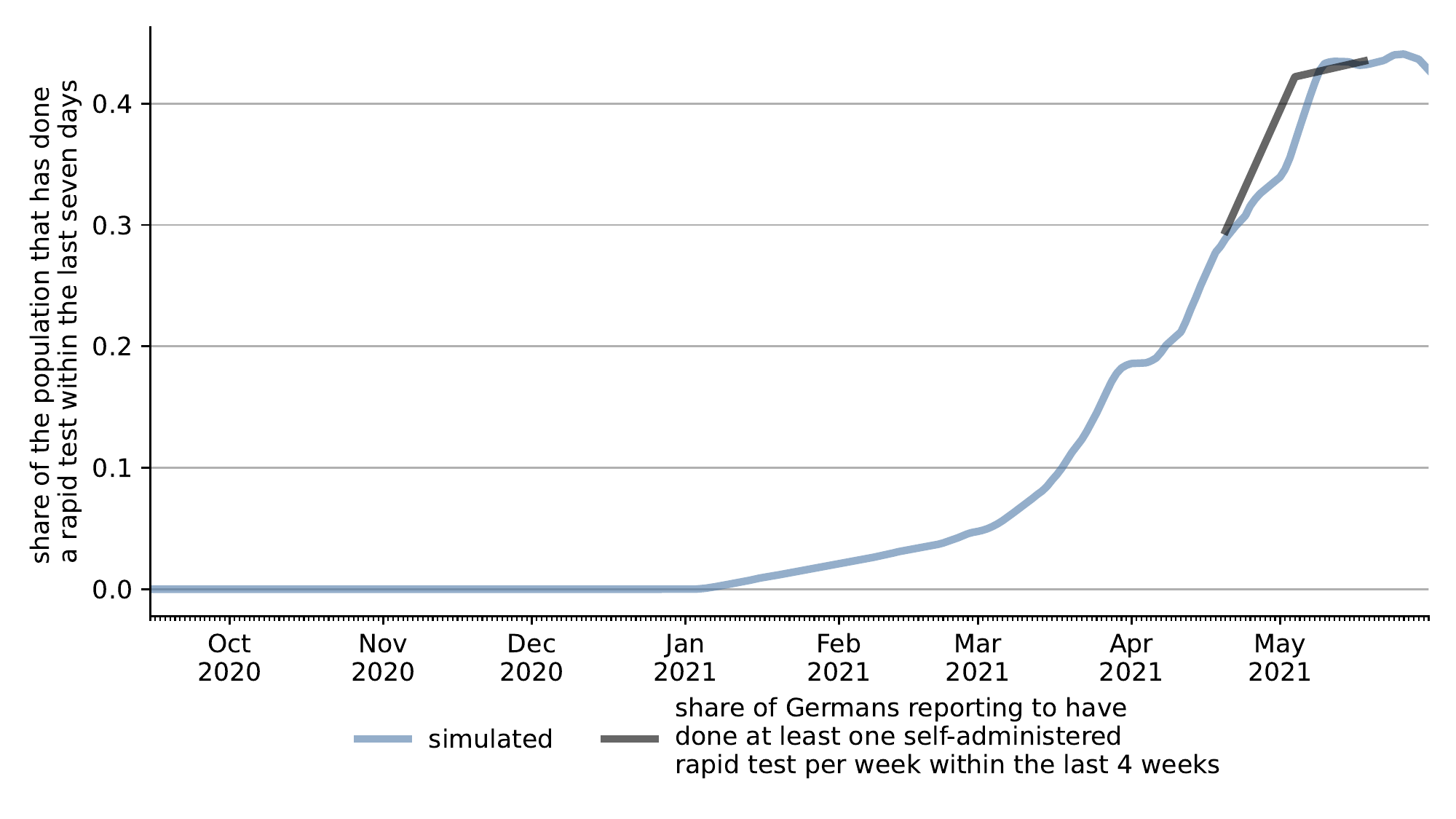}
    \end{subfigure}
    \label{fig:share_rapid_test_last_week}
    \floatfoot{\noindent \textit{Note:} The figure compares the share of individuals who
        have ever done a rapid test or done a rapid test within the last week in our
        simulations to the shares reported in the
        \href{https://projekte.uni-erfurt.de/cosmo2020/web/topic/wissen-verhalten/80-schnelltests/}{COVID-19
        Snapshot Monitoring Survey}. The left panel compares the share of individuals who
        have ever done a rapid test. The right panel compares the share of individuals
        who have done a rapid test within the last seven days in our simulation compared
        to the share reporting to have done at least weekly rapid tests in the last four
        weeks in the COSMO survey. Overall our calibration of rapid tests are slightly
        conservative. The overall share is below that in the study. We fit the share of
        weekly tests quite exactly. However, the study only covers adults while our share
        also includes children who are tested very regularly when attending school.}
\end{figure}

    \FloatBarrier

    \section{Detailed Model Description}
\label{sec:model}

\subsection{Literature Review}
\label{sec:literature_review}

We build on two strands of literature: Recent extensions of the epidemiological SEIR
model and agent-based simulation models.

The traditional SEIR model is not fine-grained enough to model nuanced policies. This
has motivated a large number of researchers to extend the standard model to allow for
more heterogeneity and flexibility. Examples are \citet{Grimm2020},
\citet{Donsimoni2020} and \citet{Acemoglu2020} who develop multi group SEIR models to
analyze the effects of targeted lockdowns and \citet{Berger2020} who extend the SEIR
model to analyze testing and conditional quarantines. For a more comprehensive review
see \citet{Avery2020}. Others have used the results of a standard SEIR model as input
for economic models that estimate the cost of policies (e.g. \citet{Dorn2020}).

While the popularity of the SEIR model is mainly due to its simplicity, the extensions
are quite complex. It is unlikely that there will be a SEIR model that combines all
proposed extensions. Moreover, the extensions do not address other key issues: The main
parameter of the SEIR model, the basic reproduction number ($R_0$), is not
policy-invariant. It is a composite of the number of contacts each person has and the
infection probability of the contacts. In fact, policy simulations are done by setting
$R_0$ to a different value but it is hard to translate a real policy into the value of
$R_0$ it will induce. In other words, SEIR models are not suited for evaluating the
effect of policies which have never been experienced before.

Another commonly used model class in epidemiology are agent-based simulation models. In
these models individuals are simulated as moving particles. Infections take place when
two particles come closer than a certain contact radius (e.g. \citet{Silva2020} and
\citet{Cuevas2020}). While the simulation approach makes it easy to incorporate
heterogeneity in disease progression, it is hard to incorporate heterogeneity in meeting
patterns. Moreover, policies are modeled as changes in the contact radius or momentum
equation of the particles. The translation from real policies to corresponding model
parameters is a hard task.

\citet{Hinch2020} is a recent extension of the prototypical agent-based simulation model
that replaces moving particles by contact networks for households, work and random
contacts. This model is similar in spirit to ours but focuses on contact tracing rather
than social distancing policies.

The above assessment of epidemiological models is not meant as a critique. We are aware
that these models were not designed to predict the effect of fine-grained social
distancing policies in real time and are very well suited to their purpose. We invite
epidemiologists to provide feedback and collaborate to improve our model.

\subsection{Summary}
\label{sub:model_summary}

To predict and quantify the effects of a wide variety of fine-grained social distancing
policies, vaccinations and rapid testing, we propose a different model structure. Our
model inherits many features from prototypical agent-based simulation models but
replaces the contacts between moving particles by contacts between individuals who work,
go to school, live in a household and enjoy leisure activities.

The structure of the model is depicted in Figure~\ref{fig:model_contacts_infections}.

We distinguish between eight types of contact
models which are all listed in Figure~\ref{fig:model_contacts_infections}: households,
recurrent and random work contacts, recurrent and random leisure contacts, and nursery,
preschool, and school contacts.

The number of contacts is translated into infections by a matching algorithm.
There are different matching algorithms for recurrent contacts (e.g. classmates, family
members) and non-recurrent contacts (e.g. clients, contacts in supermarkets).
All types of contacts can be assortative with respect to geographic and demographic
characteristics.

The infection probabilities of contacts vary with contact type, age of the susceptible
person, and the virus strain of the infected person. Moreover, they follow a seasonal
pattern. The strength of the seasonality effect is higher for contacts that are easy to
be moved to an outside location in summer (such as leisure contacts) and smaller for
contacts that take place inside even in summer (e.g. work contacts).

Once a person is infected, the disease progresses in a fairly standard way which is
depicted in Figure~\ref{fig:model_disease_progression}. Asymptomatic cases and cases with mild
symptoms are infectious for some time and recover eventually. Cases with severe symptoms
additionally require hospitalization and lead to either recovery or death.

After rapid tests become available, people who work or go to school can receive rapid
tests there. Moreover, people can decide to make a rapid test if they develop symptoms,
have many planned contacts or observe cases in their contact network. People who have a
positive rapid test demand a confirmatory PCR test with a certain probability. Moreover,
PCR tests can be demanded because of symptoms or randomly.

This rich model of PCR and rapid tests leads to a share of detected cases that varies
over time and across age groups. It also allows to quantify the effect of changes in
testing policies on the dynamic of infections.

People who have symptoms, received a positive test, or had a risk contact can reduce
their number of contacts across all contact types endogenously. The extent to which this
is done is calibrated from survey data.

The model makes it very simple to translate policies into model quantities. For example,
school closures imply the complete suspension of school contacts. A strict lockdown
implies shutting down work contacts of all people who are not employed in a systemically
relevant sector. It is also possible to have more sophisticated policies that condition
the number of contacts on observable characteristics, risk contacts or health states.

An important feature of the model is that the number of contacts an individual has of
each contact type can be calibrated from publicly available data \citep{Mossong2008}.
This in turn allows us to estimate policy-invariant infection probabilities from time
series of infection and death rates using the method of simulated moments
\citep{McFadden1989}. Since the infection probabilities are time-invariant, data
collected since the beginning of the pandemic can be used for estimation. Moreover,
since we model the testing strategies that were in place at each point in time, we can
correct the estimates for the fact that not all infections are observed.

The model has a very modular structure and can easily be extended to distinguish more
contact types, add more stages to the disease progression, implement new policies
or test demand models. The main bottleneck is not complexity or computational cost
but the availability of data to calibrate the additional model features.

\subsection{Modeling Numbers of Contacts}
\label{sec:number_of_contacts}

Consider a hypothetical population of 1,000 individuals in which 50 were infected with a
novel infectious disease. From this alone, it is impossible to say whether only
those 50 people had contact with an infectious person and the disease has an infection
probability of 1 per contact or whether everyone met an infectious person but the
disease has an infection probability of only 5 percent per contact. SEIR models do not
distinguish contact frequency from the infectiousness of each contact and
combine the two in one parameter that is not invariant to social distancing policies.

To model social distancing policies, we need to disentangle the effects of the number of
contacts of each individual and the effect of policy-invariant infection probabilities
specific to each contact type. Since not all contacts are equally infectious, we
distinguish different contact types.

The number and type of contacts in our model can be easily extended. Each type of
contacts is described by a function that maps individual characteristics, health states
and the date into a number of planned contacts for each individual. This allows to model
a wide range of contact types.

In our empirical application we distinguish the following  contact types that
are depicted in Figure~\ref{fig:model_contacts_infections} and can be further grouped
in the categories household, work, education and others.

types of contacts:\comment[id=HM]{I started updating this to understand it better myself.
In the end, I think it will be useful to group into HH / work / schools and add an
introductory sentence for each group. E.g., that work is for everybody who is working ---
except there is a different model for teachers (?), ...}

\begin{itemize}
    \item Households: Each household member meets all other household members every day.

    \item Recurrent work contacts, capturing contacts with coworkers, repeating
          clients and superiors. Some of these recurrent contacts take place on
          every workday, others just once per week.

    \item Random work contacts: Working adults have contacts with randomly drawn other
          people, which are assortative in geographical location and age.

    \item Schools: Each student meets all of his classmates every day. Class sizes are
    calibrated to be representative for Germany and students have the same age. Schools
    are closed on weekends and during vacations, which vary by states. School classes
    also meet six teachers everyday and some of the teachers meet each other.

    \item Preschools: Children who are at least three years old and younger than six may
    attend preschool. Each group of nine children interacts with the same two adults
    every day. The children in each group are of the same age. The remaining mechanics
    are similar to schools.

    \item Nurseries: Children younger than three years may attend a nursery and interact
    with one adult. The age of the children varies within groups. The remaining
    mechanics are similar to schools.

    \item Random other contacts: Contacts with randomly drawn other
    people, which are assortative with respect to geographic location and and age group.
    This contact type reflects contacts during leisure activities, grocery shopping,
    medical appointments, etc..

    \item Recurrent other contacts representing contacts with friends neighbors or
        family members who do not live in the same household. Some of these contacts
        happen daily, others only once per week.

\end{itemize}

The number of random and recurrent contacts at the workplace, during leisure activities
and at home is calibrated with data provided by \citet{Mossong2008}. For details see
Section~\ref{sec:number_of_contacts}. In particular, we sample the number of contacts or
group sizes from empirical distributions that sometimes depend on age. It would also be
possible to use economic or other behavioral models to predict the number of contacts.


\subsection{Reducing Numbers of Contacts via NPIs}
\label{sec:policies}

Our model makes it very easy to model a wide range of NPIs, either in isolation or
simultaneously. This is important for two reasons: Firstly, it allows to predict and
quantify the effect of novel NPIs. Secondly, it allows to model the actually implemented
policy environment in great detail, which is necessary to use use the full time series
of infections and fatality rates to estimate the model parameters.\footnote{
See \citet{Avery2020} for an explanation why it can be harmful to use too long time
series to estimate simple SEIR type models.}

Instead of thinking of policies as completely replacing how many contacts people have,
it is often more helpful to think of them as adjusting the pre-pandemic number of
contacts. Therefore, we implement policies as a step that happens after the number of contacts is
calculated but before individuals are matched.

On an abstract level, a policy is a functions that modifies the number of contacts of
one contact type. This function can be random or deterministic. For example, school
closures simply set all school contacts to zero. A work from home mandate leads to a
share of workers staying home every day whereas those who cannot work from home are
unaffected. Hygiene measures at work randomly reduce the number of infectious contacts
for all workers who still go to work.

Policies can also interact. For example, school vacations are temporally reducing school
contacts to zero while at the same time increasing other contacts to account for
increased leisure activities and family visits during this time. This is important to
reproduce the finding that school vacations do not reduce infection numbers even though
schools lead to infections when open \citep{Isphording2021}.

The most complex policies are typically found in the education sector. Since the
beginning of 2021 schools have switched back and fourth between full closures,
split class approaches with alternating schedules for some or all age groups and
reopening while maintaining hygiene measures. On top of that there are different
policies for allowing young students whose parents work full time to attend school
even on days where they normally would not. For details on how we calibrate these
policies see Section~\ref{subsec:policies_seasonality}.

Importantly, policies can depend on the health states of participating individuals.
This allows to quarantine entire school classes if one student tested positive or
to implement official or private contact tracing.

For some policies the exact effect on each contact type is not easy to determine. If
this refers to a policy has been active during the estimation period, it is possible to
estimate such parameters by fitting the model to time series data of infection rates.
This is only possible if the policy was not active during the whole estimation period
and thus the infection probabilities can be identified separately. We do this to account
for hygiene measures at school and in the workplace that have been in effect since
November 2020.

Not all things that reduce contacts compared to the pre-pandemic level are driven
by NPIs. Therefore, we also model endogenous contact reductions that can depend on
the health state of individuals, known risk contacts or the local incidence of
infections. Examples are strong contact reductions for symptomatic individuals or those
who have a positive PCR or rapid tests or contact reductions when a houshold member
tested positive. The extent to which contacts are reduced can be calibrated from
surveys. For an application of our model showcasing private contact tracing in the
context of the Christmas holidays see \cite{Gabler2020}.

\subsection{Matching Individuals}
\label{sec:matching}

The empirical data described above only allows to estimate the number of contacts each
person has. In order to simulate transmissions of Covid-19, the numbers of contacts has
to be translated into actual meetings between people. This is achieved by a matching
algorithm:

As described in section \ref{sec:number_of_contacts}, some contact types are recurrent
(i.e. the same people meet regularly), others are non-recurrent (i.e. it would only be
by accident that two people meet twice). The matching process is different for recurrent
and non recurrent contact models.

Recurrent contacts are described by two components: 1) A set of time invariant groups, such as school classes. The groups can be sampled from empirical data or created by
randomly matching simulated individuals into groups. 2) A deterministic or random function that takes the value 0 (non-participating) and 1
(participating) and can depend on the weekday, date and health state. This can be used
to model vacations, weekends or symptomatic people who stay home (see section
\ref{sec:policies} for details).

The matching process for recurrent contacts is then extremely simple: On each simulated
day, every person who does not stay home meets all other group members who do not stay
home. The assumption that all group members have contacts with all other group members
is not fully realistic, but a good approximation to reality, especially in
light of the suspected role of aerosol transmission for Covid-19 \citep{Morawska2020,
Anderson2020}. Alternatively, the infection probability of recurrent contact types can
be interpreted as being the product of a true infection probability and the probability
that an actual contact takes place.

The matching in non-recurrent contact models is more difficult and implemented in a two
stage sampling procedure to allow for assortative matching. Currently most contact
models are assortative with respect to age (it is more likely to meet people from the
same age group) and county (it is more likely to meet people from the same county) but
in principle any set of discrete variables can be used. This set of variables that
influence matching probabilities introduce a discrete partition of the population into
groups. The first stage of the two stage sampling process samples on the group level.
The second stage on the individual level.

The algorithm works as follows: First we randomly draw a contact type and individual.
For each contact of the drawn contact type that person has, we first draw the group of
the other person (first stage).
Next, we calculate the probability to be drawn for each member of the group, based on
the number of remaining contacts, i.e. people who have more remaining contacts are drawn
with a higher probability. The probabilities have to be re-calculated each time because
with each
matched contact, the number of remaining contacts changes. We then draw the other
individual, determine whether an infection takes place and if so update the health state
of the newly infected person. Finally, we reduce the number of remaining contacts of the
two matched individuals by one. \comment[id=J]{Add formulae here.}

The recalculation of matching probabilities in the second stage is computationally
intensive because it requires summing up all remaining contacts in that group. Using a
two stage sampling process where the first stage probabilities remain constant over time
makes the matching computationally much more tractable because the number of
computations only increases quadratically in the size of the second group and not
quadratically in the size of the entire simulated population.

\FloatBarrier

\subsection{Course of the Disease}
\label{sub:disease_progression}

The disease progression in the model is fairly standard. It is depicted in
Figure~\ref{fig:model_disease_progression} and the values and source of the relevant
parameters are describes in Section~\ref{sec:medical_params}.

First, infected individuals will become infectious after one to five days. Overall,
about one third of people remain asymptomatic. The rest develop symptoms about one to
two days after they become infectious. Modeling asymptomatic and pre-symptomatic cases
is important because those people do not reduce their contacts nor do they have an
elevated probability to demand a test. Thus they can potentially infect many other
people \citep{Donsimoni2020}. The probability to develop
symptoms with Covid-19 is highly age dependent with 75\% of children not developing
clinical symptoms \citep{Davies2020}.

A small share of symptomatic people will develop strong symptoms that require intensive
care. The exact share and time span is age-dependent. An age-dependent share of intensive
care unit (ICU) patients will die after spending up to 32 days in intensive care.
Moreover, if the ICU capacity was reached, all patients who require intensive care but do
not receive it die.

It would be easy to make the course of disease even more fine-grained. For example, we
could model people who require hospitalization but not intensive care. So far we opted
against that because only the intensive care capacities are feared to become a bottleneck
in Germany.

We allow the progression of the disease to be stochastic in two ways: Firstly, state
changes only occur with a certain probability (e.g. only a fraction of infected
individuals develops symptoms). Secondly, the number of periods for which an individual
remains in a state is drawn randomly. The parameters that govern these processes are
taken from the literature \footnote{ Detailed information on the calibration of the
disease parameters is available as part of our
\href{https://sid-dev.readthedocs.io/en/latest/reference_guides/epi_params.html}{online
documentation}.}.\comment[id=J]{Replace this by a link to the relevant table with
parameters in the data appendix, once we have that table.}

\subsection{Testing} 
\label{sub:testing}

Having a realistic model of PCR and rapid tests is crucial for two reasons: Firstly,
only via a testing model the simulated infections from the model can be made comparable
to official case numbers. Secondly, individuals with undetected or not yet detected
infections are an important driver of the pandemic.

In principle, our modeling approach is flexible enough to incorporate mechanistic test
demand, allocation and processing models. However, there is not enough data available to
calibrate such a mechanistic model.

Therefore, we aim for a simpler model of PCR and rapid tests that can be calibrated with
available data on test demand and availability and -- nevertheless -- can produce a share
of undetected cases that varies over time and across age groups and agrees with other
estimates over the time periods where they are available.

PCR tests are modeled since the beginning of the simulation and determine whether a
infections is officially recorded. Rapid tests are only added at the beginning of 2021.
Positive rapid tests do not enter official case numbers directly, but most people with a
positive rapid tests demand a confirmatory PCR test. However, positive rapid tests can
have a strong effect on the infection dynamics because they trigger contact reductions
and additional rapid tests.

During 2020 people can demand PCR tests either because they have symptoms or randomly.
The probability that a PCR test is performed in each of the two situations depends on the
number of new infections and the number of available tests. Thus, it varies strongly over
time and is unknown.

To distribute the correct number of PCR tests among symptomatic and asymptomatic
infections without knowing explicit test demand probabilities, we use the following
approach: First, we calculate the total number of positive PCR tests by multiplying the
number of newly infected individuals with an estimate of the share of detected cases from
the Dunkelzifferradar project. \footnote{The Dunkelzifferradar project publishes daily
estimates of the dark figure of infections under
\url{https://covid19.dunkelzifferradar.de/}} Next, we determine how many of these tests
should go to symptomatic and asymptomatic individuals from data by the Robert Koch
Institut. \comment[id=J]{add source}
Then, we sample the individuals to which those tests are allocated from the pools of
symptomatic and asyptomatic infected but not yet tested individuals.

Sampling uniformly from the pool of symptomatic individuals ensures that age groups who
are more likely to develop symptoms are also more likely to receive tests. Thus, the
share of detected cases is much higher for the elderly than for children in time periods
where many tests are done because of symptoms which is in line with the estimates from
the literature. \comment[id=J]{need source}

At the beginning of 2021, two challenges arise: Firstly, the externally estimated share
of detected cases from the Dunkelzifferradar can no longer be used because it is based on
case fatality rates which drastically change due to vaccinations. Secondly, rapid tests
become available at a large scale.

We solve the first challenge by assuming that the share of detected cases would have
remained at the level it reached before Christmas if rapid tests had not become
available. While this is only an approximation to reality, changes in the share of
detected cases that would have happened without rapid tests are very likely to be small
compared to the changes caused by rapid tests.

The second challenge is solved by mechanistic rapid test demand models for the workplace,
schools and by private individuals. The calibration of these models is described in
Section~\ref{subsec:rapid_test_demand}. Figure~\ref{fig:antigen_tests_vaccinations} shows
that the number of performed rapid tests in the model fits the empirical data well (where
empirical data is available).

In contrast to PCR tests, rapid tests are not perfect and can be falsely positive or
falsely negative. While the specificity of rapid tests is constant over time, their
sensitivity strongly depends on whether the tested individual is already infectious and
if so for how long he has been infectious. Before the onset of infectiousness the
sensitivity is very low (35\%). On the first day of infectiousness it is much higher
(88\%) but still lower than during the remaining infectious period (92\%). After
infectiousness stops, the sensitivity drops to 50 \%.

Modeling the diagnostic gap before and at the beginning of infectiousness is very
important to address concerns that rapid tests are too unreliable to serve as screening
devices.

We do not distinguish between self administered rapid tests and those that are
administered by medical personnel. While there were concerns that self administered tests
are less reliable, a recent study has found no basis for this concern.\comment[id=J]{cite
charite study on self administered tests}

While rapid tests do not directly enter official case numbers, many positively tested
individuals confirm their rapid test result with a PCR test.\comment[id=J]{Need to link
to data appendix once the number is there and maybe mention the number}. Importantly,
those PCR tests are made in addition to the tests that would have been done otherwise.
Section~\ref{subsec:appendix_share_known_cases} discusses the effect of rapid tests on
the share of detected cases.\comment[id=K]{Talk here about reactions to tests and
symptoms with contact reductions. (If not here, find other place where to talk about this
in detail.)}

\subsection{Seasonality}
\label{subsec:seasonality}

It is widely acknowledged that the transmission of SARS-CoV-2 is subject to seasonal
influences. Infectiousness is increased in winter when most contacts take place inside
and the immune system is weakened by low levels of vitamin D, dry air and large
temperature swings. For a detailed overview of possible drivers see
\cite{KronfeldSchor2021}.

We follow \cite{Kuehn2020} and \cite{Gavenciak2021} in modeling seasonality in the
transmission of SARS-CoV-2 as a multiplicative factor on infection probabilities. The
factor follows a sine curve that reaches its maximum at January first and its minimum
on June 30.

For simplicity we normalize the factor to reach one at its maximum. Thus the formula of
the seasonality factor is given by:

\begin{equation}
    s_k(t) = 1 + 0.5 \kappa_k  sin \left ( \pi  \left (\frac{1}{2} + \frac{t}{182.5}\right ) \right ) - 0.5 \kappa_k
\end{equation}

Where $\kappa_k$ is difference in the seasonality factor between peak infectiousness
and lowest infectiousness.

The subscript $k$ is needed because the strength of the seasonality effect differs
across contact types: Work, household and school contacts are likely to take place
inside even in summer. Thus they are only subject to seasonality due to factors that
influence the immune system. Other contacts (for example meeting friends and while doing
leisure activities) are mostly happening outside in the summer. Therefore, transmission
via those contacts should have a stronger seasonal pattern.

We calibrate $kappa_{strong}$ to 0.42 and $kappa_{weak}$ to 0.21. This is in line with
\cite{Gavenciak2021} and \cite{Kuehn2020}.

\subsection{Initial Conditions} 
\label{sub:initial_conditions}

Consider a situation where you want to start a simulation with the beginning set amidst
the pandemic. It means that several thousands of individuals should already have
recovered from the disease, be infectious, symptomatic or in intensive care at the start
of your simulation. Additionally, the sample of infectious people who will determine the
course of the pandemic in the following periods is likely not representative of the
whole population because of differences in behavior (number of contacts,
assortativity), past policies (school closures), etc.. The distribution of courses of
diseases in the population at the begin of the simulation is called initial conditions.

To come up with realistic initial conditions, we match reported infections from official
data to simulated individuals by available characteristics like age and geographic
information. The matching must be done for each day of a longer time frame like a month
to have individuals with possible health states. Then, health statuses evolve until the
begin of the simulation period without simulating infections by contacts. We also
correct reported infections for a reporting lag and scale them up to arrive at the true
number of infections.


\subsection{Estimated Parameters}
\label{subsec:estimated_params}


\FloatBarrier

We estimate parameters that cannot be calibrated outside of the model with the
method of simulated moments \citep{McFadden1989} by minimizing the distance between
simulated and observed infection rates and fatality rates (disaggregated by region
and age groups). Since our model includes a lot of randomness, we
average simulated infection rates over several model runs.

We fit our model to data for Germany from October 2020 until June 2021. We do not use
earlier periods for three reasons. Firstly, in the beginning PCR tests were highly
limited and therefore it would be difficult to find good initial conditions for our
simulations. In addition during the summer the case numbers were extremely low. This
could lead to the epidemic going extinct in our simulation. Additionally, our model does
not include international travel or other imports of cases. These would be important but
difficult to model during the summer months.


To avoid over-fitting and simplify the numerical optimization problem, we only allow for
five different infection probabilities: 1) for contacts in schools 2) contacts in
preschools and nurseries. 3) for work contacts. 4) for households. 5) for other
contacts.\comment[id=J]{Write that the exact numbers are hard to interpret and show the
infections by channel heatmap more prominently than the infection probabilities}

We also estimate a parameter that reflects the effect of hygiene measures
after November 2020 at work and in educational facilities. This parameter reduces
infectiousness of contacts by one third. In total those are 10 parameters. The
breakpoints the contact reduction changes are not determined from data but from
announcements of policy changes. Moreover, we constrain the estimated contact reduction
to follow the shape of the stringency index. The resulting contact reduction can be
seen in Figure~\ref{fig:other_multiplier}

Finally we estimate one parameter that governs the introduction of the B.1.1.7 virus
variant in January 2021. This parameter implies that at the end of January roughly one
case per 100 000 individuals per day is imported. After January we do not model imported
cases of B.1.1.7 anymore because they are negligible compared to the endogenous growth
of that virus variant.



\subsection{Shapley Values} 
\label{sub:shapley_value}

We decompose the effects of different NPIs and seasonality on the infection rates with
Shapley values. Shapley values \citep{Shapley2016} are a concept in game theory to
divide payoffs between a coalition of players. It allows to assign a single value to the
contribution of an NPI or seasonality which takes into account substitutional and
complementary effects with other factors.

More formally, define a coalitional game with $N$ players and a super-additive function
$\nu$ which maps subsets of $N$ to the real numbers. The function $\nu$ is also called
the characteristic function and assigns a value to a coalition. Then, the Shapley value
$\phi$ for player $i$ is

\begin{align*}
    \phi_i(\nu) = \frac{1}{|N| !} \sum_{S \subseteq N \setminus \{i\}} |S| ! (|N| - |S| - 1)! (\nu(S \cup \{i\}) - \nu(S))
\end{align*}

The last term $(\nu(S \cup \{i\}) - \nu(S))$ is the marginal contribution of player $i$
minus the coalition without player $i$. Then, compute the sum of marginal contributions
over all subsets $S$ of $N$ which do not include player $i$. Each marginal contribution
has to be multiplied by all combinations of other players in $S$ which precede $i$ and
all possible combinations of remaining players which follow player $i$ in the coalition.
To arrive at the Shapley value for player $i$, divide the sum by the total number of
combinations.

The Shapley value has some properties.

\begin{description}
  \item[Efficiency] The sum of Shapley values is equal to the value of a coalition
  formed by all players.
  \item[Symmetry] The Shapley does not depend on the label of a player but only on its
  position in the characteristic function.
  \item[Linearity] The Shapley value depends linearly on the values from the
  characteristic function $\nu$.
  \item[Dummy Axiom] The Shapley value of a player who contributes nothing to any
  coalition is 0.
\end{description}

To produce Figure~\ref{fig:2021_scenarios_decomposition} and
Figure~\ref{fig:2021_scenarios_decomposition_tests}, we calculate the Shapley values of
each factor in the comparison on the cumulative number of saved infections between the
main scenario and the scenario without any of the factors for every day. Then, we divide
up the saved infections on a particular day according to the Shapley values for the same
day which yields the daily saved infections for each factor.


    \section{Additional Results}
\label{sec:additional_results}

\subsection{Simulated vs. Empirical Data}
\label{subsec:fit_results}

This compares simulated data from our model with empirical data from Germany. We look at
observed infections, fatality rates, the spread of the B117 mutation, vaccinations and
rapid test demands. Where available we do not only look at aggregated statistics but also
analyze the model fit for age groups and federal states.\comment[id=J]{summarize the fit}

\begin{figure}[ht]
  \centering
  \includegraphics[width=\textwidth]{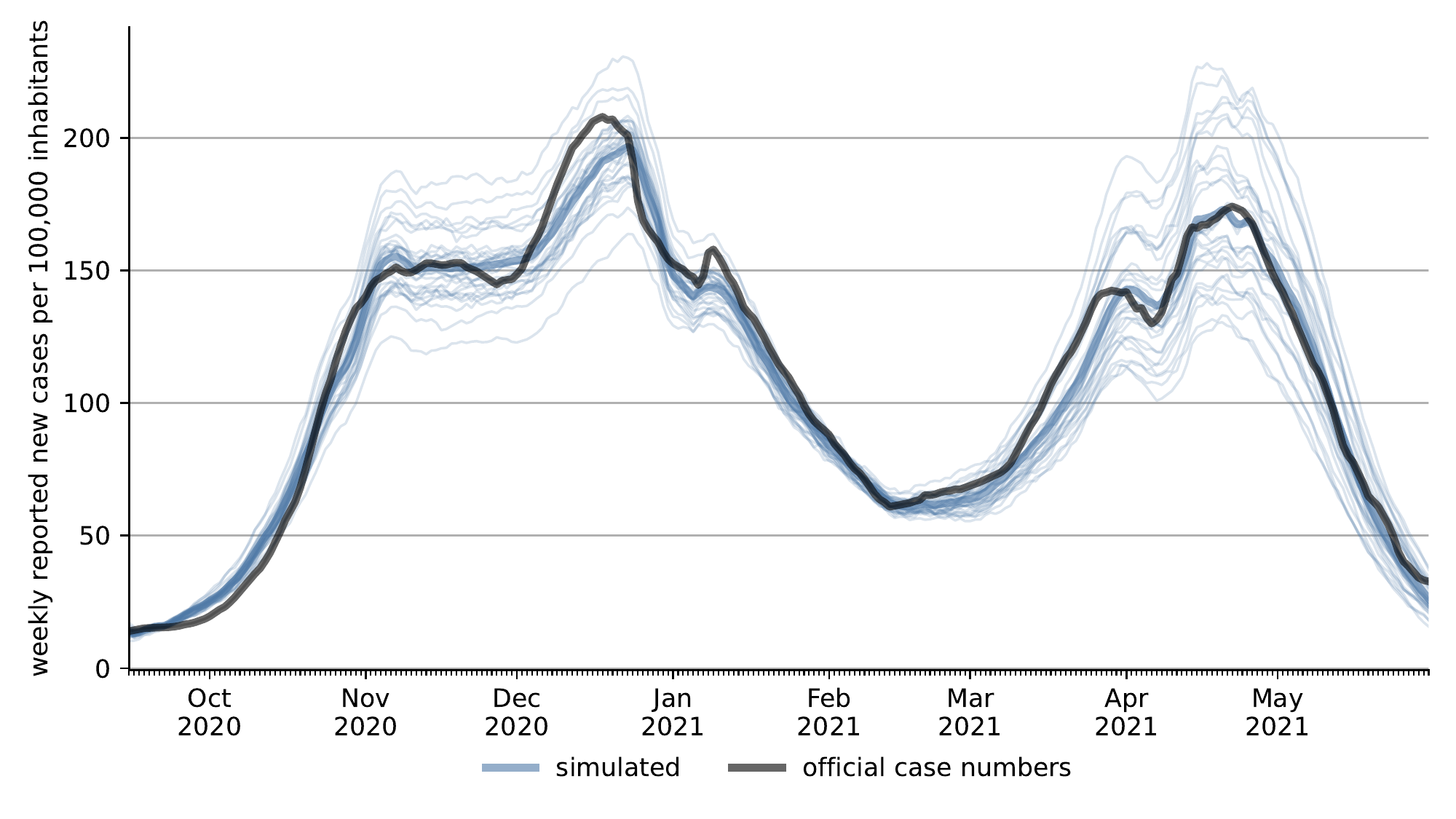}
  \caption{Fit Over the Full Simulation Time Frame with Single Simulation Runs}
  \floatfoot{\noindent \textit{Note:} The figure shows the weekly incidence rates per
  100,000 people for the reported simulated infections rates. The mean infection rate is
  the thick blue line. Single simulation runs are plotted in lighter and thinner lines.
  The official case numbers as reported by the Robert-Koch-Institut are plotted in black.
  The fit is overall very good. The higher the mean incidence and the stronger the growth
  the more variance there is between simulation runs. We averaged over 30 simulation
  runs.}
  \label{fig:aggregated_fit2}
\end{figure}

\begin{figure}[ht]
  \centering
  \includegraphics[width=\textwidth]{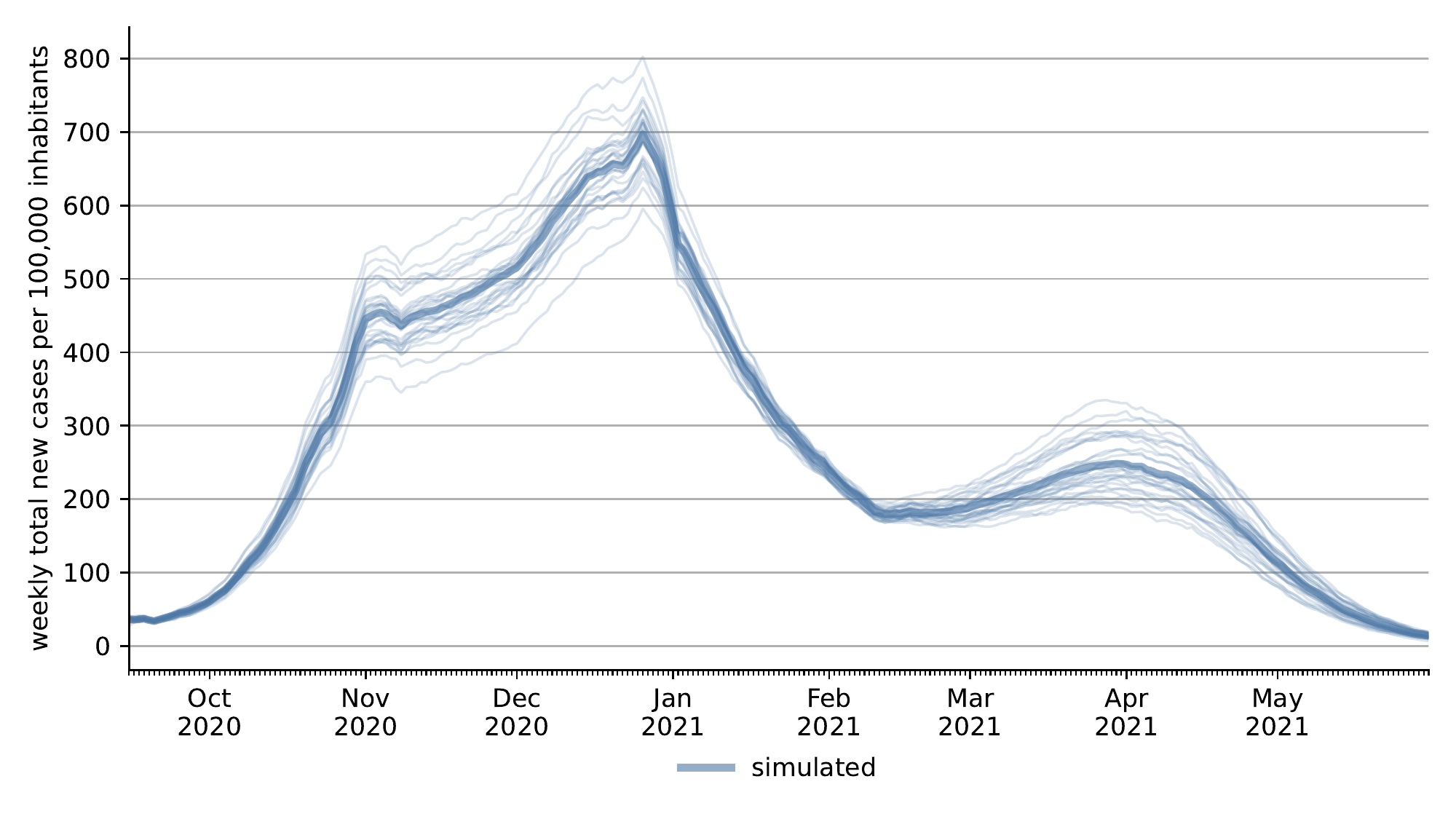}
  \caption{Development of the Total Infections Over the Full Simulation Time Frame with
  Single Simulation Runs}
  \floatfoot{\noindent \textit{Note:} The figure shows the true weekly incidence rates
  per 100,000 people, including undetected cases. The mean infection rate is the thick
  blue line. Single simulation runs are plotted in lighter and thinner lines. The higher
  the mean incidence and the stronger the growth the more variance there is between
  simulation runs. We averaged over 30 simulation runs.}
  \label{fig:newly_infected_in_baseline}
\end{figure}

\begin{figure}[ht]
  \centering
  \includegraphics[width=\textwidth]{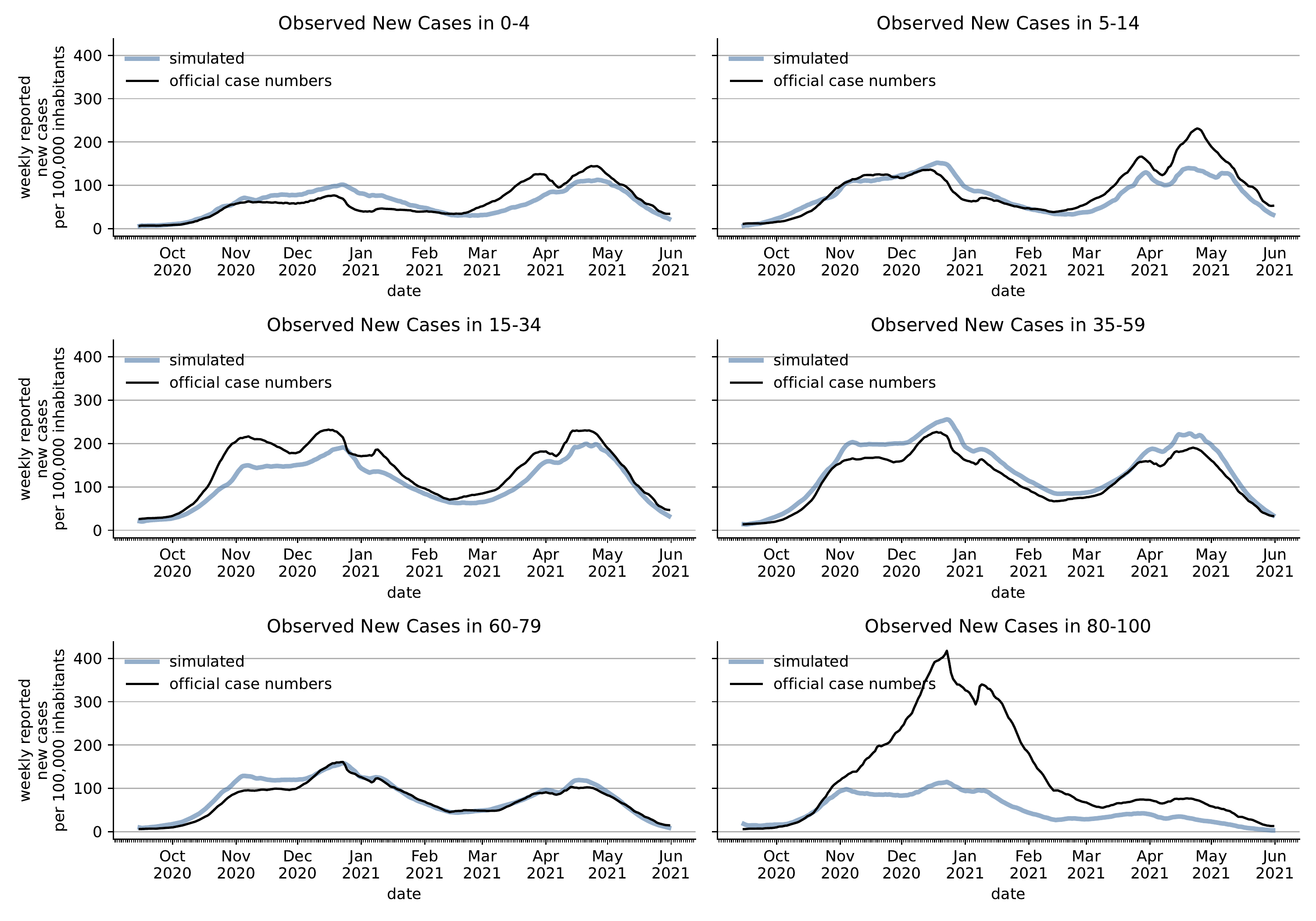}
  \caption{Simulated and Empirical Infections by Age Group}
  \floatfoot{\noindent \textit{Note:} The figure shows the weekly incidence rates per
  100,000 people for the reported versus the simulated infections rates for different age
  groups. The age group of individuals above 80 needs to be interpreted with caution
  because our synthetic population only includes private households, i.e. nursing homes
  are not represented in our model. They accounted for many cases and deaths in the
  winter of 2020 and many 80 to 100 year olds live in these facilities. However, the
  official data does not contain information on whether cases were nursing home
  inhabitants or not. We averaged over 30 simulation runs.}
  \label{fig:age_group_fit}
\end{figure}

\begin{figure}[ht]
  \centering
  \includegraphics[width=\textwidth]{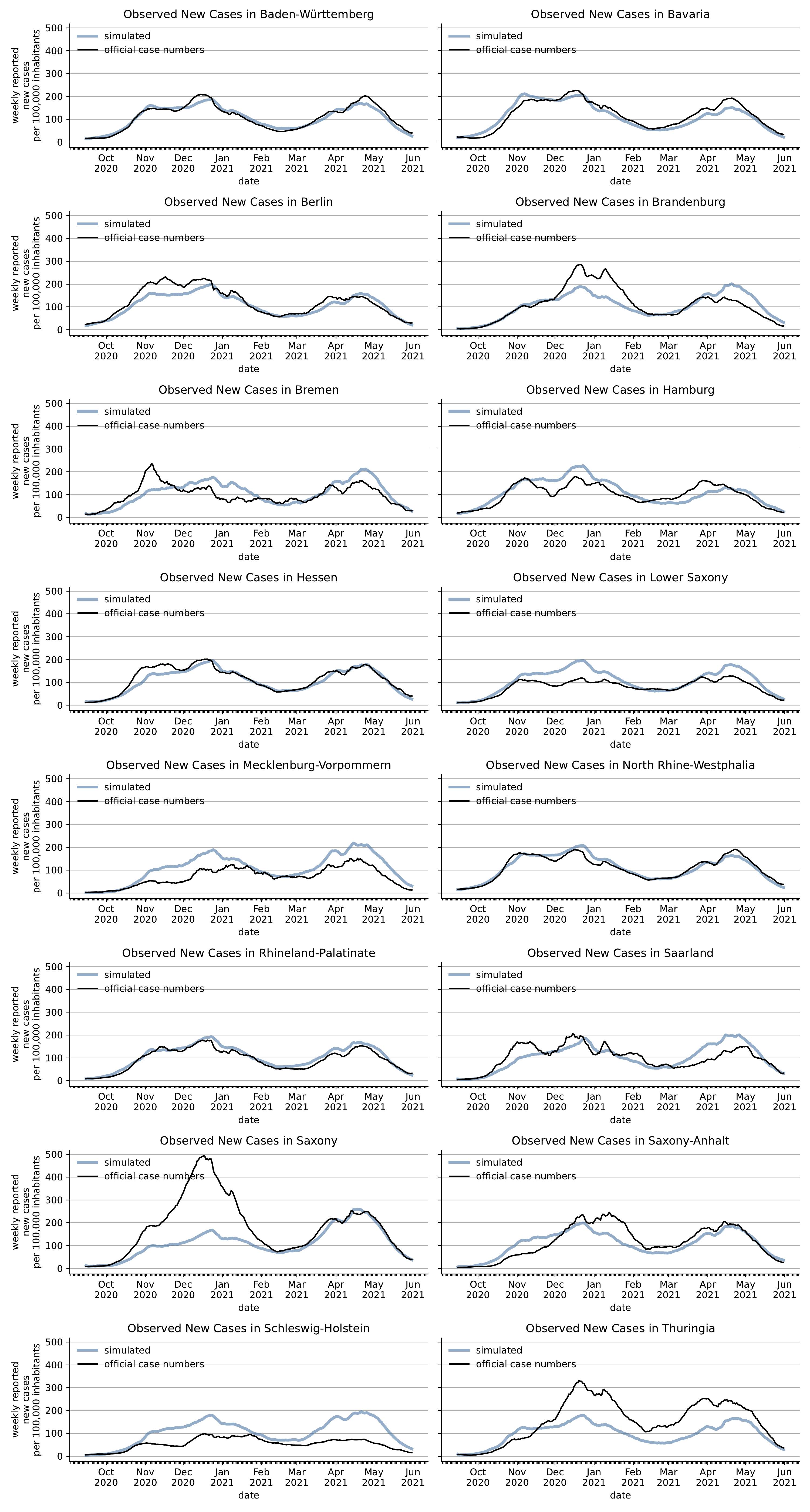}
  \caption{Simulated and Empirical Infections by Federal State}
  \floatfoot{\noindent \textit{Note:} The figure shows the weekly incidence rates per
  100,000 people for the reported versus the simulated infections rates for different
  federal states. We averaged over 30 simulation runs.}
  \label{fig:state_fit}
\end{figure}

\begin{figure}[ht]
  \centering
  \includegraphics[width=\textwidth]{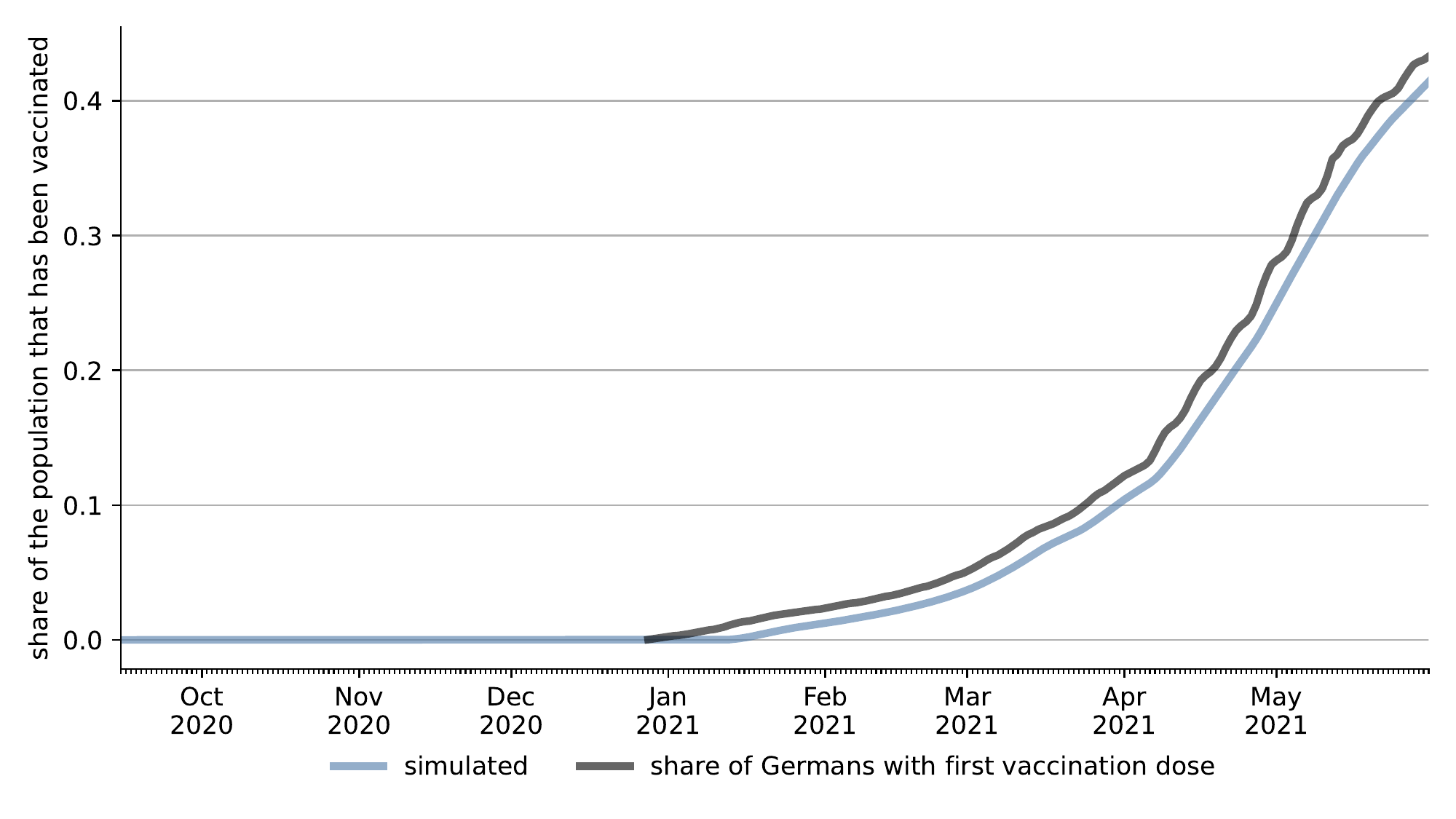}
  \caption{Share of Vaccinated Individuals}
  \floatfoot{\noindent \textit{Note:} The figure shows the rate of individuals that are
  vaccinated in our synthetic population versus in the general German population. Note
  that we excluded the vaccinations that were given to nursing homes, approximately the
  first percent of the German population that were vaccinated. Overall, our model covers
  a time frame that goes from zero vaccinated individuals to a state where over 40\% of
  the population are vaccinated. Our vaccinations work imperfectly but we do not model
  different vaccines nor do we distinguish between first and second shot.}
  \label{fig:fit_vaccinations}
\end{figure}

\begin{figure}[ht]
  \centering
  \includegraphics[width=\textwidth]{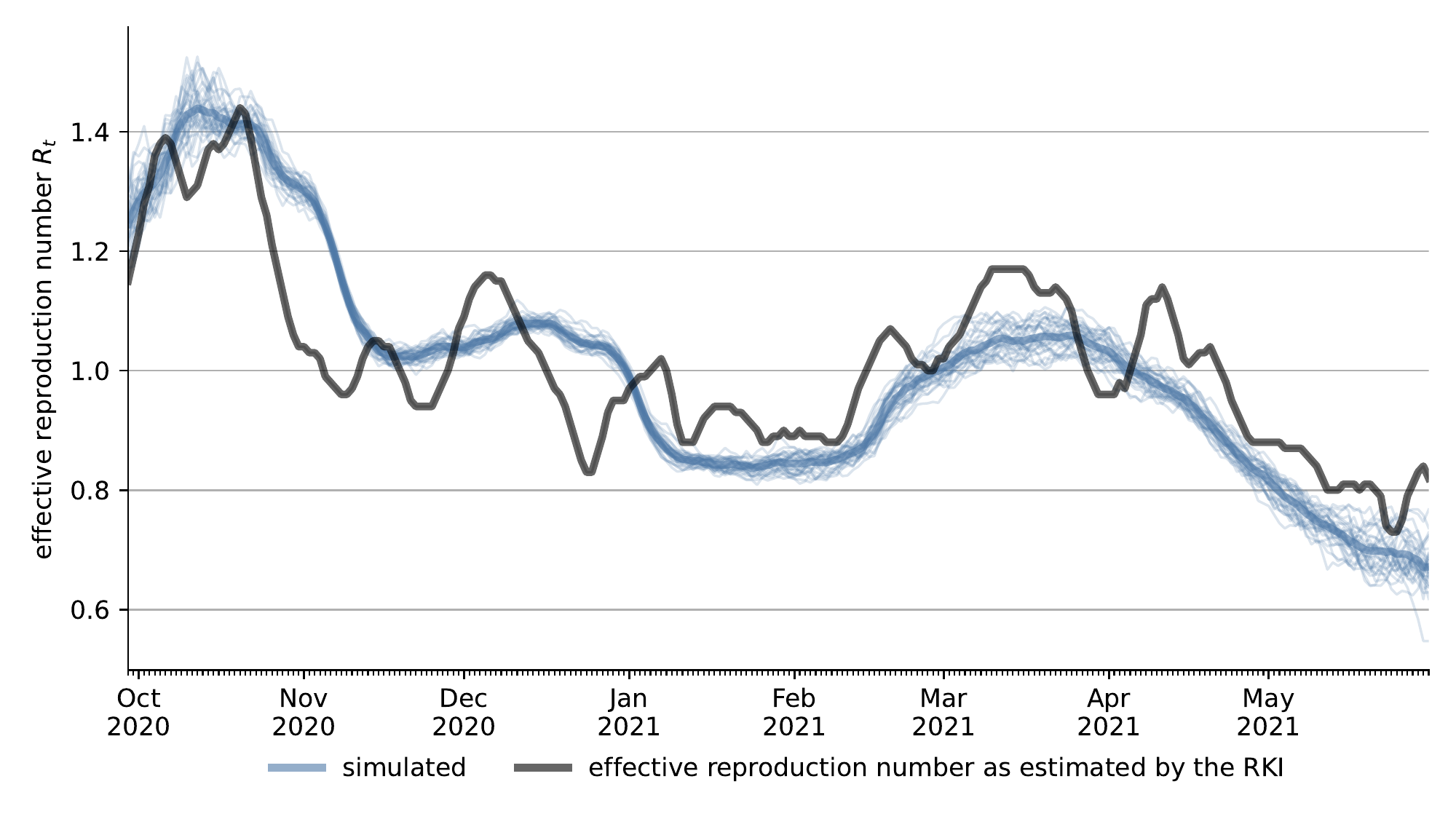}
  \caption{Effective Replication Number $R_t$ in the Model and as Reported by the
  Robert-Koch-Institute}
  \floatfoot{\noindent \textit{Note:} The figure shows the effective replication number
  ($R_t$) as reported by the RKI and as calculated in our model. The $R_t$ gives the
  average number of new infections caused by one infected individual. The $R_t$ in our
  model broadly follows the $R_t$ reported by the RKI. Two trends stand out. Firstly, the
  RKI's $R_t$ drops faster in November. This could be due to a change in the testing
  policy that focused tests on the elderly when the second wave hit Germany and led to a
  decline in the overall share of detected cases. The second difference is from mid
  February to mid March where the RKI's reported $R_t$ increased more rapidly than that
  in our model. Here the opposite effect can be expected. During this time rapid tests
  increased strongly leading to more cases being detected. In the short term this leads
  an $R_t$ estimation that is based on detected cases to overestimate the replication
  number.}
  \label{fig:fit_r_effective}
\end{figure}

\begin{figure}[ht]
  \centering
  \includegraphics[width=\textwidth]{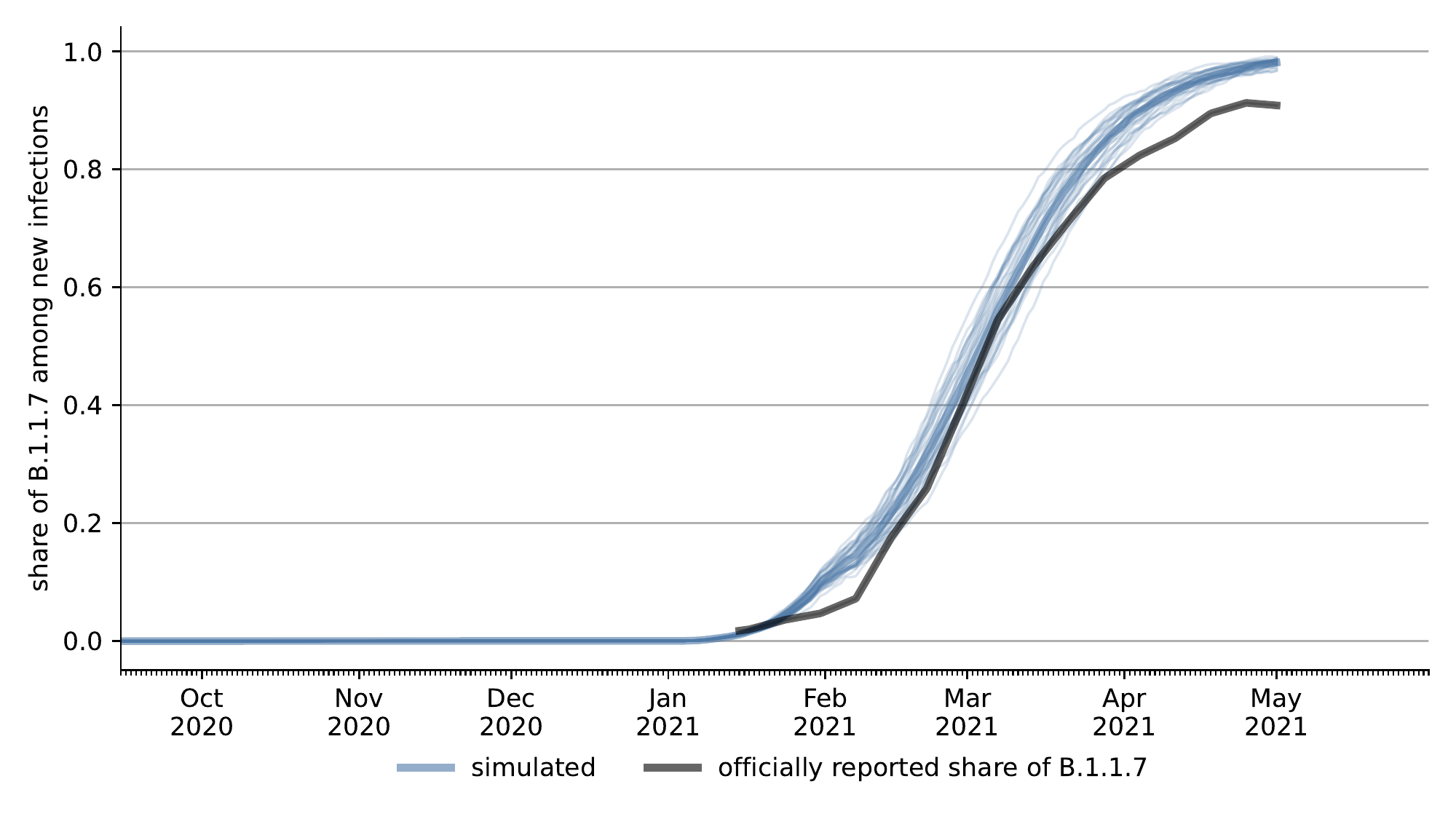}
  \caption{Share of B.1.1.7 in the Model and as Reported by the Robert-Koch-Institute}
  \floatfoot{\noindent \textit{Note:} The figure shows the share of B.1.1.7 as
  reported by the RKI and as calculated in our model. We only introduce a few cases over
  the cause of January. From then B.1.1.7 takes over endogenously through its increased
  infectiousness. We model no other features of B.1.1.7. At most we introduce 0.75 cases per 100,000 inhabitants.}
  \label{fig:fit_share_b117}
\end{figure}

\FloatBarrier

\subsection{Share of Cases that are Detected}
\label{subsec:appendix_share_known_cases}

\begin{figure}[ht]
  \centering
  \includegraphics[width=\textwidth]{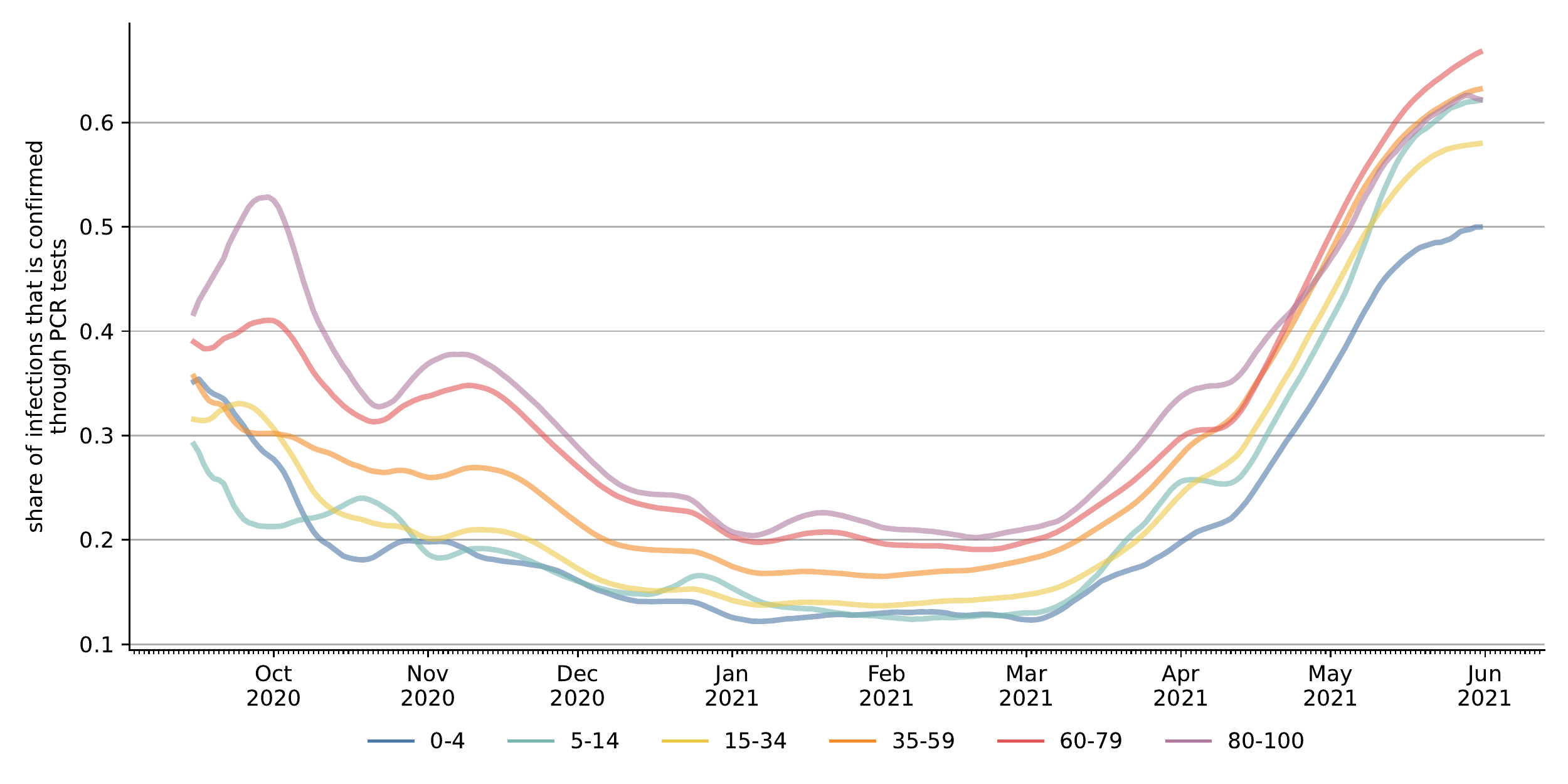}
  \caption{Share of Detected Cases by Age Group}
  \label{fig:share_known_cases_by_age_group}
  \floatfoot{\noindent \textit{Note:} The figure shows the share of cases that is
  reported as an official case via PCR confirmation. We use the overall share of known
  cases that was estimated through the case fatality ratio by the
  \href{https://covid19.dunkelzifferradar.de/}{Dunkelzifferradar} for all of 2020 and
  then assume it to be constant as vaccinations of the elderly strongly affect the case
  fatality rate which the project does not account for. To get from an overall share of
  detected cases to the share of cases that is detected in each age group we use that
  asymptomatic cases are much less likely to be detected. As our model covers age
  specific asymptomatic rates this endogenously leads to group specific share known cases
  that verify that infections in younger age groups are under-detected. Starting in 2021
  in addition to the overall numbers of detected cases through symptoms and the share
  known cases, cases are also detected through confirmation of positive rapid tests. This
  leads to an increase in the share of known cases for all age groups but in particular
  for the younger age groups that are covered extensively with rapid tests through the
  rapid test requirement for participating in school.}
\end{figure}

It's noteworthy that the share of detected cases increases rapidly in May for the five to
fourteen year olds. This is a direct result of the mandatory tests in
school.

\subsection{Rapid Tests}
\label{subsec:appendix_rapid_tests}

\begin{figure}[ht]
  \centering
  \begin{subfigure}[b]{0.425\textwidth}
    \centering
    \includegraphics[width=\textwidth]{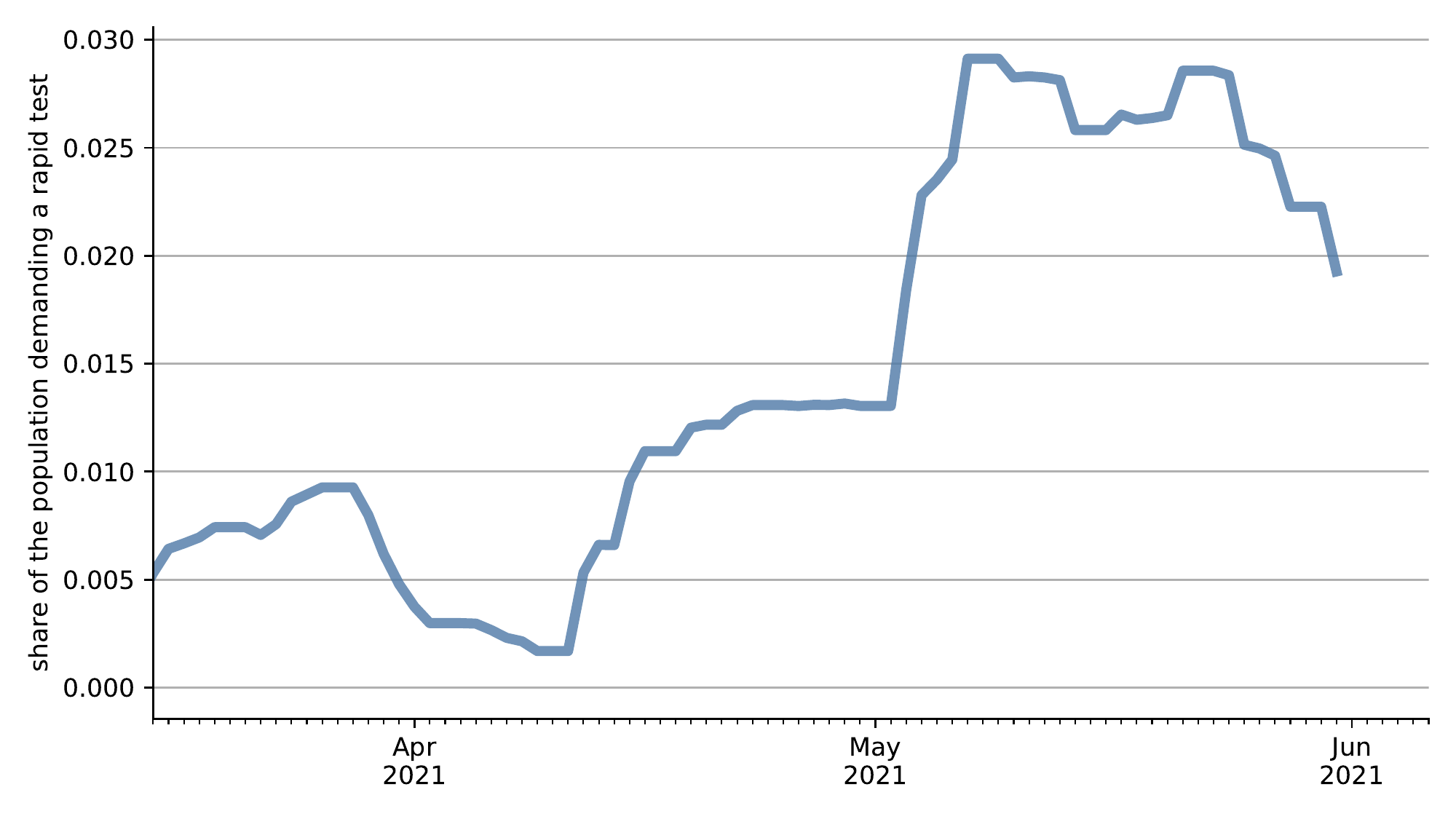}
    \caption{Share of the Population Demanding a Rapid Test in an Education Setting}
    \label{fig:educ_rapid_test_demand}
  \end{subfigure}
  \hfill
  \begin{subfigure}[b]{0.425\textwidth}
    \centering
    \includegraphics[width=\textwidth]{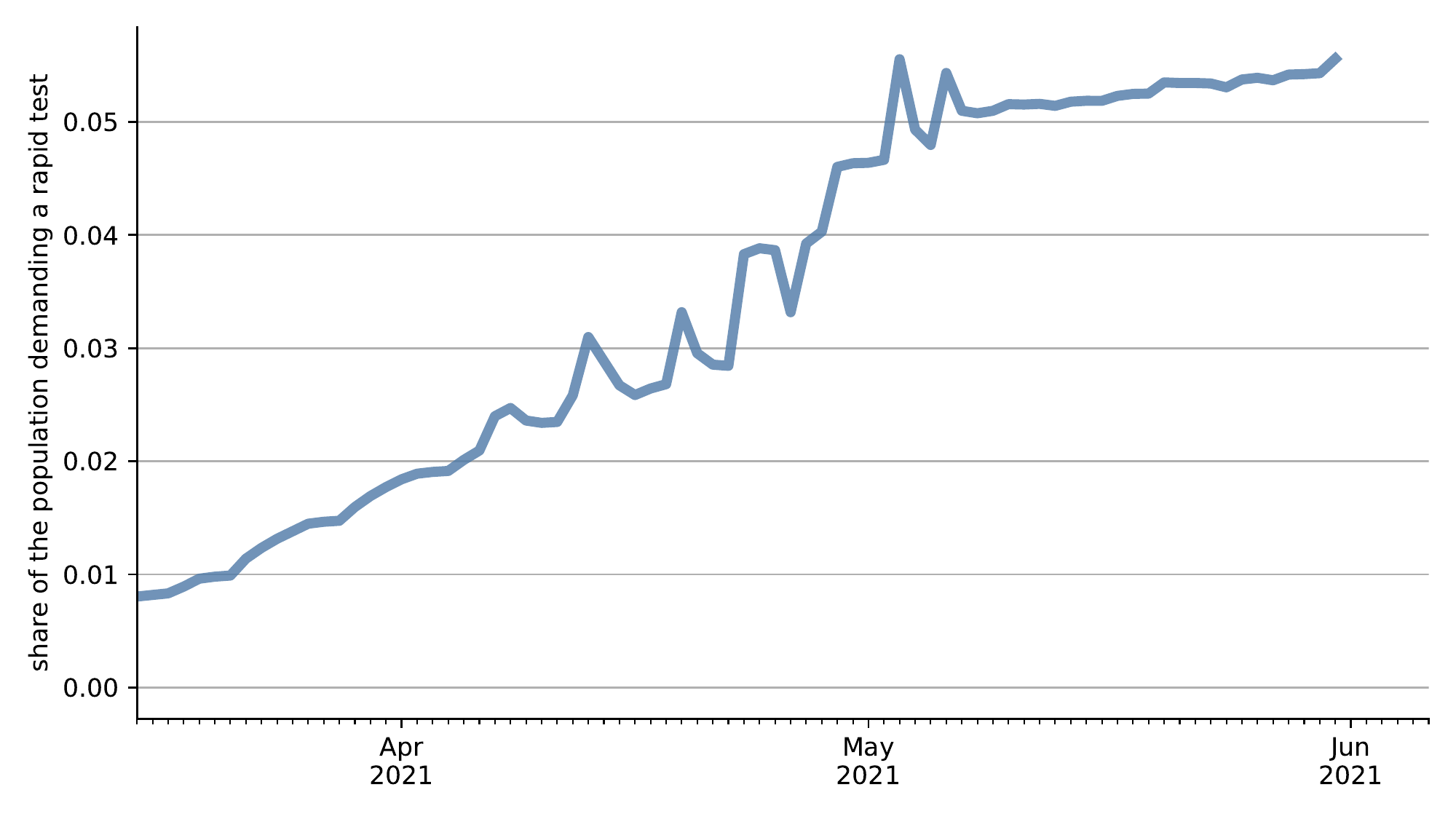}
    \caption{Share of the Population Demanding a Rapid Test due to Work}
    \label{fig:work_rapid_test_demand}
  \end{subfigure}
  \hfill
  \begin{subfigure}[b]{0.425\textwidth}
    \centering
    \includegraphics[width=\textwidth]{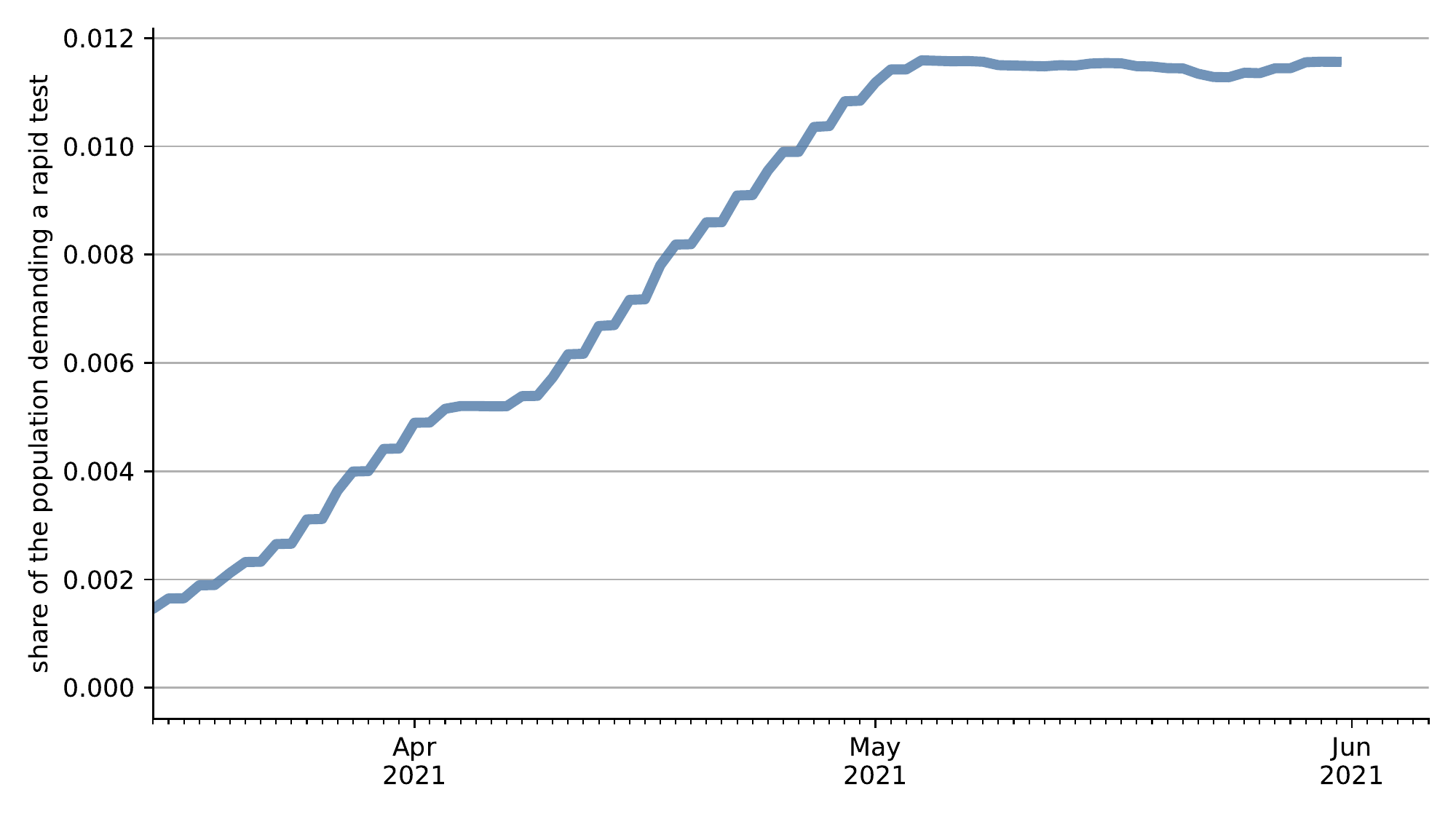}
    \caption{Share of the Population Demanding a Rapid Test for Private Reasons}
  \label{fig:private_rapid_test_demand}
  \end{subfigure}
  \caption{Rapid Test Demand by Reason}
  \label{fig:rapid_test_demand_by_reason}
  \floatfoot{\noindent \textit{Note:}
    Rapid tests in the education setting are demanded by teachers (nursery, preschool and
    school) as well as school pupils. After Easter the required frequency of tests is
    increased from once per week to twice per week. Work rapid tests are demanded by
    individuals that still have work contacts, i.e. do not work from home. The share of
    employers offering rapid tests increases over the time frame and the frequency of
    testing is also increased. Tests are demanded by individuals for one of three private
    reasons: having developed symptoms without access to a PCR test, having a household
    member that has tested positive or developed symptoms or having planned weekly
    meeting with friends.
  }
\end{figure}

\begin{figure}[ht]
  \centering
  \begin{subfigure}[b]{0.425\textwidth}
    \centering
    \includegraphics[width=\textwidth]{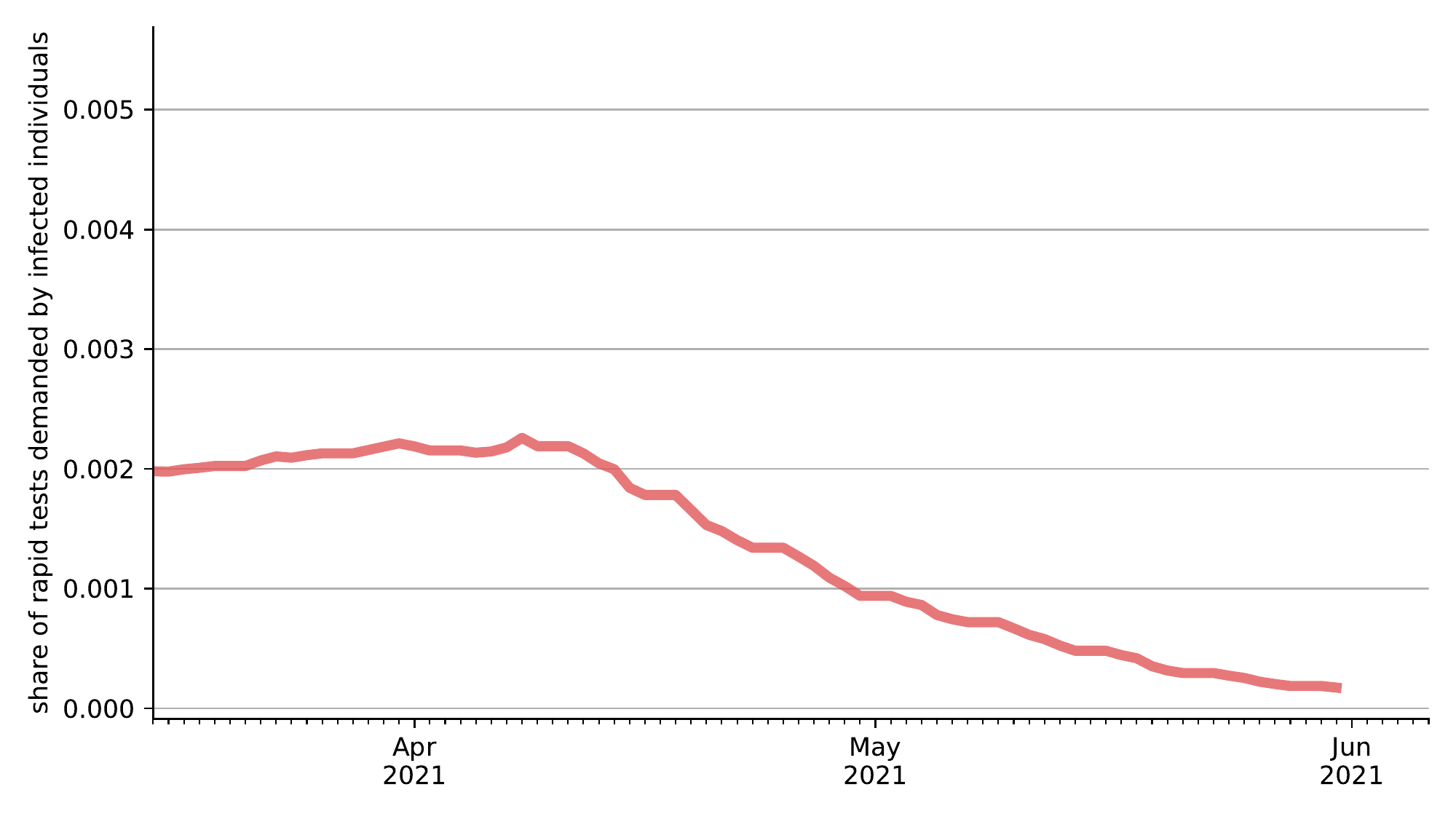}
    \caption{Share of Rapid Tests in the Educational Setting Demanded by Infected
    Individuals}
    \label{fig:share_infected_among_educ_rapid_test_demand}
  \end{subfigure}
  \hfill
  \begin{subfigure}[b]{0.425\textwidth}
    \centering
    \includegraphics[width=\textwidth]{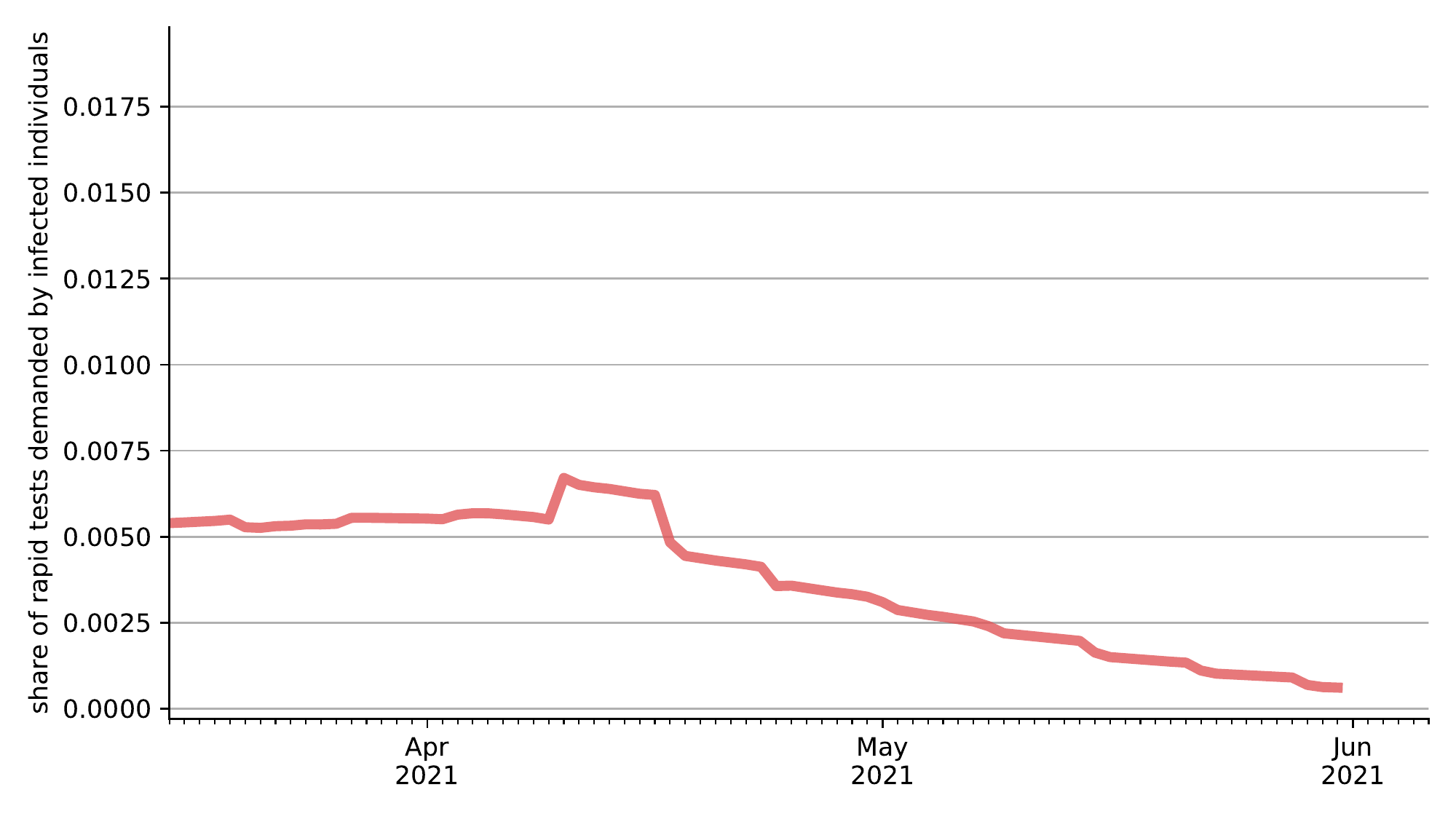}
    \caption{Share of Work Rapid Tests Demanded by Infected Individuals}
    \label{fig:share_infected_among_work_rapid_test_demand}
  \end{subfigure}
  \hfill
  \begin{subfigure}[b]{0.425\textwidth}
    \centering
    \includegraphics[width=\textwidth]{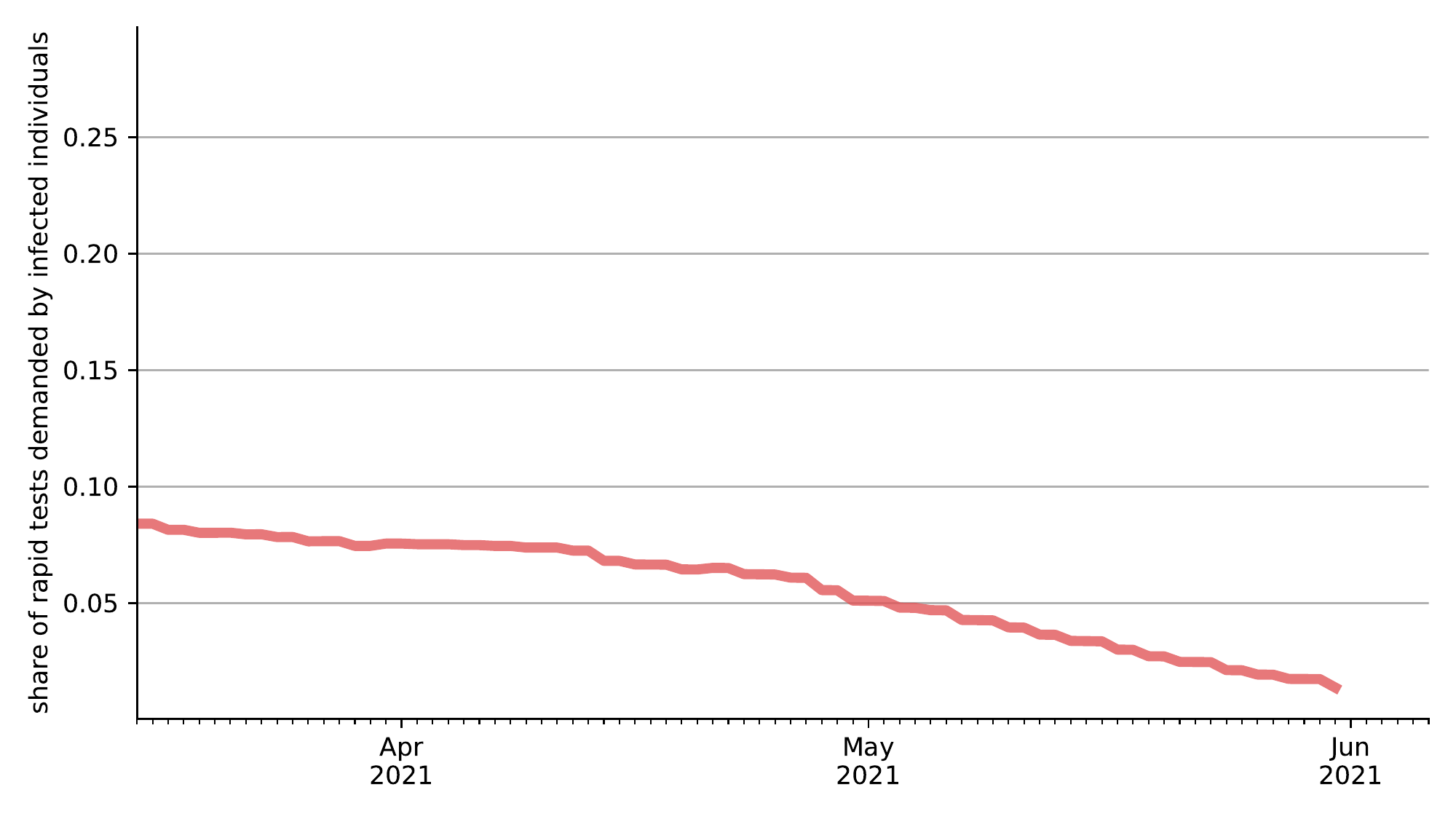}
    \caption{Share of Private Rapid Tests Demanded by Infected Individuals}
  \label{fig:share_infected_among_private_rapid_test_demand}
  \end{subfigure}
  \caption{Share of Rapid Tests Demanded by Infected Individuals by Reason}
  \label{fig:share_infected_among_rapid_test_demand_by_reason}
  \floatfoot{\noindent \textit{Note:}
    Rapid tests in the education setting are demanded by teachers (nursery, preschool and
    school) as well as school pupils. After Easter the required frequency of tests is
    increased from once per week to twice per week. Work rapid tests are demanded by
    individuals that still have work contacts, i.e. do not work from home. The share of
    employers offering rapid tests increases over the time frame and the frequency of
    testing is also increased. Tests are demanded by individuals for one of three private
    reasons: having developed symptoms without access to a PCR test, having a household
    member that has tested positive or developed symptoms or having planned weekly
    meeting with friends. Private rapid tests have a much higher share of infected
    individuals because they are mostly triggered by events that make an infection
    likely. Remember however, that one reason is that a household member has a positive
    rapid test. This means that work and education rapid tests which have a low rate of
    infected individuals trigger more targeted rapid tests through the household demand.
    They also have a much higher volume of tests. As can be seen in the decomposition
    (Figure~\ref{fig:2021_scenarios_decomposition_tests} every category of tests is
    important for the overall effect. It appears that the combination of wide testing to
    find infection chains plus targeted tests to break those chains taking together
    is important.}
\end{figure}

\FloatBarrier

\subsection{Scenarios}
\label{subsec:appendix_scenarios}

\begin{figure}[ht]
  \centering
  \begin{subfigure}{.6\textwidth}
    \includegraphics[width=0.9 \textwidth]{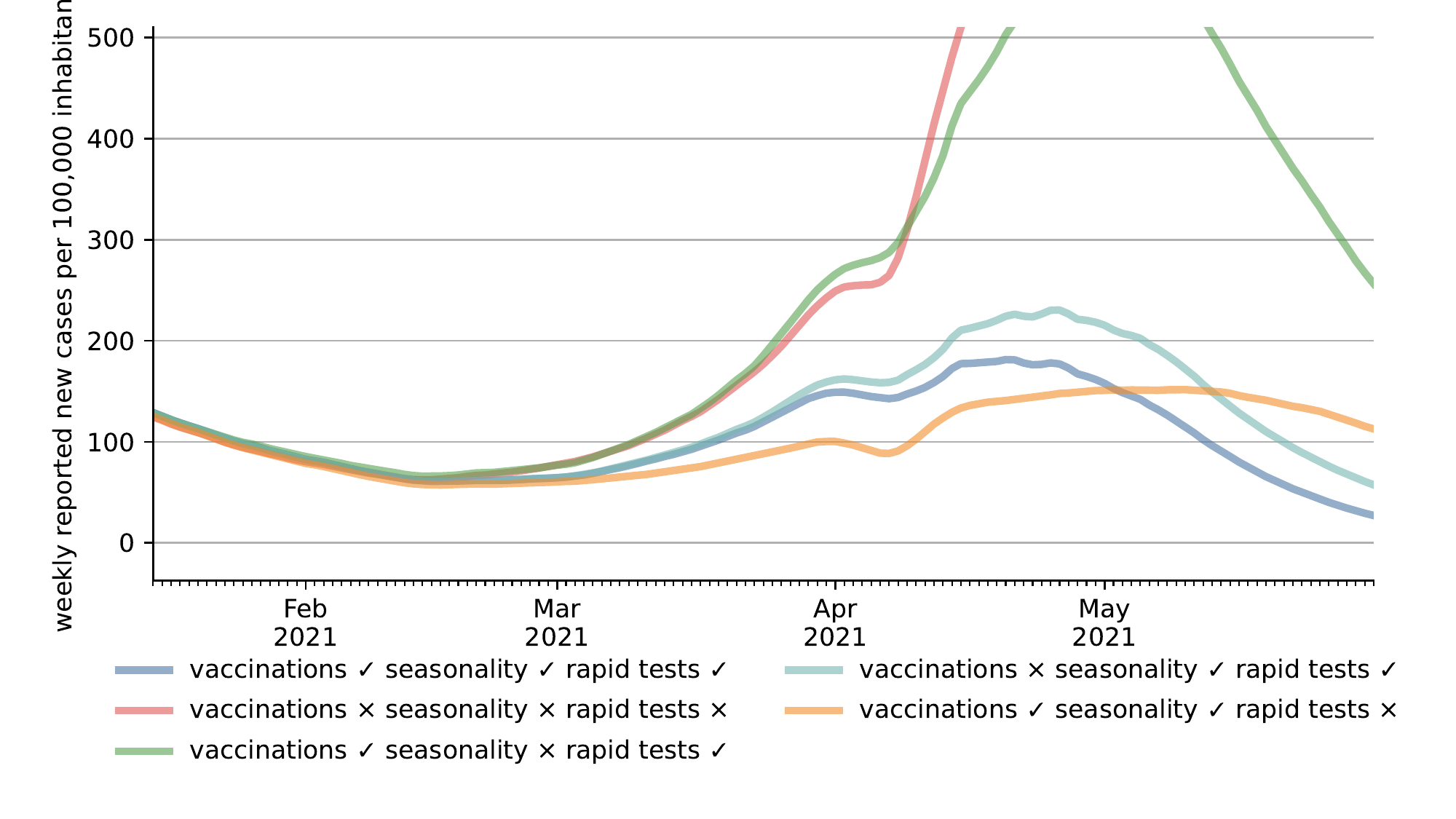}
  \end{subfigure}%
  \begin{subfigure}{.6\textwidth}
    \includegraphics[width=0.9 \textwidth]{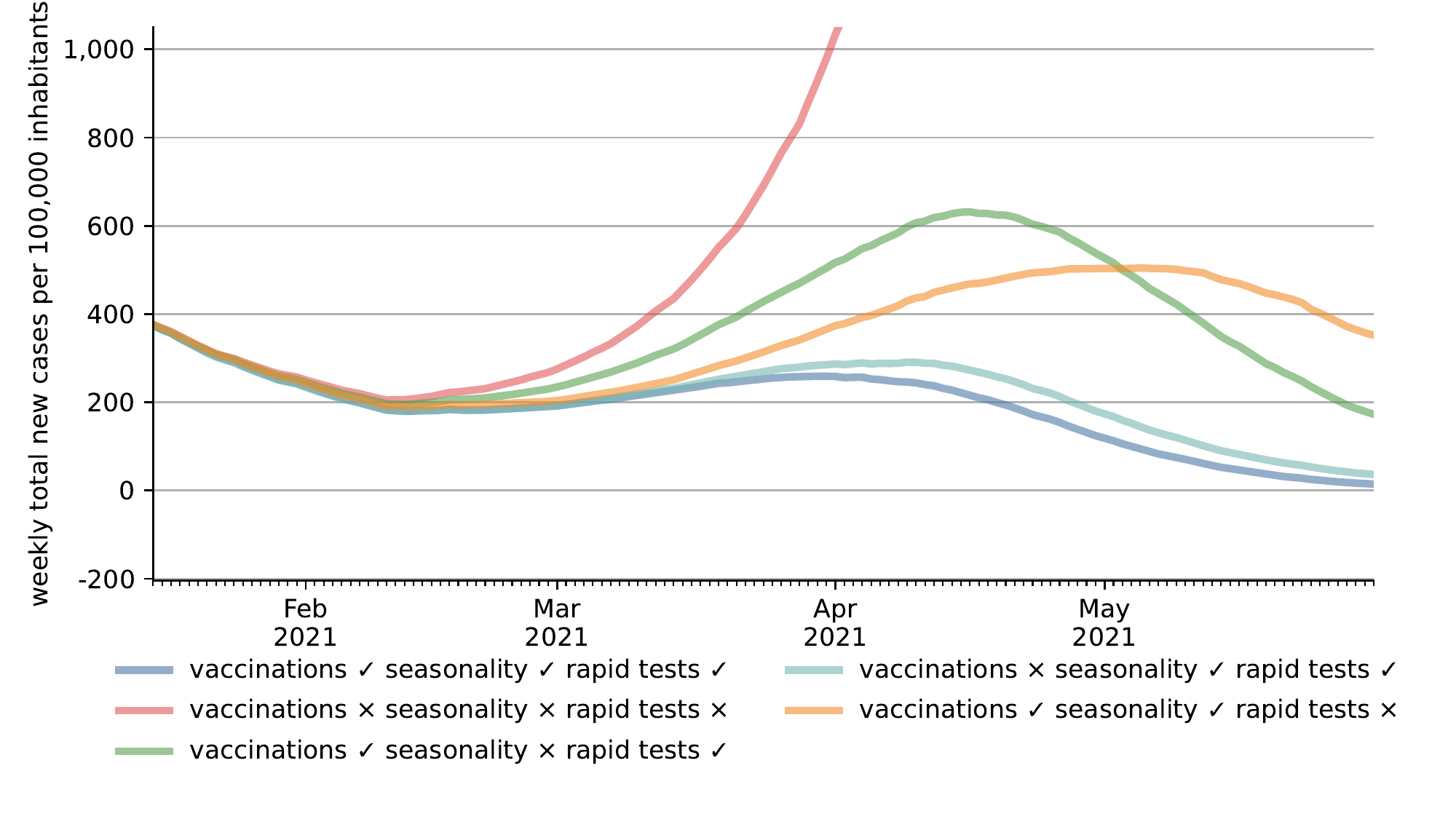}
  \end{subfigure}
  \caption{The Effect of Policies on Observed and Unobserved Cases}
  \label{fig:explain_decline}
  \floatfoot{\noindent \textit{Note:} \ldots}
\end{figure}

\begin{figure}[ht]
  \centering
  \begin{subfigure}{.6\textwidth}
    \includegraphics[width=0.9 \textwidth]{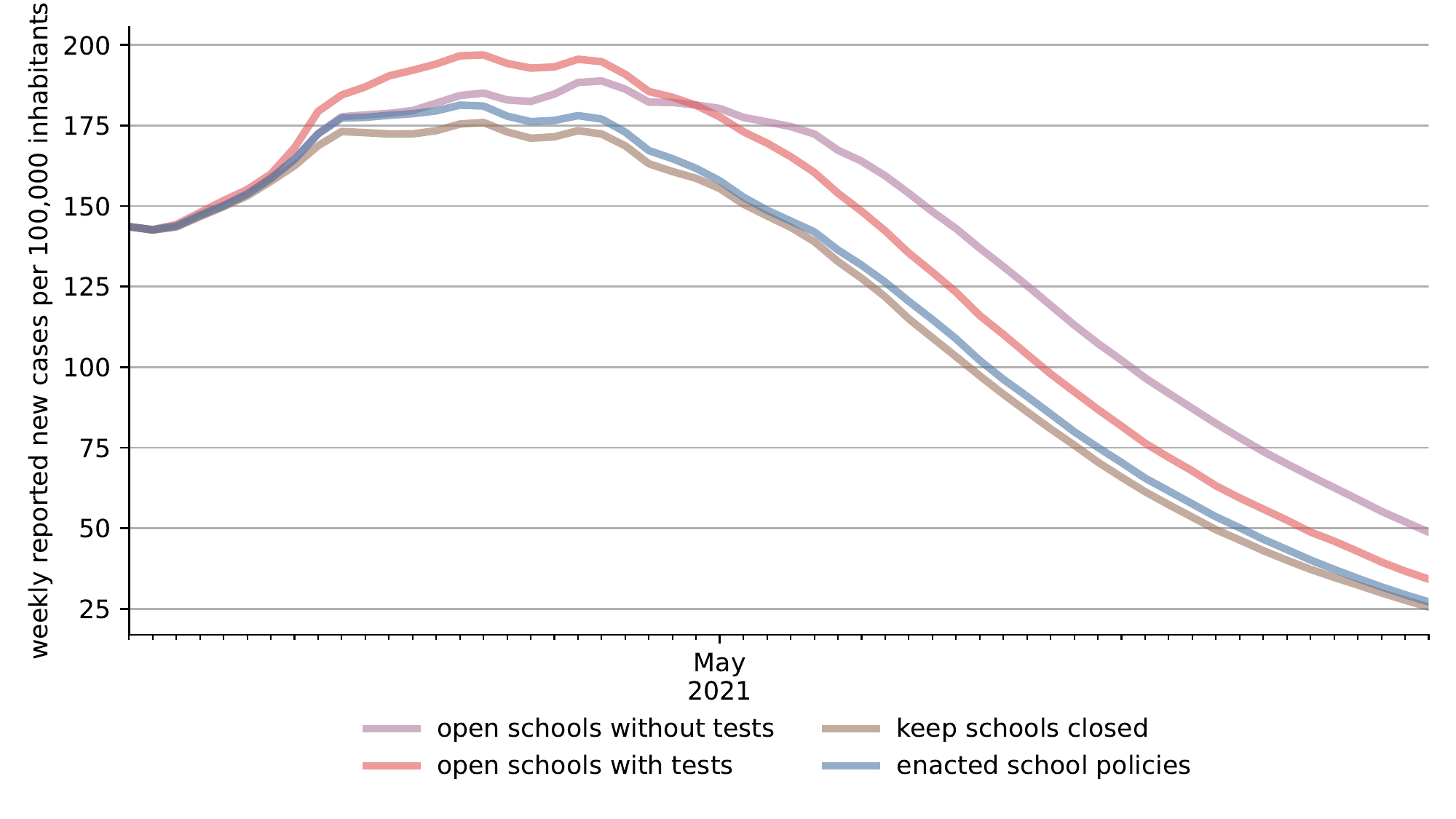}
  \end{subfigure}%
  \begin{subfigure}{.6\textwidth}
    \includegraphics[width=0.9 \textwidth]{figures/results/figures/scenario_comparisons/school_scenarios/full_newly_infected}
  \end{subfigure}
  \caption{The Effect of Different School Scenarios on Observed and Unobserved Cases}
  \label{fig:school_scenarios_detailed}
\end{figure}

\FloatBarrier

    \printbibliography[heading=bibintoc]

    \end{refsection}
\end{appendices}

\end{document}